\documentclass{aa}  

\usepackage{graphicx}
\usepackage{txfonts}
\usepackage{xcolor}
\usepackage{hyperref}
\newcommand{\mbol}{m_{\mathrm{bol}}}
\newcommand{\JE}{J_{\rm E}}
\newcommand{\BCJ}{\mathrm{BC}_J}
\newcommand{\flam}{f_\lambda}\newcommand{\fnu}{f_\nu}
\begin{document} 

   \title{Euclid DR1 Legacy Science on Ultracool Dwarfs: New URLD candidates in Euclid and SPHEREx}
%
   %\subtitle{}
%
   \author{N.\ Lodieu \inst{1,2}
       \and
       M.\ {\v{Z}}erjal \inst{1,2}
       \and
       N.\ Vitas \inst{1,2}
       \and
       C.\ Dom\'inguez Tagle \inst{1,2}
       \and
       E.\ L.\ Mart\'in \inst{1,2}
       \and
       J.-Y.\ Zhang \inst{3}
       \and
       S.\ Tsilia \inst{1,2}
        }

    \institute{Instituto de Astrof\'isica de Canarias (IAC), Calle V\'ia L\'actea s/n, E-38200 La Laguna, Tenerife, Spain \\
    \email{nlodieu@iac.es}
    \and
    Departamento de Astrof\'isica, Universidad de La Laguna (ULL), E-38206 La Laguna, Tenerife, Spain
    \and
    Department of Physics and Astronomy, The University of Western Ontario, 1151 Richmond St, London, ON N6A 3K7, Canada
       }

   \date{\today{},\today{}}

% \abstract{}{}{}{}{} 
% 5 {} token are mandatory
 
  \abstract
   {Unusually red L dwarfs (URLDs) probe extreme atmospheric conditions in the substellar regime, where enhanced dust opacity, cloud structure, and surface gravity effects challenge standard brown dwarf models. As free-floating analogues of young giant exoplanets, they provide key benchmarks for interpreting low-resolution infrared spectra from current and future wide-field space missions.
   }
   {The goal is to identify new red L dwarf candidates both photometrically and spectroscopically in the first data release (DR1) of the {\it{Euclid}} space mission. The area covered is about 1,900 and 50 square degrees in the wide and deep survey, respectively. 
   }
   {We used colour cuts in multiple colour-colour diagrams combined with point-source and quality criteria to identify candidate red L dwarfs in both the {\it{Euclid}} wide and deep survey fields. We cross-matched photometric candidates with the SPHEREx mission and {\it{Euclid}} red grism spectra to confirm spectroscopic properties and evaluate bolometric magnitudes and colours.
   }
   {We identified 1202 and 38 red L dwarf candidates in the {\it{Euclid}} wide and deep surveys, respectively. Spectroscopic follow-up and archival cross-matching confirm a substantial new population of red L dwarfs. Distance-independent diagnostics show that these objects occupy distinct loci characterised by enhanced condensate opacity and significant redness.  
   }
   {{\it{Euclid}} vastly expands the census of unusually red L dwarfs. While preliminary spectral classifications suggest many of these new candidates are intermediate-age (Pleiades-like) objects, a subset exhibits extremely red signatures indicative of very young, low-gravity ($\leq$\,10\,Myr) atmospheres, providing vital benchmarks for young exoplanet research.}

               
\maketitle

%
%%%%%%%%%%%%%%%%%%%%%%%%%%%%%%%%%%%%%%%%%%%%
%%%%% Introduction %%%%%
%%%%%%%%%%%%%%%%%%%%%%%%%%%%%%%%%%%%%%%%%%%%
%
\section{Introduction}
\label{redLdwarfs:intro}

Ultracool dwarfs, spanning spectral types from late-M through L, T, and Y \citep{kirkpatrick99,martin99a,burgasser02,kirkpatrick05,cushing11}, represent the lowest-mass products of star formation, bridging the gap between hydrogen-burning stars and giant planets \citep{chabrier00a,burrows01,baraffe03}. These objects have effective temperatures below about 2500 K \citep{basri00,golimowski04a,stephens09,filippazzo15} and emit most of their flux in the infrared. Wide-field optical and infrared surveys have uncovered a large number of field L, T, and Y dwarfs \citep{delfosse97,kirkpatrick00,kirkpatrick11}, and have led to
well-characterised volume-limited samples of the solar neighbourhood
\citep{reyle21,kirkpatrick21a,best21,smart21a}. Within this population, an intriguing subset known as unusually URLDs (hereafter URLDs) stands out due to their near-infrared colours that are significantly redder than typical field L dwarfs of the same spectral type \citep{kirkpatrick06,gizis12,marocco14,schneider17a}. These URLDs are rare but scientifically valuable because they challenge standard models of ultracool atmospheres and offer unique insights into substellar atmospheric physics and evolutionary processes.

In this context, certain extreme cases such as the L7pec object ULAS J222711$-$004547 \citep{marocco14} have revealed that enhanced dust opacity and non-standard cloud properties likely drive the very red photometric and spectroscopic characteristics of some URLDs. Other well-studied examples include the very young field L dwarf 2MASS\,J01415823$-$4633574
\citep{kirkpatrick06} and the unusually red, dusty L dwarf WISEP\,J004701.06$+$680352.1 \citep{gizis12,gizis15a}. Detailed spectral analysis showed that differences in cloud optical depth and sub-micron grain populations can well explain their infrared excesses relative to "normal" L dwarfs \citep{marocco14,hiranaka16}, posing challenges for traditional atmospheric models which typically include only larger cloud grains and simpler cloud prescriptions \citep{ackerman01,marley02}.

The first systematic search for unusually URLDs was presented by \citet{schneider17a}, who exploited combined photometry from 2MASS and AllWISE to identify 50 new late-type L dwarf candidates, 47 of which were spectroscopically confirmed. This study not only expanded the sample of known URLDs substantially but also identified likely members of young stellar associations such as Beta Pictoris, AB Doradus, Tucana-Horologium, and Argus \citep{zuckerman04,torres08}, reinforcing the view that red colours may be linked to youth and low surface gravity in some cases. This link had already emerged from the identification of young L dwarfs in the field \citep{kirkpatrick08,cruz09,allers13}, several of which were subsequently associated with young moving groups \citep{schneider14a,gagne15c,best17a} or were shown to be free-floating planetary-mass members of nearby associations \citep{schneider16b}. Parallax programmes have since provided distances, luminosities, and temperatures for a number of these objects \citep{zapatero14a,faherty16a,liu16a}. Additional peculiar L dwarfs were identified in other public wide-field surveys \citep{zhang22a}, and follow-up spectroscopy has confirmed a handful of ultra-red objects spanning a range of ages, gravities, and kinematic properties \citep{marocco15,scholz20a,bickle24}.

The scientific importance of URLDs can be summarised through three key aspects. First, URLDs provide empirical constraints on the formation, composition, and vertical extent of condensate clouds in cool, low-temperature atmospheres \citep{stephens09,hiranaka16}. Traditional cloud models developed for L dwarfs often underpredict the reddening seen in these objects \citep{marocco14}. Models that incorporate both large and sub-micron dust grains have been shown to reproduce observed spectral energy distributions more effectively, motivating improvements in atmospheric modelling. Second, the association of some URLDs with young moving groups suggests that their spectral peculiarities could be signatures of lower gravity \citep{cruz09,allers13,liu16a}. However, not all URLDs show unambiguous indicators of youth, implying that multiple physical parameters (e.g.\ cloud opacity, metallicity) play a role \citep{looper08b,burgasser08b,liu13a} play a role \citep{liu13a}. Third, URLDs share photometric and spectroscopic similarities with directly imaged giant exoplanets, which also display very red near-infrared colours \citep{chauvin04,chauvin05c,marois08,lagrange09c,marois10,barman11b,macintosh15,chauvin15,gauza15a}. As free-floating analogues, they represent valuable benchmarks for exoplanet atmospheric models and for interpreting photometry and spectra of young giant planets \citep{faherty16a,liu13a,bowler16}, including those now being observed with the James Webb Space telescope \citep{miles23a,petrus24}. The spectral sequence of young planetary-mass objects in star-forming regions provides a natural extension of this comparison to even lower masses and ages \citep{lodieu18a,luhman18a}.

The Euclid mission \citep{laureijs11,mellier25} and the SPHEREx all-sky spectral survey \citep{bock26a} open a new window on this population. Early works have already demonstrated the potential of Euclid for ultracool-dwarf science, from the identification of candidates in deep extragalactic fields \citep{solano21} to spectroscopic reconnaissance in the Euclid Deep Fields \citep{zhang24a} and photometric searches in the Quick Data Release \citep{zerjal26}, while SPHEREx now provides a homogeneous near- to mid-infrared spectral library of ultracool dwarfs \citep{gagne26b}.

This paper is organised as follows. 
In Sect.\ \ref{redLdwarfs:datasets}, we present the photometric and spectroscopic datasets. 
In Sect.\ \ref{redLdwarfs:sample} we describe the reference samples and new URLD candidates identified in the {\it{Euclid}} Data Release 1 (DR1).
In Sect.\ \ref{redLdwarfs:method} we expose the methodology to classify and derive luminosities for the candidates.
In Sect.\ \ref{redLdwarfs:discussion} we put our results into context to look into a possible age dependence for this population.
In Sect.\ \ref{redLdwarfs:conclusions}, we highlight the main conclusion of this work.

%
%%%%%%%%%%%%%%%%%%%%%%%%%%%%%%%%%%%%%%%%%%%%%%%%%%%%%
%%%%% Datasets %%%%%
%%%%%%%%%%%%%%%%%%%%%%%%%%%%%%%%%%%%%%%%%%%%%%%%%%%%%
%
\section{Datasets}
\label{redLdwarfs:datasets}
\subsection{{\it{Euclid}} Data Release 1}
\label{redLdwarfs:datasets_Euclid}
The {\it{Euclid}} space telescope is designed to explore the geometry of the dark Universe by mapping billions of galaxies across 14,000 square degrees \citep{laureijs11,mellier16a,mellier25}. The Data Release 1 (DR1) provides the first major milestone, covering approximately 2,100 square degrees of the Wide Survey alongside deep, and calibration fields. Optical photometry is delivered by the VIS instrument via a single broad band \citep[$I_E$, 530$-$920\,nm;][]{cropper25}. Near-infrared photometry is provided by the NISP instrument across three broad filters: $Y_E$, $J_E$, and $H_E$ spanning the 920$-$2000\,nm wavelength range \citep{jahnke25,hormuth25}. In addition to imaging, NISP operates in slitless spectroscopy mode providing infrared spectra in the 1200$-$1890\,nm wavelength range at low-resolution of $\sim$550$-$740 \citep{schirmer22,copin26}.

\subsection{SPHEREx spectroscopy}
\label{redLdwarfs:datasets_SPHEREx}

The Spectro-Photometer for the History of the Universe, Epoch of Reionization and Ices Explorer (\textit{SPHEREx}) is a NASA Medium Explorer (MIDEX) mission designed to perform the first all-sky near-infrared spectral survey \citep{bock26a}. Over its nominal two-year mission lifetime, \textit{SPHEREx} will map the entire sky four times, complemented by two highly sampled $\sim$100~deg$^2$ deep fields located at the ecliptic poles. The payload utilises Linear Variable Filters (LVFs) to cover a continuous near-infrared wavelength range from 0.75 to 5.0\,$\mu$m across 102 narrow spectral channels, achieving a pixel scale of 6.2 arcsec. The instrument delivers a dual-resolution setup with a resolving power of about 40 and 120 below and above 3.8\,$\mu$m. Upon completion of its four-survey co-add, \textit{SPHEREx} achieves a global 5$\sigma$ point-source sensitivity depth in the 19.5$-$19.9 mag range in the short-wavelength bands (0.75--3.8\,$\mu$m) and between 17.8 and 18.8 mag at longer wavelengths ($\geq$\,3.8\,$\mu$m). Spectra are calibrated in flux with units of W\,m$^{-2}\,\mu$m$^{-1}$ with their uncertainties \citep{gagne26a}.

%
%%%%%%%%%%%%%%%%%%%%%%%%%%%%%%%%%%%%%%%%%%%%%%%%%%%%%
%%%%% Samples %%%%%
%%%%%%%%%%%%%%%%%%%%%%%%%%%%%%%%%%%%%%%%%%%%%%%%%%%%%
%
\section{Samples}
\label{redLdwarfs:sample}
\subsection{Reference samples of ultracool dwarfs}
\label{redLdwarfs:sample_UCDs}

The main reference samples used in this work come from the UltraCool Dwarf (UCDs) compilation of Jonathan Gagn\'e publicly available on his website\footnote{\url{https://jgagneastro.com/}}. The full compendium includes 2304 UCDs\footnote{\url{https://jgagneastro.com/list-of-ultracool-dwarfs/}}. We have extracted several subsamples (Table \ref{tab_redLdwarfs:samples}). First, a subsample of 425 "normal" L dwarfs contains L dwarfs without any special note in the compilation and acts as the field reference 
We retrieved six intermediate-age L dwarfs also known as $\beta$ L dwarfs 
%(orange dots in Fig.\ \ref{fig_redLdwarfs:all_diagrams}) 
following the definition of \citet{allers13}, which are young L dwarfs with an age comparable to the Pleiades cluster \citep[130$\pm$20 Myr;][]{stauffer98,barrado04b}. We also retrived 16 very young L-type objects with ages below about 10 Myr are called $\gamma$ L dwarfs 
We added to this sample the 11 L-type members in $\sigma$\,Orionis. 

We cross-matched this compendium of UCDs with the SPHEREx database with a matching radius of 5 arcsec (Table \ref{tab_redLdwarfs:samples}). These samples are used in the subsequent analysis. We found 344 normal L dwarfs, only 318 are useful spectra with sufficient signal-to-noise ratio for subsequent analysis. We retrieved all $\beta$ L dwarfs and 15 out of 16 $\gamma$ L dwarfs, while only two $\sigma$\,Ori L-type members have SPHEREx spectra. The availability of the large wavelength coverage of SPHEREx spectra is important for the determination of the flux of these reference UCDs, as described in Sect. \ref{redLdwarfs:method_bol}. 

%
%%%%%%%%%%%%%%%%%%%%%%%%%%%%%%%%%%%%%%%%%%%%%%%%%%%%%
%%%%% Table samples %%%%%
%%%%%%%%%%%%%%%%%%%%%%%%%%%%%%%%%%%%%%%%%%%%%%%%%%%%%
%
\begin{table}
\caption{Samples used in this work.}
\label{tab_redLdwarfs:samples}
\centering
\begin{tabular}{lccc}
\hline\hline
Sample & UCDs & SPHEREx & Used \\
\hline
normal L (reference)                      & 425 & 344 & 318 \\
L$\beta$ (INT-G)                          &   6 &   6 &   6 \\
L$\gamma$ (VL-G)                          &  16 &  15 &  15 \\
young red L (Gagn\'e)                     &   4 &   4 &   4 \\
$\sigma$\,Ori L                           &  11 &   2 &   1 \\
Known URLDs                               &  24 &   7 &   6 \\
New {\it{Euclid}} URLDs                   & --- &   6 &  12 \\
\hline
\end{tabular}
\tablefoot{
``UCDs'' is the number of objects in the compendium of UCDs. \newline
``SPHEREx'' corresponds to the number of sources with a SPHEREx spectrum. \newline
``Used'' gives the number of objects used in this analysis (Sect.~\ref{redLdwarfs:method}).}
\end{table}
\subsection{Known URLDs in {\it{Euclid}} DR1}
\label{redLdwarfs:sample_redL_known}
We compiled the list of known URLDs published in the literature by various authors \citep{schneider17a,scholz20a,zhang22a,bickle24}. We cross-matched this sample with the {\it{Euclid}} DR1 using a matching radius of ten arcsec to take into account the proper motions of these objects. We find several URLDs in {\it{Euclid}} DR1 listed in Table \ref{tab_redLdwarfs:Euclid_phot_known_redL} with their {\it{Euclid}} photometry and spectra types from the literature. We describe some of their properties and peculiarities below. 

\begin{itemize}
\item [$\bullet$] WISEA\,J00581169$-$5653321 is classifed as a slightly red L7 with photometric distance is 22$\pm$2 pc \citep{schneider17a}, in disagreement with its trigonometric parallax \citep[35.3$\pm$4.1 pc;][]{kirkpatrick21a}. Its moving-group membership remains ambiguous with modest-probability membership to $\beta$\,Pic ($\sim$27\%) and Argus ($\sim$31\%) based on BANYAN II \citep{malo13,gagne14a}, and TucHor ($\sim$48\%) and ABDor ($\sim$40\%) according to LACEwING \citep{riedel17}.
\item [$\bullet$] 2MASSI\,J0318540$-$342129 is classified as a L6.5$-$L7 dwarf \citep{schneider14a} with a $Gaia$ distance of 14.09$\pm$0.39 pc \citep{Gaia_Brown21}. \citet{burgasser10a} suggests that this object is an unresolved binary made of a late-L primary and a T5.5 secondary.
\item [$\bullet$] CWISE\,J04250660$-$4255101 is a young L8 dwarf located a photometric distance of 22.0$\pm$3.9 pc \citep{schneider17a}. This object is member of the Ursa Major corona association whose mean age is 414 Myr \citep{jones15} and the physical group is not fully confirmed as it consists of several Theia groups \citep{kounkel19a,gagne20b}.
\item [$\bullet$] WISEA\,J04371877$-$5509440 a red L5 dwarf at 35$\pm$7 pc \citep{schneider17a}. According to BANYAN II and the convergent point method of \citet{rodriguez13}, this object is a high-probability ($>$90\%) member of $\beta$\,Pic. However, its photometric distance conflicts with its kinematic membership to $\beta$\,Pic \citep[17$-$19 pc;][]{miret_roig24}, so its youth and membership remains unresolved at this stage. 
\item [$\bullet$] 2MASS\,J04590034$-$2853396 is classified as a red L7: at 38$\pm$8 pc \citep{schneider17a}. The spectral type has a ±1 subtype uncertainty because the spectrum is low signal-to-noise ($\sim$\,4). Its membership to young moving is unclear due to discrepant kinematic and photometric distance estimates and its location in the sky. Note that no SPHEREx spectrum is not available for this source.
\item [$\bullet$] WISEA\,J09025899$+$6708331 is classified as a slightly red L7 dwarf at a distance of 23$\pm$2 pc \citep{schneider17a}. It is a medium-to-high probability AB Dor member with an age of about 150Myr \citep{bell15a}, with a predicted membership distance matches the photometric one.
\item [$\bullet$] CWISE\,J09581211$+$7151441 is classified as a L5 $\beta$,$\gamma$ peculiar object \citep{gagne26a} at a distance of 45.46$\pm$1.03 pc with a 90\% probability membership to the Crius\,198 moving group \citep{moranta22}. This association has an age of 100$-$700\,Myr based on isochrone fitting.
\item [$\bullet$] WISE\,J17385927$+$6142421: this object is described as a late-L/early-T dwarf \citep{mace13} at a distance of 44.5$\pm$3 pc \citep{kirkpatrick21a}, whose unusual spectrum defies standard classification \citep{schneider17a}. It is extremely red with $J-K_{s}$\,$\sim$\,2.55 mag with a strongly varying infrared flux reported. Moreover, excess flux on the blue side of the $H$ and $K$ passbands possibly hint at the presence of a T dwarf secondary companion \citep{mace13}.
\end{itemize}
%

%
%%%%%%%%%%%%%%%%%%%%%%%%%%%%%%%%%%%%%%%%%%%%%%%%%%%%%
%%%%% Table: photometry KNOWN URLDs %%%%%
%%%%%%%%%%%%%%%%%%%%%%%%%%%%%%%%%%%%%%%%%%%%%%%%%%%%%
%
\begin{table*}[!h]
\centering
\caption{{\it{Euclid}} photometry of known URLDs (Sect.\ \ref{redLdwarfs:sample_redL_known}).}
%\begin{tabular}{lccccccccc}
\begin{tabular}{@{\hspace{0mm}}l@{\hspace{2mm}}c@{\hspace{2mm}}c@{\hspace{2mm}}c@{\hspace{2mm}}c@{\hspace{2mm}}c@{\hspace{2mm}}c@{\hspace{2mm}}l@{\hspace{0mm}}}
\hline
\hline
Name & SpT & ID & $I_E$ & $Y_E$ & $J_E$ & $H_E$ & Note \\
\hline
WISEA\,J00581169$-$5653321 & L7 & $-$145501753568921593 & 21.805 & 18.517 & 17.881 & 17.203 & sl.\ red; ambiguous YMG \\
%2MASSI\,J031854.08$-$342129.1 & L6.5$-$L7$+$T5.5 &  $-$497285970343578614 & 20.705 & 17.367 & 16.704 & 16.092 & spec binary \\
CWISE\,J04250660$-$4255101 & L8 & $-$662770206429197561 & 21.953 & 18.509 & 17.886 & 17.219 & young \\
WISEA\,J04371877$-$5509440 & L5 & $-$693289112551618214 & 22.169 & 18.807 & 18.033 & 17.382 & red; $\beta$\,Pic? \\
2MASS\,J04590034$-$2853396 & L7$\pm$1 & $-$747520247288936772 & 22.941 & 19.687 & 18.902 & 18.061 & red; uncertain distance \\
WISE\,J09025899$+$6708331  & L7 & 1357448441671416566 & 21.815 & 18.397 & 17.570 & 16.845 & sl.\ red; AB Dor? \\
CWISE\,J09581211$+$7151441 & L5 & 1495492300718620996 & 23.047 & 19.647 & 19.000 & 18.283 & $\beta$,$\gamma$ \\
WISE\,J17385927$+$6142421  & L7.5$\pm$2 & 2647470739617127439 & 23.371 & 19.817 & 18.995 & 18.109 & red, binary ? \\
\hline
\end{tabular}
\label{tab_redLdwarfs:Euclid_phot_known_redL}
\end{table*}
\subsection{New URLDs in {\it{Euclid}} DR1}
\label{redLdwarfs:sample_redL_new}

Thanks to the recovery of known URLDs in the {\it{Euclid}} DR1 (Sect.\ \ref{redLdwarfs:sample_redL_known}) and the sequence of UCDs from the Q1 \citep{zerjal26} and DR1 ({\v Z}erjal et al.\ 2027, subm) releases, we applied the broad photometric criteria to identify new URLD candidates in both the wide and deep surveys: 
$I_{E}-H_{\rm E}$\,$\geq$\,4.4 mag, 
$J_{E}-H_{\rm E}$\,=\,0.5$-$1.0 mag, 
$Y_{E}-H_{\rm E}$\,$\geq$\,1.2 mag, 
$Y_{E}-J_{\rm E}$\,$\geq$\,0.6 mag, and
$I_{E}-J_{\rm E}$\,$>$\,3.8 mag (AB system). 
We imposed a detection in the visible band, good quality and point source criteria to remove galaxies and false positive as much as a possible and trained the code to learn about the colour-colour loci of known UCDs, as described in (Zerjal et al.\ 2026, subm): signal to noise ratio larger than 3 in the $J$ photometry, detection quality flag equal to 0\footnote{A value of 0 (or clean status) generally means the source passed all basic validation checks without detected flaws}, and the {\tt{mumax\_minus\_mag}} parameter lies in the $-$3.5 to $-$2.5 interval\footnote{Larger negative values below $-$2.6 indicate a high peak-to-total ratio typical of point-like sources (such as stars, quasars/AGNs, or ultra-cool brown dwarfs)}. This selection procedure led to 1202 and 38 URLD candidates in the wide and deep surveys, respectively. Among them, five have also SPHEREx spectra whose coordinates and {\it{Euclid}} photometry is given in Table \ref{tab_redLdwarfs:Euclid_phot_new_redL}.

\begin{figure*}[!h]
   \centering
   \includegraphics[width=\linewidth, angle=0]{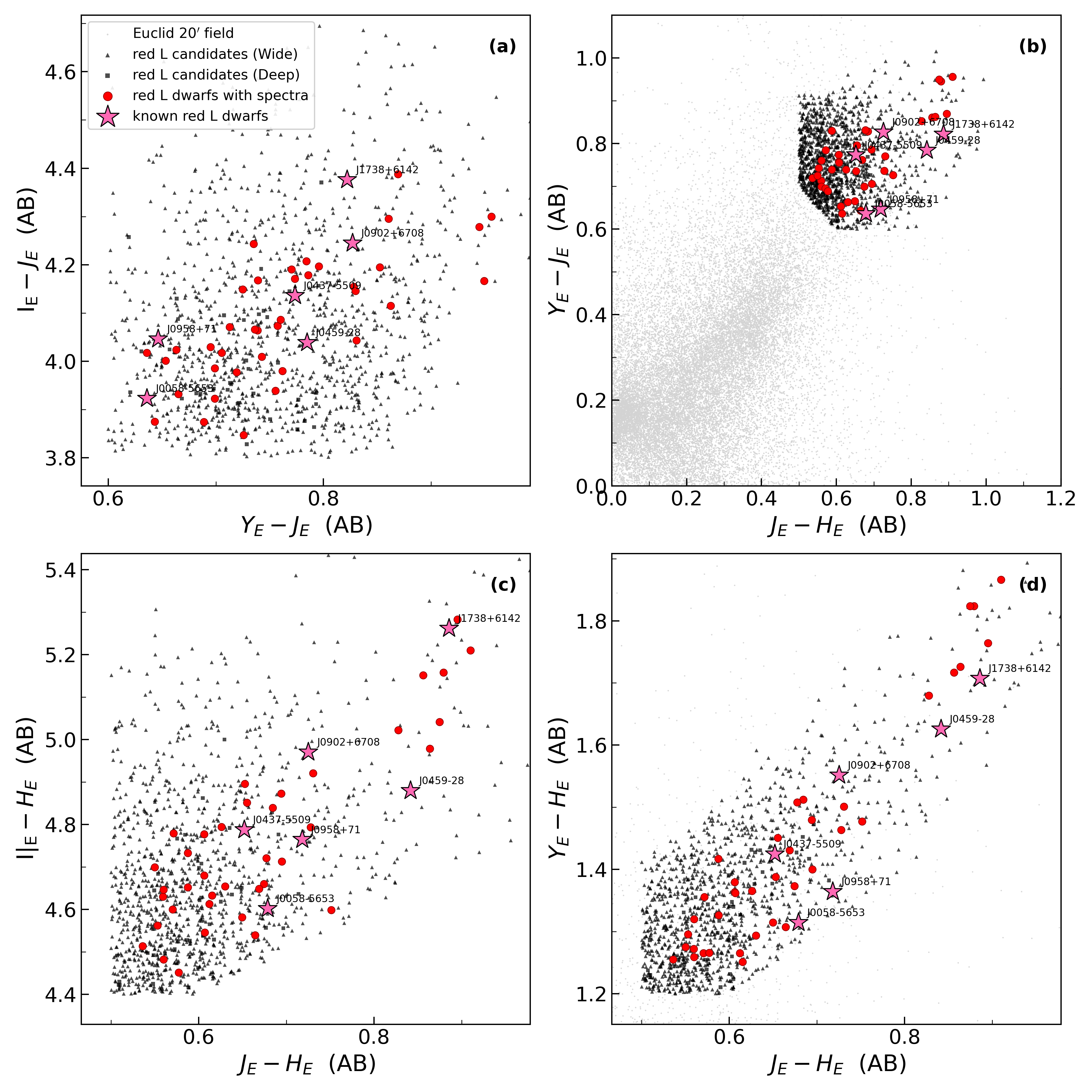}
   \caption{Colour-colour diagrams depicting the location of URLDs.
Light grey dots are the detections of a 20 arcmin random {\it{Euclid}} field. 
Black triangles and squares are the red L photometric candidates of the
Wide and Deep surveys, respectively. Red dots are the URLDs with {\it{Euclid}} spectra and pink stars the known URLDs covered by {\it{Euclid}}.}
   \label{fig_redLdwarfs:colcol_diagrams}
\end{figure*}

A sub-sample of candidates in the wide and deep surveys with their coordinates and photometry are provided in Table \ref{tab_redLdwarfs:Euclid_cand_redL_WIDE} and Table \ref{tab_redLdwarfs:Euclid_cand_redL_DEEP} in Appendix \ref{tab_redLdwarfs:Euclid_cand_redL}, respectively. The full list will be available electronically online in Vizier. Candidates in the deep survey are fainter on average than in the wide survey, as expected. We display several colour-colour diagrams in Fig.\ \ref{fig_redLdwarfs:colcol_diagrams} to locate these new candidates with {\it{Euclid}} photometry along with known URLDs.

Among these photometrically-selected candidates, the clustering technique developed by Vitas et al.\ (2026, accepted) suggest that 45 might be URLDs based on their low-resolution red grism spectra covering the 1.3$-$1.8\,$\mu$m wavelength range (Table \ref{tab_redLdwarfs:Euclid_45_redL} in Appendix \ref{tab_redLdwarfs:Euclid_cand_redL}). We have compared the spectra of these 45 candidates with field L dwarf templates from the SpeX library\footnote{\url{https://svod.cab.inta-csic.es/spex/}} \citep{burgasser14} for spectral classification. We show a subsample of images in Fig.\ \ref{fig_redLdwarfs:Euclid_images} and spectra in Fig.\ \ref{fig_redLdwarfs:Euclid_spectra} in Appendix \ref{tab_redLdwarfs:Euclid_cand_redL}. All images and spectra with the comparison with field L dwarf templates will be made public for all URLD candidates either through the Spanish virtual observatory or Vizier on the Centre de Donn\'ees de Strasbourg.

%
%%%%%%%%%%%%%%%%%%%%%%%%%%%%%%%%%%%%%%%%%%%%
%%%%% Figure: {\it{Euclid}} diagrams %%%%%
%%%%%%%%%%%%%%%%%%%%%%%%%%%%%%%%%%%%%%%%%%%%
%
%

%
%%%%%%%%%%%%%%%%%%%%%%%%%%%%%%%%%%%%%%%%%%%%%%%%%%%%%
%%%%% Table: photometry NEW URLDs %%%%%
%%%%%%%%%%%%%%%%%%%%%%%%%%%%%%%%%%%%%%%%%%%%%%%%%%%%%
%
\begin{table*}[!h]
\centering
\caption{
{\it{Top:}} {\it{Euclid}} photometry of the new URLDs with SPHEREx spectra.
{\it{Bottom:}} Coordinates and photometry of URLD candidates identified by the spectroscopic search.
}
%\begin{tabular}{lccccccccc}
\begin{tabular}{@{\hspace{0mm}}l@{\hspace{2mm}}c@{\hspace{2mm}}c@{\hspace{2mm}}c@{\hspace{2mm}}c@{\hspace{2mm}}c@{\hspace{2mm}}c@{\hspace{2mm}}l@{\hspace{0mm}}}
\hline
\hline
Name & SpT & ID & $I_E$ & $Y_E$ & $J_E$ & $H_E$ & Note \\
\hline
WISEA\,J022908.28$-$530517.6 & L7 & $-$372851399530883137 & 25.022 & 21.587 & 20.726 & 19.870 & --- \\
WISEP\,J031854.37$-$342128.7 & L9 & $-$497285970343578614 & 20.705 & 17.357 & 16.704 & 16.092 & --- \\
WISEA\,J051249.55$-$433842.6 & L7 & $-$782067599436445676 & 22.749 & 19.415 & 18.720 & 18.150 & --- \\
WISEA\,J051447.17$-$385454.7 & L9 & $-$786968280389151351 & 23.166 & 19.746 & 18.975 & 18.245 & --- \\
WISEA\,J053710.26$-$405011.0 & L7 & $-$842932247408362401 & 22.470 & 19.246 & 18.547 & 17.987 & --- \\
\hline
Euclid\,J035230.2$-$492547 & L9 & $-$581256705494296969 & 25.744 & 22.402 & 21.715 & 20.950 & possible \\
Euclid\,J055917.8$-$493057 & L8 & $-$898243280495157949 & 24.796 & 21.187 & 20.338 & 19.373 & $\beta$ ? \\
Euclid\,J092823.4$+$682404 & L5 & 1420972919684011689 & 23.701 & 20.339 & 19.576 & 18.790 & $\beta$ ? \\
Euclid\,J235742.2$+$641514 & T0 & $-$3594259130642538812 & 24.589 & 21.163 & 20.422 & 19.571 & possible \\
Euclid\,J050118.5$-$451214 & T2 & $-$753271732452039163 & 24.827 & 21.144 & 20.595 & 20.112 & possible \\
Euclid\,J233047.9$-$330351 & L6 & $-$3526994577330641981 & 23.377 & 20.013 & 19.221 & 18.442 & $\gamma$ \\
Euclid\,J044147.1$-$212717 & L3 & $-$704464473214546194 & 24.142 & 20.910 & 20.910 & 19.645 & " \\\hline
\end{tabular}
\label{tab_redLdwarfs:Euclid_phot_new_redL}
\end{table*}

We display the SPHEREx spectra of the known and new URLD candidates in {\it{Euclid}} in black in Figs.\ \ref{fig_redLdwarfs:SPHEREX_spectra_known} and \ref{fig_redLdwarfs:SPHEREX_spectra_new} in Appendix \ref{redLdwarfs:append_SPHEREx}, respectively. Each spectrum was compared with a $\chi^{2}$ method to SPHEREx spectra of the compendium of 2304 UCDs \citep{gagne26a} and best two fits are overplotted on each plot for the new URLD candidates. The best fit is not shown for the known URLDs displayed in Fig.\ \ref{fig_redLdwarfs:SPHEREX_spectra_known} because the $\chi^{2}$ procedure found the best fit for the object itself in the compendium of UCDs, validating the method.

To complement the aforementioned search, we have applied the spectral index technique described in \citet{dominguez_tagle25} and Dominguez-Tagle et al.\ (2027, subm) to identify any additional candidates of interest independently of photometric cuts. We add seven sources in the bottom part of Table \ref{tab_redLdwarfs:Euclid_phot_new_redL} with their tentative classification. One of them is in the SPHEREx MOCA database: CWISE\,J233047.79$-$330351.2, classified as a L7$\gamma$ \citep{gagne26a,schneider26}.

%
%%%%%%%%%%%%%%%%%%%%%%%%%%%%%%%%%%%%%%%%%%%%
%%%%% Figure: diagrams %%%%%
%%%%%%%%%%%%%%%%%%%%%%%%%%%%%%%%%%%%%%%%%%%%
%
\begin{figure*}[!h]
   \centering
   \includegraphics[width=\linewidth, angle=0]{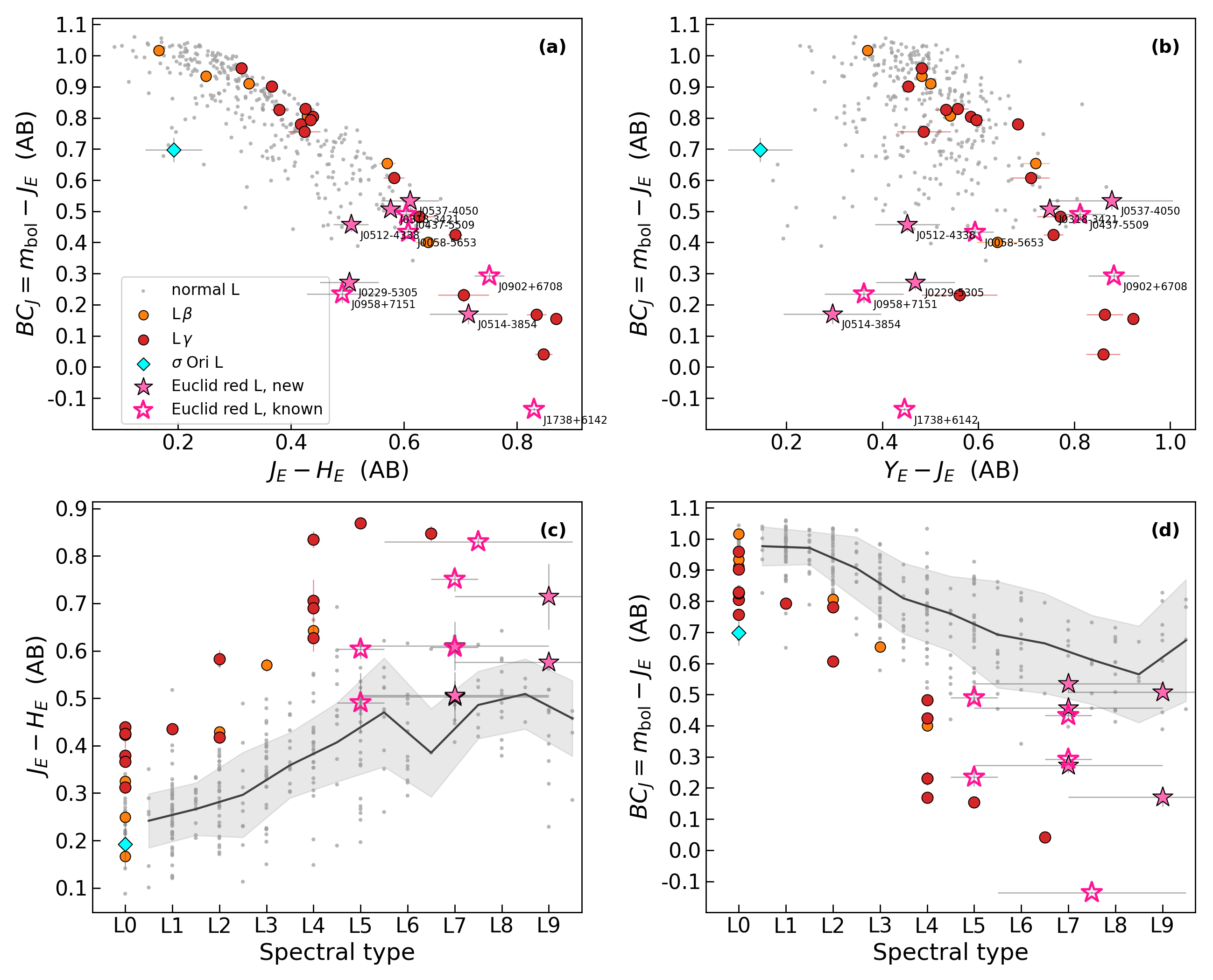}
   \caption{
   Distance-independent diagnostics of the URLDs. 
   Grey dots are normal L dwarfs, orange circles L$\beta$, red circles L$\gamma$ and cyan diamonds
   $\sigma$\,Ori L dwarfs. Pink stars are the {\it{Euclid}} URLDs with SPHEREx spectra: filled
   for the new candidates and open for those already known. Pink diamonds are the
   {\it{Euclid}} candidates with measured {\it{Euclid}} colours lacking SPHEREx spectra (Sect.\ \ref{redLdwarfs:sample_UCDs}; Table \ref{tab_redLdwarfs:samples}). The black line and grey
   band show the median and $\pm1\sigma$ of the normal L dwarfs per spectral subtype. 
}
   \label{fig_redLdwarfs:all_diagrams}
\end{figure*}
%

%
%%%%%%%%%%%%%%%%%%%%%%%%%%%%%%%%%%%%%%%%%%%%
%%%%% Methodology %%%%%
%%%%%%%%%%%%%%%%%%%%%%%%%%%%%%%%%%%%%%%%%%%%
%
\section{Methodology}
\label{redLdwarfs:method}
\subsection{SPHEREx spectra}
\label{redLdwarfs:method_prep}
For the samples described in Sect.\ \ref{redLdwarfs:sample_UCDs} and listed in Table \ref{tab_redLdwarfs:samples}, we kept only spectra with signal-to-noise ratios larger than 3 and 10 at 1.1\,$\mu$m and 2.2\,$\mu$m, respectively. The signal-to-noise ratio (S/N) of each spectrum was measured as the median of the ratio of the flux to its uncertainty in each channel.
Moreover, we applied some removal of spikes affected a small number of high-quality spectra, mainly normal field L dwarfs. Isolated outliers channels were removed
if their flux deviate more than 10$\sigma$ the median absolute deviation (MAD) of seven contiguous channels, hence removing about 0.33\% of the channels in more than 26\% (559 out of 2304) of the spectra.
This retains all ten {\it{Euclid}} URLDs, all L$\beta$ and L$\gamma$ dwarfs, one $\sigma$\,Ori member, and 92\% of the normal L dwarfs.

\subsection{Bolometric flux and magnitude}
\label{redLdwarfs:method_bol}
For each spectrum the observed flux was obtained by integrating $F_\lambda$ with the trapezoidal rule over the observed range, with the channel uncertainties $\sigma_i$ propagated in quadrature $\sigma_F^2=\sum_i (w_i\sigma_i)^2$, where $w_i$ are the trapezoid weights. Outside the SPHEREx range we added two model-dependent extrapolations. Blueward of $\lambda_{\min}$\,=\,0.744\,$\mu$m the flux density was assumed to fall linearly to zero at $\lambda$\,=\,0.
For wavelengths longwards of 5\,$\mu$m, we adopted a Rayleigh--Jeans tail, $F_\lambda$\,=\,$A$\,$\times$\,$\lambda^{-4}$, where $A$ is the weighted mean of $F_\lambda\times\lambda^{4}$ over the eight reddest channels.
The bolometric flux is the sum of the integration of the SPHEREx spectrum plus the blue and red tails.
For the normal field L dwarfs this range contains about 87$-$90\% of the bolometric flux. The apparent bolometric magnitude follows the IAU 2015 Resolution B2 zero point \citep{andersen99a}\footnote{\url{https://www.pas.rochester.edu/~emamajek/IAUres_B2.pdf}}.
\begin{equation}
M_{bol}\,=\,-2.5\log(flux) - 18.9986
\end{equation}
The quoted errors come from the channel uncertainties, while the extrapolated tails are model dependent and their systematic effect is not included. We note that no luminosities are derived in this analysis, because distances are not available for the objects of interest of this work, the URLDs.

\subsection{Synthetic {\it{Euclid}} photometry}
\label{redLdwarfs:method_synth}

We computed the magnitudes that {\it{Euclid}} would measure for each SPHEREx spectrum by integrating the spectrum over the response curves of the NISP $Y_{\rm E}$, $J_{\rm E}$ and $H_{\rm E}$ filters. The $I_E$ filter is not computed. The SPHEREx spectra are given as $\flam$, whereas AB magnitudes are defined for the flux density per unit frequency.
The mean flux density in each band $X$, where $X$\,=\,$Y_E$, $J_E$, $H_E$, is the response-weighted average
\begin{equation}
\langle\fnu\rangle_X=\frac{\int \fnu(\lambda)\,S_X(\lambda)\,\lambda^{-1}\,d\lambda}{\int S_X(\lambda)\,\lambda^{-1}\,d\lambda}
=\frac{1}{c\,D_X}\int \flam(\lambda)\,S_X(\lambda)\,\lambda\,d\lambda,
\label{redLdwarfs_eq:meanfnu}
\end{equation}
where $D_X=\int S_X(\lambda)\,\lambda^{-1}\,d\lambda$ is the normalisation. The AB magnitude is
\begin{equation}
m_X=-2.5\log_{10}\!\left(\frac{\langle\fnu\rangle_X}{3631\,\mathrm{Jy}}\right)
=-2.5\log_{10}\langle\fnu\rangle_X-48.60,
\label{redLdwarfs_eq:ab_mag}
\end{equation}
with $\langle\fnu\rangle_X$ in erg\,s$^{-1}$\,cm$^{-2}$\,Hz$^{-1}$.The weight $\lambda^{-1}$ arises because the {\it{Euclid}} detectors count photons and not energy.

In practice the spectrum is known at the SPHEREx channel wavelengths $\lambda_i$ only. We interpolated $\flam$ linearly between channels onto the (finer) wavelength grid of the filter curve, so that the sharp edges of the filters are resolved, and evaluated the integrals with the trapezoidal rule. This operation is linear in the channel fluxes.

We define the distance-independent bolometric correction
\begin{equation}
\BCJ=\mbol-\JE ,
\end{equation}
with $\JE$ the synthetic AB magnitude.

A flat spectrum $\fnu=3631$\,Jy returns a null magnitude in all three bands to better than $10^{-3}$\,mag. For the 12 {\it{Euclid}} URLDs with both SPHEREx spectra and measured Euclid photometry, the synthetic colours differ from the measured ones by $-0.05$\,mag in $J-H$ (median; scatter $0.12$\,mag) and $-0.14$\,mag in $Y-J$ (scatter $0.24$\,mag). 
These offsets are small compared with the differences between the samples in $J-H$ but not negligible in $Y-J$, thus we adopt as an estimate of the systematic uncertainty of the synthetic colours.

%
%%%%%%%%%%%%%%%%%%%%%%%%%%%%%%%%%%%%%%%%%%%%
%%%%% Discussion %%%%%
%%%%%%%%%%%%%%%%%%%%%%%%%%%%%%%%%%%%%%%%%%%%
%
\section{Discussion}
\label{redLdwarfs:discussion}
\subsection{Expanding the census of URLDs}
\label{redLdwarfs:discussion_census}
Searches for URLDs have relied on targeted or all-sky cross-matches between optical, near-infrared, and mid-infrared surveys-such as the Two Micron All-Sky Survey \citep[2MASS;][]{skrutskie06}, the Deep Near Infrared Survey of the Southern Sky \citep[DENIS;][]{delfosse97}, Sloan Digital Sky Survey \citep[SDSS;][]{york00}, UKIRT Infrared Deep Sky Survey \citep[UKIDSS;][]{lawrence07}, the All Wide-field Infrared Survey Explorer \citep[AllWISE;][]{wright10,cutri13} and its successor CatWISE2020 \citep{eisenhardt20,marocco21}. These efforts yielded initial collections of peculiar L dwarfs \citep{gizis00a,cruz03,cruz07,kirkpatrick06,kirkpatrick08,looper08b,cruz09,kirkpatrick10,schmidt10a,gizis12,mace13}, culminated by systematic searches that expanded the sample to a few tens or hundreds of objects \citep{schneider14a,marocco14,marocco15,gagne15c,best17a,schneider17a,scholz20a,zhang22a,bickle24}. Trigonometric parallaxes of young and red L dwarfs \citep{zapatero14a,faherty16a,liu16a,best20} have subsequently placed these
objects on absolute colour-magnitude and bolometric-luminosity sequences
\citep{filippazzo15}, but such measurements remain available for only a small
fraction of the known population.

Our photometric selection criteria identifies more than one thousand candidates in {\it{Euclid}} DR1, including about 40 confirmed spectroscopically thanks to the slitless spectroscopic survey down to about $J_E$\,=\,19$-$20 mag. The other object look like normal field L dwarfs in the 1.2$-$1.8 micron range but we cannot discard that they are red at longer wavelengths. By combining {\it{Euclid}}'s high spatial resolution and multi-band infrared photometry with SPHEREx spectroscopy and {\it{Euclid}} NISP red grism spectra, we expand the candidate pool by an order of magnitude and double the spectroscopically confirmed sample. This substantial increase demonstrates {\it{Euclid}}'s unmatched discovery power for ultracool dwarfs, building on the first Euclid-based searches in the Deep Fields and in the
Quick Data Release \citep{zhang24a,dominguez_tagle25,zerjal26}, and providing the large statistical sample required to study the low-temperature, highly dust-obscured tail of the brown dwarf population. A second epoch in one infrared band in a few years would be extremely beneficial to measure proper motions of members of this population with sufficient accuracy to assign preliminary membership to nearby young moving groups. The brightest and reddest objects of our sample are also natural targets for higher-quality spectroscopy, including with JWST \citep{miles23a,petrus24}.

\subsection{Bolometric corrections, colours, and spectral types}
\label{redLdwarfs:discussion_SpT}
The distance-independent diagnostics presented in Fig.\ \ref{fig_redLdwarfs:all_diagrams} provide key physical insights into the atmospheric structures of these URLDs.

%{\it{Bolometric Corrections ($BC_J$) vs. Colours and Spectral Types:}} 
The $BC_J$ values for URLDs display systematic offsets compared to normal field L dwarfs of equivalent spectral types \citep{cushing05,filippazzo15,golimowski04a,vrba04,stephens09}, in line with the behaviour reported for young, low-gravity objects \citep{filippazzo15,faherty16a,liu16a}. In the $BC_J$ vs $J_E-H_E$ and $Y_E-J_E$ diagrams, URLDs occupy a distinct locus characterised by lower bolometric corrections due to the larger flux beyond $\sim$1.2 microns. This photometric behaviour is most likely driven by enhanced condensate cloud opacity, thick cloud decks, and sub-micron grain populations \citep{ackerman01,allard01,marley02,tsuji02,tsuji04,tsuji05,burrows06,hellin08a,
cushing08,saumon08,madhusudhan11a,marocco14,hiranaka16,manjavacas14}.

%{\it{Spectral Type:}} 
The enhanced reddening is most striking among late-L subtypes (L5$-$L9), starting in the $K$- or $H$-band regions for slightly red and very red, respectively. The spectral classification is usually done at earlier wavelengths, optical when available or in the $Y$ and $J$ regions. While normal field L dwarfs follow a well-known colour sequence with a plateau that turns bluer at the L/T transition \citep{knapp04,chiu06,dupuy12,faherty12,faherty16a}, the URLD sequence exhibits marked red excursions. These trends align with dynamic cloud models incorporating non-equilibrium chemistry, altered grain size distributions, and varying cloud scale heights rather than a uniform cloud deck
\citep{adams23a,barman11a,barman11b,burningham17,lam26a,marley10,morley12,
phillips24,tremblin15,suarez22}. The importance of heterogeneous cloud cover is
also supported by the photometric and spectroscopic variability detected in L and T dwarfs \citep{artigau09b,biller13b,crossfield14,gizis15b}, which suggests that time-domain follow-up of the brightest URLDs in our sample would
be valuable. At the youngest ages, the similarity with directly imaged planets
such as VHS\,1256$-$1257\,b, whose JWST spectra have been analysed with both
self-consistent models and retrievals \citep{gauza15a,miles23a,petrus24,lueber24}, further illustrates how these
objects can serve as laboratories for cloud physics at low surface gravity.

\subsection{Age dependence}
\label{redLdwarfs:discussion_age}
A critical question in ultracool dwarf astrophysics is whether unusually red colours are uniquely driven by low surface gravity ($\log g$) and youth, or if secondary parameters like high metallicity or cloud geometry play a role \citep{allers13,looper08b,stephens09,burgasser08b,marocco14}.

%Intermediate-Age ($\beta$ / INT-G) analogues: 
Many of our new and recovered URLDs show photometric and spectroscopic properties typical of intermediate-age ($\beta$; INT-G) objects, whose age are comparable to the Pleiades \citep[$\sim 130$ Myr;][]{stauffer98,barrado04b}, a
cluster whose substellar sequence extends down to the planetary-mass regime
\citep{zapatero14b,zapatero17}. Their classification relies on the gravity-sensitive features calibrated on young field and cluster dwarfs
\citep{kirkpatrick08,cruz09,allers13,manjavacas20a}. These objects are frequently associated with young moving groups such as $\beta$~Pictoris
\citep[21--24\,Myr;][]{gagne18a,gagne18b,mamajek14,miret_roig24,zuckerman04,zuckerman01b},
AB~Doradus \citep[100--150\,Myr;][]{luhman05b,torres06,zuckerman04},
Tucana-Horologium \citep[$\sim$40\,Myr;][]{zuckerman01a}, and Argus \citep[40--50\,Myr;][]{gagne14d,gagne18a,malo13,torres08}. The membership of young L dwarfs
to these groups has been assessed with the BANYAN tools
\citep{gagne14a,gagne15c} and, more recently, with the MOCA database
\citep{gagne26a}. Lower surface gravity suppresses atmospheric settling, extending cloud scale heights and keeping fine dust suspended in the upper photosphere \citep{cruz09,metchev06,manjavacas14,suarez22}.

%Very Young ($\gamma$ / VL-G) Signatures: 
A subset of our URLD sample exhibits extreme reddening ($J-K_{s}$\,$>$2.0 mag) and spectral signatures indicative of very young ($\gamma$; VL-G) ages below 10 Myr, resembling low-mass members of star-forming regions like $\sigma$ Orionis \citep{zapatero99a,zapatero00,barrado01c,bejar99,bejar01,martin01a,martin03b,
lodieu09e,penya12a,zapatero17,barrado03a}, Upper Scorpius \citep{bouy22,lodieu13c,lodieu13d,lodieu21,lodieu11a,lodieu08a,lodieu18a,luhman18a,miret_roig22,penya16a}, and Taurus \citep{luhman00a,luhman04a,luhman18b,luhman03a,luhman17a,martin01b,slesnick06b,white03,esplin17a,todorov10a}. These ultra-red dwarfs serve as direct, free-floating empirical benchmarks for directly imaged young giant exoplanets,
%such as HR 8799 bcd, 2168-3832 b, or HIP 65426 b, 
which display remarkably similar red infrared spectra and low-gravity atmospheres \citep{chauvin04,marois08,marois10,barman11b,bowler16,chauvin18}.

%
%%%%%%%%%%%%%%%%%%%%%%%%%%%%%%%%%%%%%%%%%%%%
%%%%% Conclusions %%%%%
%%%%%%%%%%%%%%%%%%%%%%%%%%%%%%%%%%%%%%%%%%%%
%
\section{Conclusions}
\label{redLdwarfs:conclusions}

Our search for URLD candidates in the first data release of {\it{Euclid}} and associated photometric plus spectroscopic analysis can be summarised as follows:
\begin{itemize}
    \item [$\bullet$] {\it{Euclid}} will soon provide a massive expansion of the URLD census: with its first data release, we identified 1202 and 38 candidates in the wide and deep survey fields, respectively, representing a major increase over previous samples.
    \item [$\bullet$] Thanks to the {\it{Euclid}} red grism spectroscopic capabilities complemented by SPHEREx infrared spectra, we confirmed about 40 candidates as URLDs, significantly growing the pool of such objects verified spectroscopically.
    \item [$\bullet$] Distance-independent diagnostics reveal that these objects occupy a distinct locus due to the suppressed flux at shorter optical and near-infrared wavelengths relative to $J$, supporting atmospheric models driven by enhanced condensate opacity and thick sub-micron dust clouds.
    \item [$\bullet$] The population of URLDs exhibits diverse evolutionary characteristics from intermediate-age (Pleiades-like) to extremely red signatures typical of very young ($\leq$\,10\,Myr) low-gravity objects, making them exceptional free-floating benchmarks for young giant exoplanets.
\end{itemize}

%
%%%%%%%%%%%%%%%%%%%%%%%%%%%%%%%%%%%%%%%%%%%%
%%%%% Data availability %%%%%
%%%%%%%%%%%%%%%%%%%%%%%%%%%%%%%%%%%%%%%%%%%%
%
\section{Data availability}
{\it{Euclid}} photometry, image cutouts, and spectra will be made available either in Vizier hosted by the Centre de Donn\'ees de Strasbourg or the Spanish Virtual Observatory. Data can also be requested to the lead author in the meantime.

%
%%%%%%%%%%%%%%%%%%%%%%%%%%%%%%%%%%%%%%%%%%%%
%%%%% Acknowledgements %%%%%
%%%%%%%%%%%%%%%%%%%%%%%%%%%%%%%%%%%%%%%%%%%%
%
\begin{acknowledgements}
% National project
NL acknowledge support from the Agencia  Estatal de Investigaci\'on del Ministerio de Ciencia e Innovaci\'on (AEI-MCINN) under grants PID2022-137241NB-C41 and PID2025-176415NB-I00\@.
The authors acknowledge financial support from the Agencia Estatal de Investigación (AEI) through the Severo Ochoa Centre of Excellence accreditation awarded to the Instituto de Astrofísica de Canarias, grant CEX2025-001609-S, funded by MICIU/AEI/10.13039/501100011033.
% Thanks Adam Schneider
We thank Adam Schneider for kindly providing the spectra of URLDs published by him and his team. 
% ERC Substellar
Funding for M{\v Z}, NV, JYZ, NS, ELM, and ST was provided by the European Union (ERC Advanced Grant, SUBSTELLAR, project number 101054354; PI ELM).
% {\it{Euclid}}
This work has made use of the Euclid Data Release 1 (DR1) data from the Euclid mission of the European Space Agency (ESA), 2026 \url{https://doi.org/10.57780/esa-ab289b7}
% SPHEREx
This publication makes use of data products from the Spectro-Photometer for the History of the Universe, Epoch of Reionization and Ices Explorer (SPHEREx), which is a joint project of the Jet Propulsion Laboratory and the California Institute of Technology, and is funded by the National Aeronautics and Space Administration.
% SpeX
This study is partially based on data from the SpeX Prism Library service developed by the Spanish Virtual Observatory in the framework of the IAU Comission G5 Work-
ing Group: Spectral Stellar Libraries.
% CDS+ADS
This research has made use of the Simbad and Vizier databases, and the Aladin sky atlas operated at the centre de Donn\'ees Astronomiques de Strasbourg (CDS), and of NASA's Astrophysics Data System Bibliographic Services (ADS).
\end{acknowledgements}


%
%%%%%%%%%%%%%%%%%%%%%%%%%%%%%%%%%%%%%%%%%%%%
%%%%% Bibliography %%%%%
%%%%%%%%%%%%%%%%%%%%%%%%%%%%%%%%%%%%%%%%%%%%
%
%\bibliographystyle{aa} % style aa.bst
%\bibliography{mnemonic,biblio_old} 
\bibliography{biblio_old} 

\begin{thebibliography}{177}
\expandafter\ifx\csname natexlab\endcsname\relax\def\natexlab#1{#1}\fi

\bibitem[{{Ackerman} \& {Marley}(2001)}]{ackerman01}
{Ackerman}, A.~S. \& {Marley}, M.~S. 2001, ApJ, 556, 872

\bibitem[{{Adams} {et~al.}(2023){Adams}, {Meyer}, {Howe}, {Burningham},
  {Daemgen}, {Fortney}, {Line}, {Marley}, {Quanz}, \& {Todorov}}]{adams23a}
{Adams}, A.~D., {Meyer}, M.~R., {Howe}, A.~R., {et~al.} 2023, 166, 192

\bibitem[{{Allard} {et~al.}(2001){Allard}, {Hauschildt}, {Alexander},
  {Tamanai}, \& {Schweitzer}}]{allard01}
{Allard}, F., {Hauschildt}, P.~H., {Alexander}, D.~R., {Tamanai}, A., \&
  {Schweitzer}, A. 2001, ApJ, 556, 357

\bibitem[{{Allers} \& {Liu}(2013)}]{allers13}
{Allers}, K.~N. \& {Liu}, M.~C. 2013, ApJ, 772, 79

\bibitem[{{Andersen}(1999)}]{andersen99a}
{Andersen}, J. 1999, Transactions of the International Astronomical Union,
  Series B, 23B

\bibitem[{{Artigau} {et~al.}(2009){Artigau}, {Bouchard}, {Doyon}, \&
  {Lafreni{\`e}re}}]{artigau09b}
{Artigau}, {\'E}., {Bouchard}, S., {Doyon}, R., \& {Lafreni{\`e}re}, D. 2009,
  ApJ, 701, 1534

\bibitem[{{Baraffe} {et~al.}(2003){Baraffe}, {Chabrier}, {Barman}, {Allard}, \&
  {Hauschildt}}]{baraffe03}
{Baraffe}, I., {Chabrier}, G., {Barman}, T.~S., {Allard}, F., \& {Hauschildt},
  P.~H. 2003, A\&A, 402, 701

\bibitem[{{Barman} {et~al.}(2011{\natexlab{a}}){Barman}, {Macintosh},
  {Konopacky}, \& {Marois}}]{barman11a}
{Barman}, T.~S., {Macintosh}, B., {Konopacky}, Q.~M., \& {Marois}, C.
  2011{\natexlab{a}}, 733, 65

\bibitem[{{Barman} {et~al.}(2011{\natexlab{b}}){Barman}, {Macintosh},
  {Konopacky}, \& {Marois}}]{barman11b}
{Barman}, T.~S., {Macintosh}, B., {Konopacky}, Q.~M., \& {Marois}, C.
  2011{\natexlab{b}}, 735, L39

\bibitem[{{Barrado y Navascu{\' e}s} {et~al.}(2004){Barrado y Navascu{\' e}s},
  {Stauffer}, \& {Jayawardhana}}]{barrado04b}
{Barrado y Navascu{\' e}s}, D., {Stauffer}, J.~R., \& {Jayawardhana}, R. 2004,
  ApJ, 614, 386

\bibitem[{{Barrado y Navascu{\'e}s} {et~al.}(2003){Barrado y Navascu{\'e}s},
  {B{\' e}jar}, {Mundt}, {Mart{\'{\i}}n}, {Rebolo}, {Zapatero Osorio}, \&
  {Bailer-Jones}}]{barrado03a}
{Barrado y Navascu{\'e}s}, D., {B{\' e}jar}, V.~J.~S., {Mundt}, R., {et~al.}
  2003, A\&A, 404, 171

\bibitem[{{Barrado y Navascu{\'e}s} {et~al.}(2001){Barrado y Navascu{\'e}s},
  {Zapatero Osorio}, {B{\' e}jar}, {Rebolo}, {Mart{\'{\i}}n}, {Mundt}, \&
  {Bailer-Jones}}]{barrado01c}
{Barrado y Navascu{\'e}s}, D., {Zapatero Osorio}, M.~R., {B{\' e}jar},
  V.~J.~S., {et~al.} 2001, A\&A, 377, L9

\bibitem[{{Basri} {et~al.}(2000){Basri}, {Mohanty}, {Allard}, {Hauschildt},
  {Delfosse}, {Mart{\'{\i}}n}, {Forveille}, \& {Goldman}}]{basri00}
{Basri}, G., {Mohanty}, S., {Allard}, F., {et~al.} 2000, ApJ, 538, 363

\bibitem[{{B{\'e}jar} {et~al.}(2001){B{\'e}jar}, {Mart{\'{\i}}n}, {Zapatero
  Osorio}, {Rebolo}, {Barrado y Navascu{\' e}s}, {Bailer-Jones}, {Mundt},
  {Baraffe}, {Chabrier}, \& {Allard}}]{bejar01}
{B{\'e}jar}, V.~J.~S., {Mart{\'{\i}}n}, E.~L., {Zapatero Osorio}, M.~R.,
  {et~al.} 2001, ApJ, 556, 830

\bibitem[{{B{\'e}jar} {et~al.}(1999){B{\'e}jar}, {Zapatero Osorio}, \&
  {Rebolo}}]{bejar99}
{B{\'e}jar}, V.~J.~S., {Zapatero Osorio}, M.~R., \& {Rebolo}, R. 1999, ApJ,
  521, 671

\bibitem[{{Bell} {et~al.}(2015){Bell}, {Mamajek}, \& {Naylor}}]{bell15a}
{Bell}, C. P.~M., {Mamajek}, E.~E., \& {Naylor}, T. 2015, 454, 593

\bibitem[{{Best} {et~al.}(2017){Best}, {Liu}, {Magnier}, {Bowler}, {Aller},
  {Zhang}, {Kotson}, {Burgett}, {Chambers}, {Draper}, {Flewelling}, {Hodapp},
  {Kaiser}, {Metcalfe}, {Wainscoat}, \& {Waters}}]{best17a}
{Best}, W.~M.~J., {Liu}, M.~C., {Magnier}, E.~A., {et~al.} 2017, ApJ, 837, 95

\bibitem[{{Best} {et~al.}(2020){Best}, {Liu}, {Magnier}, \& {Dupuy}}]{best20}
{Best}, W. M.~J., {Liu}, M.~C., {Magnier}, E.~A., \& {Dupuy}, T.~J. 2020, AJ,
  159, 257

\bibitem[{{Best} {et~al.}(2021){Best}, {Liu}, {Magnier}, \& {Dupuy}}]{best21}
{Best}, W. M.~J., {Liu}, M.~C., {Magnier}, E.~A., \& {Dupuy}, T.~J. 2021, AJ,
  161, 42

\bibitem[{{Bickle} {et~al.}(2024){Bickle}, {Schneider}, {Gagn{\'e}}, {Faherty},
  {Rothermich}, {Vos}, {Su{\'a}rez}, {Kirkpatrick}, {Meisner}, {Kuchner},
  {Burgasser}, {Marocco}, {Casewell}, {Caselden}, {Bardalez Gagliuffi}, \& {The
  Backyard Worlds: Planet 9 Collaboration}}]{bickle24}
{Bickle}, T.~P., {Schneider}, A.~C., {Gagn{\'e}}, J., {et~al.} 2024, 168, 66

\bibitem[{{Biller} {et~al.}(2013){Biller}, {Crossfield}, {Mancini}, {Ciceri},
  {Southworth}, {Kopytova}, {Bonnefoy}, {Deacon}, {Schlieder}, {Buenzli},
  {Brandner}, {Allard}, {Homeier}, {Freytag}, {Bailer-Jones}, {Greiner},
  {Henning}, \& {Goldman}}]{biller13b}
{Biller}, B.~A., {Crossfield}, I.~J.~M., {Mancini}, L., {et~al.} 2013, ApJL,
  778, L10

\bibitem[{{Bock} {et~al.}(2026){Bock}, {Aboobaker}, {Adamo}, {Akeson}, {Alred},
  {Alibay}, {Ashby}, {Bach}, {Bleem}, {Bolton}, {Braun}, {Bruton}, {Bryan},
  {Chang}, {Chen}, {Cheng}, {Cheshire}, {Chiang}, {Choppin de Janvry},
  {Condon}, {Cook}, {Cooray}, {Crill}, {Cukierman}, {Dor{\'e}}, {Dowell},
  {Dubois-Felsmann}, {Eifler}, {Everett}, {Fabinsky}, {Faisst}, {Fanson},
  {Farrington}, {Fatahi}, {Fazar}, {Feder}, {Frater}, {Grasshorn Gebhardt},
  {Giri}, {Goldina}, {Gorjian}, {Habib}, {Hart}, {Heinrich}, {Hora}, {Huai},
  {Hui}, {Jo}, {Jeong}, {Kang}, {Kang}, {Kecman}, {Kim}, {Kim}, {Kim}, {Kim},
  {Kim}, {Kirkpatrick}, {Kobayashi}, {Korngut}, {Krause}, {Lee}, {Lee}, {Lee},
  {Lee}, {Lisse}, {Mariani}, {Masters}, {Mauskopf}, {Melnick}, {Minasyan},
  {Mirocha}, {Miyasaka}, {Moore}, {Moore}, {Murgia}, {Naylor}, {Nelson},
  {Nguyen}, {Nguyen}, {Noh}, {Padin}, {Paladini}, {Park}, {Penanen}, {Putnam},
  {Pyo}, {Ramachandra}, {Ramanathan}, {Rustamkulov}, {Reiley}, {Rice}, {Rocca},
  {Seok}, {Smith}, {Stober}, {Susca}, {Teplitz}, {Thelen}, {Tolls}, {Torrini},
  {Trangsrud}, {Unwin}, {Velicheti}, {Wang}, {Wen}, {Werner}, {Williams},
  {Williamson}, {Wincentsen}, {Windhorst}, {Yang}, {Yang}, \&
  {Zemcov}}]{bock26a}
{Bock}, J.~J., {Aboobaker}, A.~M., {Adamo}, J., {et~al.} 2026, 999, 139

\bibitem[{{Bouy} {et~al.}(2022){Bouy}, {Tamura}, {Barrado}, {Motohara}, {Castro
  Rodr{\'\i}guez}, {Miret-Roig}, {Konishi}, {Koyama}, {Takahashi},
  {Hu{\'e}lamo}, {Bertin}, {Olivares}, {Sarro}, {Berihuete}, {Cuillandre},
  {Galli}, {Yoshii}, \& {Miyata}}]{bouy22}
{Bouy}, H., {Tamura}, M., {Barrado}, D., {et~al.} 2022, 664, A111

\bibitem[{{Bowler}(2016)}]{bowler16}
{Bowler}, B.~P. 2016, PASP, 128, 102001

\bibitem[{{Burgasser}(2014)}]{burgasser14}
{Burgasser}, A.~J. 2014, in Astronomical Society of India Conference Series,
  Vol.~11, Astronomical Society of India Conference Series, 7--16

\bibitem[{{Burgasser} {et~al.}(2010){Burgasser}, {Cruz}, {Cushing}, {Gelino},
  {Looper}, {Faherty}, {Kirkpatrick}, \& {Reid}}]{burgasser10a}
{Burgasser}, A.~J., {Cruz}, K.~L., {Cushing}, M., {et~al.} 2010, ApJ, 710, 1142

\bibitem[{{Burgasser} {et~al.}(2002){Burgasser}, {Kirkpatrick}, {Brown},
  {Reid}, {Burrows}, {Liebert}, {Matthews}, {Gizis}, {Dahn}, {Monet}, {Cutri},
  \& {Skrutskie}}]{burgasser02}
{Burgasser}, A.~J., {Kirkpatrick}, J.~D., {Brown}, M.~E., {et~al.} 2002, ApJ,
  564, 421

\bibitem[{{Burgasser} {et~al.}(2008){Burgasser}, {Looper}, {Kirkpatrick},
  {Cruz}, \& {Swift}}]{burgasser08b}
{Burgasser}, A.~J., {Looper}, D.~L., {Kirkpatrick}, J.~D., {Cruz}, K.~L., \&
  {Swift}, B.~J. 2008, ApJ, 674, 451

\bibitem[{{Burningham} {et~al.}(2017){Burningham}, {Marley}, {Line}, {Lupu},
  {Visscher}, {Morley}, {Saumon}, \& {Freedman}}]{burningham17}
{Burningham}, B., {Marley}, M.~S., {Line}, M.~R., {et~al.} 2017, 470, 1177

\bibitem[{{Burrows} {et~al.}(2001){Burrows}, {Hubbard}, {Lunine}, \&
  {Liebert}}]{burrows01}
{Burrows}, A., {Hubbard}, W.~B., {Lunine}, J.~I., \& {Liebert}, J. 2001,
  Reviews of Modern Physics, 73, 719

\bibitem[{{Burrows} {et~al.}(2006){Burrows}, {Sudarsky}, \&
  {Hubeny}}]{burrows06}
{Burrows}, A., {Sudarsky}, D., \& {Hubeny}, I. 2006, ApJ, 640, 1063

\bibitem[{{Chabrier} \& {Baraffe}(2000)}]{chabrier00a}
{Chabrier}, G. \& {Baraffe}, I. 2000, ARA\&A, 38, 337

\bibitem[{{Chauvin} {et~al.}(2018){Chauvin}, {Gratton}, {Bonnefoy}, {Lagrange},
  {de Boer}, {Vigan}, {Beust}, {Lazzoni}, {Boccaletti}, {Galicher}, {Desidera},
  {Delorme}, {Keppler}, {Lannier}, {Maire}, {Mesa}, {Meunier}, {Kral},
  {Henning}, {Menard}, {Moor}, {Avenhaus}, {Bazzon}, {Janson}, {Beuzit},
  {Bhowmik}, {Bonavita}, {Borgniet}, {Brandner}, {Cheetham}, {Cudel}, {Feldt},
  {Fontanive}, {Ginski}, {Hagelberg}, {Janin-Potiron}, {Lagadec}, {Langlois},
  {Le Coroller}, {Messina}, {Meyer}, {Mouillet}, {Peretti}, {Perrot}, {Rodet},
  {Samland}, {Sissa}, {Olofsson}, {Salter}, {Schmidt}, {Zurlo}, {Milli}, {van
  Boekel}, {Quanz}, {Feautrier}, {Le Mignant}, {Perret}, {Ramos}, \&
  {Rochat}}]{chauvin18}
{Chauvin}, G., {Gratton}, R., {Bonnefoy}, M., {et~al.} 2018, 617, A76

\bibitem[{{Chauvin} {et~al.}(2004){Chauvin}, {Lagrange}, {Dumas}, {Zuckerman},
  {Mouillet}, {Song}, {Beuzit}, \& {Lowrance}}]{chauvin04}
{Chauvin}, G., {Lagrange}, A.-M., {Dumas}, C., {et~al.} 2004, A\&A, 425, L29

\bibitem[{{Chauvin} {et~al.}(2005){Chauvin}, {Lagrange}, {Zuckerman}, {Dumas},
  {Mouillet}, {Song}, {Beuzit}, {Lowrance}, \& {Bessell}}]{chauvin05c}
{Chauvin}, G., {Lagrange}, A.-M., {Zuckerman}, B., {et~al.} 2005, A\&A, 438,
  L29

\bibitem[{{Chauvin} {et~al.}(2015){Chauvin}, {Vigan}, {Bonnefoy}, {Desidera},
  {Bonavita}, {Mesa}, {Boccaletti}, {Buenzli}, {Carson}, {Delorme},
  {Hagelberg}, {Montagnier}, {Mordasini}, {Quanz}, {Segransan}, {Thalmann},
  {Beuzit}, {Biller}, {Covino}, {Feldt}, {Girard}, {Gratton}, {Henning},
  {Kasper}, {Lagrange}, {Messina}, {Meyer}, {Mouillet}, {Moutou}, {Reggiani},
  {Schlieder}, \& {Zurlo}}]{chauvin15}
{Chauvin}, G., {Vigan}, A., {Bonnefoy}, M., {et~al.} 2015, 573, A127

\bibitem[{{Chiu} {et~al.}(2006){Chiu}, {Fan}, {Leggett}, {Golimowski}, {Zheng},
  {Geballe}, {Schneider}, \& {Brinkmann}}]{chiu06}
{Chiu}, K., {Fan}, X., {Leggett}, S.~K., {et~al.} 2006, AJ, 131, 2722

\bibitem[{{Crossfield} {et~al.}(2014){Crossfield}, {Biller}, {Schlieder},
  {Deacon}, {Bonnefoy}, {Homeier}, {Allard}, {Buenzli}, {Henning}, {Brandner},
  {Goldman}, \& {Kopytova}}]{crossfield14}
{Crossfield}, I.~J.~M., {Biller}, B., {Schlieder}, J.~E., {et~al.} 2014, 505,
  654

\bibitem[{{Cruz} {et~al.}(2009){Cruz}, {Kirkpatrick}, \& {Burgasser}}]{cruz09}
{Cruz}, K.~L., {Kirkpatrick}, J.~D., \& {Burgasser}, A.~J. 2009, AJ, 137, 3345

\bibitem[{{Cruz} {et~al.}(2007){Cruz}, {Reid}, {Kirkpatrick}, {Burgasser},
  {Liebert}, {Solomon}, {Schmidt}, {Allen}, {Hawley}, \& {Covey}}]{cruz07}
{Cruz}, K.~L., {Reid}, I.~N., {Kirkpatrick}, J.~D., {et~al.} 2007, AJ, 133, 439

\bibitem[{{Cruz} {et~al.}(2003){Cruz}, {Reid}, {Liebert}, {Kirkpatrick}, \&
  {Lowrance}}]{cruz03}
{Cruz}, K.~L., {Reid}, I.~N., {Liebert}, J., {Kirkpatrick}, J.~D., \&
  {Lowrance}, P.~J. 2003, AJ, 126, 2421

\bibitem[{{Cushing} {et~al.}(2011){Cushing}, {Kirkpatrick}, {Gelino},
  {Griffith}, {Skrutskie}, {Mainzer}, {Marsh}, {Beichman}, {Burgasser},
  {Prato}, {Simcoe}, {Marley}, {Saumon}, {Freedman}, {Eisenhardt}, \&
  {Wright}}]{cushing11}
{Cushing}, M.~C., {Kirkpatrick}, J.~D., {Gelino}, C.~R., {et~al.} 2011, ApJ,
  743, 50

\bibitem[{{Cushing} {et~al.}(2008){Cushing}, {Marley}, {Saumon}, {Kelly},
  {Vacca}, {Rayner}, {Freedman}, {Lodders}, \& {Roellig}}]{cushing08}
{Cushing}, M.~C., {Marley}, M.~S., {Saumon}, D., {et~al.} 2008, ApJ, 678, 1372

\bibitem[{{Cushing} {et~al.}(2005){Cushing}, {Rayner}, \& {Vacca}}]{cushing05}
{Cushing}, M.~C., {Rayner}, J.~T., \& {Vacca}, W.~D. 2005, ApJ, 623, 1115

\bibitem[{{Cutri} {et~al.}(2013){Cutri}, {Wright}, {Conrow}, {Fowler},
  {Eisenhardt}, {Grillmair}, {Kirkpatrick}, {Masci}, {McCallon}, \&
  {Wheelock}}]{cutri13}
{Cutri}, R.~M., {Wright}, E.~L., {Conrow}, T., {et~al.} 2013, {Explanatory
  Supplement to the AllWISE Data Release Products}, Explanatory Supplement to
  the AllWISE Data Release Products

\bibitem[{{Delfosse} {et~al.}(1997){Delfosse}, {Tinney}, {Forveille},
  {Epchtein}, {Bertin}, {Borsenberger}, {Copet}, {de Batz}, {Fouque},
  {Kimeswenger}, {Le Bertre}, {Lacombe}, {Rouan}, \& {Tiphene}}]{delfosse97}
{Delfosse}, X., {Tinney}, C.~G., {Forveille}, T., {et~al.} 1997, A\&A, 327, L25

\bibitem[{{Dominguez-Tagle} {et~al.}(2025){Dominguez-Tagle}, {{\v{Z}}erjal},
  {Sedighi}, {Mas-Buitrago}, {Martin}, {Zhang}, {Vitas}, {B{\'e}jar}, {Tsilia},
  {Mu{\~n}oz Torres}, \& et~al.}]{dominguez_tagle25}
{Dominguez-Tagle}, C., {{\v{Z}}erjal}, M., {Sedighi}, N., {et~al.} 2025, 991,
  84

\bibitem[{{Dupuy} \& {Liu}(2012)}]{dupuy12}
{Dupuy}, T.~J. \& {Liu}, M.~C. 2012, ApJS, 201, 19

\bibitem[{{Eisenhardt} {et~al.}(2020){Eisenhardt}, {Marocco}, {Fowler},
  {Meisner}, {Kirkpatrick}, {Garcia}, {Jarrett}, {Koontz}, {Marchese},
  {Stanford}, {Caselden}, {Cushing}, {Cutri}, {Faherty}, {Gelino}, {Gonzalez},
  {Mainzer}, {Mobasher}, {Schlegel}, {Stern}, {Teplitz}, \&
  {Wright}}]{eisenhardt20}
{Eisenhardt}, P. R.~M., {Marocco}, F., {Fowler}, J.~W., {et~al.} 2020, ApJS,
  247, 69

\bibitem[{{Esplin} \& {Luhman}(2017)}]{esplin17a}
{Esplin}, T.~L. \& {Luhman}, K.~L. 2017, AJ, 154, 134

\bibitem[{{Euclid Collaboration} {et~al.}(2026){Euclid Collaboration}, {Copin},
  {Fumana}, {Mancini}, {Appleton}, {Chary}, {Conseil}, {Faisst}, {Hemmati},
  {Masters}, {Scarlata}, {Scodeggio}, {Alavi}, {Carle}, {Casenove}, {Contini},
  {Das}, {Gillard}, {Herzog}, {Jacobson}, {Le Brun}, {Maino}, {Setnikar},
  {Stickley}, {Tavagnacco}, {Xie}, {Aghanim}, {Altieri}, {Amara}, {Andreon},
  {Auricchio}, {Aussel}, {Baccigalupi}, {Baldi}, {Balestra}, {Bardelli},
  {Basset}, {Battaglia}, {Belikov}, {Biviano}, {Bonchi}, {Branchini},
  {Brescia}, {Brinchmann}, {Camera}, {Ca{\~n}as-Herrera}, {Capobianco},
  {Carbone}, {Carretero}, {Casas}, {Castander}, {Castellano}, {Castignani},
  {Cavuoti}, {Chambers}, {Cimatti}, {Colodro-Conde}, {Congedo}, {Conselice},
  {Conversi}, {Courbin}, {Courtois}, {Da Silva}, {da Silva}, {Degaudenzi}, {de
  la Torre}, {De Lucia}, {Di Giorgio}, {Dole}, {Dubath}, {Dupac}, {Dusini},
  {Ealet}, {Escoffier}, {Farina}, {Farinelli}, {Ferriol}, {Finelli},
  {Fotopoulou}, {Fourmanoit}, {Frailis}, {Franceschi}, {Franzetti}, {Galeotta},
  {George}, {Gillis}, {Giocoli}, {Gracia-Carpio}, {Granett}, {Grazian},
  {Grupp}, {Guzzo}, {Haugan}, {Hoar}, {Hoekstra}, {Holmes}, {Hook}, {Hormuth},
  {Hornstrup}, {Hudelot}, {Jahnke}, {Jhabvala}, {Joachimi}, {Keih{\"a}nen},
  {Kermiche}, {Kiessling}, {Kubik}, {Kuijken}, {K{\"u}mmel}, {Kunz},
  {Kurki-Suonio}, {Le Boulc'h}, {Le Brun}, {Le Mignant}, {Ligori}, {Lilje},
  {Lindholm}, {Lloro}, {Mainetti}, {Maiorano}, {Mansutti}, {Marcin},
  {Marggraf}, {Markovic}, {Martinelli}, {Martinet}, {Marulli}, {Massey},
  {Maurogordato}, {Medinaceli}, {Mei}, {Melchior}, {Mellier}, {Meneghetti},
  {Merlin}, {Meylan}, {Mora}, {Moresco}, {Morris}, {Moscardini}, {Nakajima},
  {Neissner}, {Nichol}, {Niemi}, {Nightingale}, {Padilla}, {Paltani}, {Pasian},
  {Pedersen}, {Percival}, {Pettorino}, {Pires}, {Polenta}, {Poncet}, {Popa},
  {Pozzetti}, {Racca}, {Raison}, {Rebolo}, {Renzi}, {Rhodes}, {Riccio},
  {Romelli}, {Roncarelli}, {Rossetti}, {Saglia}, {Sakr}, {S{\'a}nchez},
  {Sapone}, {Sartoris}, {Schewtschenko}, {Schirmer}, {Schneider}, {Schrabback},
  {Secroun}, {Sefusatti}, {Seidel}, {Serrano}, {Simon}, {Sirignano}, {Sirri},
  {Spurio Mancini}, {Stanco}, {Steinwagner}, {Tallada-Cresp{\'\i}}, {Taylor},
  {Teplitz}, {Tereno}, {Tessore}, {Toft}, {Toledo-Moreo}, {Torradeflot},
  {Tutusaus}, {Valenziano}, {Valiviita}, {Vassallo}, {Verdoes Kleijn},
  {Veropalumbo}, {Wang}, \& {Weller}}]{copin26}
{Euclid Collaboration}, {Copin}, Y., {Fumana}, M., {et~al.} 2026, 711, A6

\bibitem[{{Euclid Collaboration} {et~al.}(2025{\natexlab{a}}){Euclid
  Collaboration}, {Cropper}, {Al-Bahlawan}, {Amiaux}, {Awan}, {Azzollini},
  {Benson}, {Berthe}, {Boucher}, {Bozzo}, {Brockley-Blatt}, {Candini}, {Cara},
  {Chaudery}, {Cole}, {Danto}, {Denniston}, {Di Giorgio}, {Dryer}, {Dubois},
  {Endicott}, {Farina}, {Galli}, {Genolet}, {Gow}, {Guttridge}, {Hailey},
  {Hall}, {Harper}, {Hoekstra}, {Holland}, {Horeau}, {Hu}, {James}, {Khalil},
  {King}, {Kitching}, {Kohley}, {Larcheveque}, {Lawrenson}, {Liebing}, {Liu},
  {Martignac}, {Massey}, {McCracken}, {Miller}, {Murray}, {Nakajima}, {Niemi},
  {Nightingale}, {Paltani}, {Pendem}, {Philippon}, {Plana}, {Pool},
  {Pottinger}, {Racca}, {Rhodes}, {Rousseau}, {Ruane}, {Salatti}, {Salvignol},
  {Sciortino}, {Short}, {Skottfelt}, {Smit}, {Swindells}, {Szafraniec},
  {Thomas}, {Thomas}, {Tommasi}, {Tosti}, {Visticot}, {Walton}, {Willis},
  {Winter}, {Aghanim}, {Altieri}, {Amara}, {Andreon}, {Auricchio}, {Aussel},
  {Baccigalupi}, {Baldi}, {Balestra}, {Bardelli}, {Basset}, {Bender},
  {Bernardeau}, {Bodendorf}, {Boenke}, {Bonino}, {Branchini}, {Brescia},
  {Brinchmann}, {Camera}, {Capobianco}, {Carbone}, {Cardone}, {Carretero},
  {Casas}, {Casas}, {Castander}, {Castellano}, {Castignani}, {Cavuoti},
  {Cimatti}, {Colodro-Conde}, {Congedo}, {Conselice}, {Conversi}, {Copin},
  {Courbin}, {Courtois}, {Crocce}, {Cuby}, {Cuillandre}, {Da Silva},
  {Degaudenzi}, {De Lucia}, {Dinis}, {Dolding}, {Douspis}, {Duncan}, {Dupac},
  {Dusini}, {Ealet}, {Fabricius}, {Farrens}, {Ferriol}, {Fosalba},
  {Fotopoulou}, {Frailis}, {Franceschi}, {Franzetti}, {Frugier}, {Fumana},
  {Galeotta}, {Garilli}, {George}, {Gillard}, {Gillis}, {Giocoli},
  {G{\'o}mez-Alvarez}, {Granett}, {Grazian}, {Grupp}, {Guzzo}, {Haugan},
  {Herent}, {Hoar}, {Holliman}, {Holmes}, {Hook}, {Hormuth}, {Hornstrup},
  {Hudelot}, {Ili{\'c}}, {Jahnke}, {Jhabvala}, {Joachimi}, {Keih{\"a}nen},
  {Kermiche}, {Kiessling}, {Kilbinger}, {Kubik}, {Kuijken}, {K{\"u}mmel},
  {Kunz}, {Kurki-Suonio}, {Lahav}, {Laureijs}, {Ligori}, {Lilje}, {Lindholm},
  {Lloro}, {Lorenzo Alvarez}, {Mainetti}, {Maino}, {Maiorano}, {Mansutti},
  {Marcin}, {Marggraf}, {Markovic}, {Martinelli}, {Martinet}, {Marulli},
  {Masters}, {Maurogordato}, {Medinaceli}, {Mei}, {Melchior}, {Mellier},
  {Meneghetti}, {Merlin}, {Meylan}, {Mohr}, {Moresco}, {Moscardini}, \&
  {Neissner}}]{cropper25}
{Euclid Collaboration}, {Cropper}, M.~S., {Al-Bahlawan}, A., {et~al.}
  2025{\natexlab{a}}, 697, A2

\bibitem[{{Euclid Collaboration} {et~al.}(2025{\natexlab{b}}){Euclid
  Collaboration}, {Hormuth}, {Jahnke}, {Schirmer}, {Lee}, {Scott}, {Barbier},
  {Ferriol}, {Gillard}, {Grupp}, {Holmes}, {Holmes}, {Kubik}, {Macias-Perez},
  {Laurent}, {Marpaud}, {Marton}, {Medinaceli}, {Morgante}, {Toledo-Moreo},
  {Trifoglio}, {Rix}, {Secroun}, {Seiffert}, {Stassi}, {Wachter}, {Gutierrez},
  {Vescovi}, {Amara}, {Andreon}, {Auricchio}, {Baccigalupi}, {Baldi},
  {Balestra}, {Bardelli}, {Battaglia}, {Bender}, {Bodendorf}, {Bonino},
  {Branchini}, {Brescia}, {Brinchmann}, {Camera}, {Capobianco}, {Carbone},
  {Cardone}, {Carretero}, {Casas}, {Casas}, {Castellano}, {Castignani},
  {Cavuoti}, {Cimatti}, {Colodro-Conde}, {Congedo}, {Conselice}, {Conversi},
  {Copin}, {Corcione}, {Courbin}, {Courtois}, {Da Silva}, {Degaudenzi}, {De
  Lucia}, {Dinis}, {Douspis}, {Dubath}, {Ducret}, {Dupac}, {Dusini},
  {Fabricius}, {Farina}, {Farrens}, {Faustini}, {Fotopoulou}, {Fourmanoit},
  {Frailis}, {Franceschi}, {Franzetti}, {Fumana}, {Galeotta}, {Garilli},
  {George}, {Gillis}, {Giocoli}, {Grazian}, {Guzzo}, {Haugan}, {Hoekstra},
  {Hook}, {Hornstrup}, {Hudelot}, {Jhabvala}, {Keih{\"a}nen}, {Kermiche},
  {Kiessling}, {Kilbinger}, {Kitching}, {Kohley}, {K{\"u}mmel}, {Kunz},
  {Kurki-Suonio}, {Le Mignant}, {Ligori}, {Lilje}, {Lindholm}, {Lloro},
  {Mainetti}, {Maiorano}, {Mansutti}, {Marcin}, {Marggraf}, {Markovic},
  {Martinelli}, {Martinet}, {Marulli}, {Massey}, {Maurogordato}, {McCracken},
  {Mei}, {Melchior}, {Mellier}, {Meneghetti}, {Merlin}, {Meylan}, {Mohr},
  {Moresco}, {Morris}, {Moscardini}, {Munari}, {Nakajima}, {Neissner},
  {Nichol}, {Niemi}, {Nightingale}, {Padilla}, {Paech}, {Paltani}, {Pasian},
  {Pedersen}, {Percival}, {Pettorino}, {Pires}, {Polenta}, {Poncet}, {Popa},
  {Raison}, {Rebolo}, {Renzi}, {Rhodes}, {Riccio}, {Romelli}, {Roncarelli},
  {Rossetti}, {Rusholme}, {Saglia}, {Sakr}, {S{\'a}nchez}, {Sapone},
  {Sartoris}, {Sauvage}, {Schewtschenko}, {Schneider}, {Schrabback},
  {Sefusatti}, {Seidel}, {Serrano}, {Sirignano}, {Sirri}, {Smadja}, {Stanco},
  {Steinwagner}, {Tallada-Cresp{\'\i}}, {Tavagnacco}, {Taylor}, {Teplitz},
  {Tereno}, {Torradeflot}, {Tutusaus}, {Valenziano}, {Vassallo}, {Veropalumbo},
  {Wang}, {Weller}, {Zacchei}, {Zamorani}, {Zerbi}, {Zucca}, {Biviano},
  {Bolzonella}, {Boucaud}, {Bozzo}, {Burigana}, {Calabrese}, {Di Ferdinando},
  {Escartin Vigo}, {Farinelli}, {Gracia-Carpio}, {Kazandjian}, \&
  {Mauri}}]{hormuth25}
{Euclid Collaboration}, {Hormuth}, F., {Jahnke}, K., {et~al.}
  2025{\natexlab{b}}, 697, A4

\bibitem[{{Euclid Collaboration} {et~al.}(2025{\natexlab{c}}){Euclid
  Collaboration}, {Jahnke}, {Gillard}, {Schirmer}, {Ealet}, {Maciaszek},
  {Prieto}, {Barbier}, {Bonoli}, {Corcione}, {Dusini}, {Grupp}, {Hormuth},
  {Ligori}, {Martin}, {Morgante}, {Padilla}, {Toledo-Moreo}, {Trifoglio},
  {Valenziano}, {Bender}, {Castander}, {Garilli}, {Lilje}, {Rix}, {Andersen},
  {Auricchio}, {Balestra}, {Barriere}, {Battaglia}, {Berthe}, {Bodendorf},
  {Boenke}, {Bon}, {Bonnefoi}, {Caillat}, {Capobianco}, {Carle}, {Casas},
  {Cho}, {Costille}, {Ducret}, {Ferriol}, {Franceschi}, {Gimenez}, {Holmes},
  {Hornstrup}, {Jhabvala}, {Kohley}, {Kubik}, {Laureijs}, {Le Mignant},
  {Lloro}, {Medinaceli}, {Mellier}, {Polenta}, {Racca}, {Renzi}, {Salvignol},
  {Secroun}, {Seidel}, {Seiffert}, {Sirignano}, {Sirri}, {Strada}, {Smadja},
  {Stanco}, {Wachter}, {Anselmi}, {Borsato}, {Caillat}, {Cogato},
  {Colodro-Conde}, {Crouzet}, {Conforti}, {D'Alessandro}, {Copin},
  {Cuillandre}, {Davies}, {Davini}, {Derosa}, {Diaz}, {Di Domizio}, {Di
  Ferdinando}, {Farinelli}, {Ferrari}, {Fornari}, {Gabarra}, {Garcia},
  {Gutierrez}, {Giacomini}, {Lagier}, {Gianotti}, {Krause}, {Madrid},
  {Laudisio}, {Macias-Perez}, {Naletto}, {Niclas}, {Marpaud}, {Mauri}, {da
  Silva}, {Passalacqua}, {Paterson}, {Patrizii}, {Risso}, {Solheim},
  {Scodeggio}, {Stassi}, {Steinwagner}, {Tenti}, {Testera}, {Travaglini},
  {Tosi}, {Troja}, {Tubio}, {Valieri}, {Vescovi}, {Ventura}, {Aghanim},
  {Altieri}, {Amara}, {Amiaux}, {Andreon}, {Appleton}, {Aussel}, {Baccigalupi},
  {Baldi}, {Bardelli}, {Basset}, {Bonchi}, {Bonino}, {Branchini}, {Brescia},
  {Brinchmann}, {Camera}, {Carbone}, {Cardone}, {Carretero}, {Casas},
  {Castellano}, {Castignani}, {Cavuoti}, {Chabaud}, {Cimatti}, {Congedo},
  {Conselice}, {Conversi}, {Courbin}, {Courtois}, {Crocce}, {Cropper}, {Cuby},
  {Da Silva}, {Degaudenzi}, {De Lucia}, {Di Giorgio}, {Dinis}, {Douspis},
  {Dubath}, {Duncan}, {Dupac}, {Fabricius}, {Farina}, {Farrens}, {Faustini},
  {Fosalba}, {Fotopoulou}, {Fourmanoit}, {Frailis}, {Franzetti}, {Galeotta},
  {George}, {Gillis}, {Giocoli}, {G{\'o}mez-Alvarez}, {Granett}, {Grazian},
  {Guzzo}, {Hailey}, {Haugan}, {Hoar}, {Hoekstra}, {Hook}, {Hudelot},
  {Ili{\'c}}, {Joachimi}, {Keih{\"a}nen}, {Kermiche}, {Kiessling}, {Kilbinger},
  {Kitching}, {K{\"u}mmel}, {Kunz}, {Kurki-Suonio}, {Lahav}, {Liebing},
  {Lindholm}, {Lorenzo Alvarez}, \& {Mainetti}}]{jahnke25}
{Euclid Collaboration}, {Jahnke}, K., {Gillard}, W., {et~al.}
  2025{\natexlab{c}}, 697, A3

\bibitem[{{Euclid Collaboration} {et~al.}(2025{\natexlab{d}}){Euclid
  Collaboration}, {Mellier}, {Abdurro'uf}, {Acevedo Barroso}, {Ach{\'u}carro},
  {Adamek}, {Adam}, {Addison}, {Aghanim}, {Aguena}, {Ajani}, {Akrami},
  {Al-Bahlawan}, {Alavi}, {Albuquerque}, {Alestas}, {Alguero}, {Allaoui},
  {Allen}, {Allevato}, {Alonso-Tetilla}, {Altieri}, {Alvarez-Candal}, {Alvi},
  {Amara}, {Amendola}, {Amiaux}, {Andika}, {Andreon}, {Andrews}, {Angora},
  {Angulo}, {Annibali}, {Anselmi}, {Anselmi}, {Arcari}, {Archidiacono},
  {Aric{\`o}}, {Arnaud}, {Arnouts}, {Asgari}, {Asorey}, {Atayde}, {Atek},
  {Atrio-Barandela}, {Aubert}, {Aubourg}, {Auphan}, {Auricchio}, {Aussel},
  {Aussel}, {Avelino}, {Avgoustidis}, {Avila}, {Awan}, {Azzollini},
  {Baccigalupi}, {Bachelet}, {Bacon}, {Baes}, {Bagley}, {Bahr-Kalus},
  {Balaguera-Antolinez}, {Balbinot}, {Balcells}, {Baldi}, {Baldry}, {Balestra},
  {Ballardini}, {Ballester}, {Balogh}, {Ba{\~n}ados}, {Barbier}, {Bardelli},
  {Baron}, {Barreiro}, {Barrena}, {Barriere}, {Barros}, {Barthelemy},
  {Bartolo}, {Basset}, {Battaglia}, {Battisti}, {Baugh}, {Baumont},
  {Bazzanini}, {Beaulieu}, {Beckmann}, {Belikov}, {Bel}, {Bellagamba}, {Bella},
  {Bellini}, {Benabed}, {Bender}, {Benevento}, {Bennett}, {Benson},
  {Bergamini}, {Bermejo-Climent}, {Bernardeau}, {Bertacca}, {Berthe},
  {Berthier}, {Bethermin}, {Beutler}, {Bevillon}, {Bhargava}, {Bhatawdekar},
  {Bianchi}, {Bisigello}, {Biviano}, {Blake}, {Blanchard}, {Blazek}, {Blot},
  {Bosco}, {Bodendorf}, {Boenke}, {B{\"o}hringer}, {Boldrini}, {Bolzonella},
  {Bonchi}, {Bonici}, {Bonino}, {Bonino}, {Bonvin}, {Bon}, {Booth}, {Borgani},
  {Borlaff}, {Borsato}, {Bose}, {Botticella}, {Boucaud}, {Bouche}, {Boucher},
  {Boutigny}, {Bouvard}, {Bouwens}, {Bouy}, {Bowler}, {Bozza}, {Bozzo},
  {Branchini}, {Brando}, {Brau-Nogue}, {Brekke}, {Bremer}, {Brescia}, {Breton},
  {Brinchmann}, {Brinckmann}, {Brockley-Blatt}, {Brodwin}, {Brouard}, {Brown},
  {Bruton}, {Bucko}, {Buddelmeijer}, {Buenadicha}, {Buitrago}, {Burger},
  {Burigana}, {Busillo}, {Busonero}, {Cabanac}, {Cabayol-Garcia}, {Cagliari},
  {Caillat}, {Caillat}, {Calabrese}, {Calabro}, {Calderone}, {Calura}, {Camacho
  Quevedo}, {Camera}, {Campos}, {Ca{\~n}as-Herrera}, {Candini}, {Cantiello},
  {Capobianco}, {Cappellaro}, {Cappelluti}, {Cappi}, {Caputi}, {Cara},
  {Carbone}, {Cardone}, {Carella}, {Carlberg}, {Carle}, {Carminati}, {Caro},
  {Carrasco}, {Carretero}, {Carrilho}, {Carron Duque}, \& {Carry}}]{mellier25}
{Euclid Collaboration}, {Mellier}, Y., {Abdurro'uf}, {et~al.}
  2025{\natexlab{d}}, 697, A1

\bibitem[{{Euclid Collaboration} {et~al.}(2022){Euclid Collaboration},
  {Schirmer}, {Jahnke}, {Seidel}, {Aussel}, {Bodendorf}, {Grupp}, {Hormuth},
  {Wachter}, {Appleton}, {Barbier}, {Brinchmann}, {Carrasco}, {Castander},
  {Coupon}, {De Paolis}, {Franco}, {Ganga}, {Hudelot}, {Jullo}, {Lan{\c{c}}on},
  {Nucita}, {Paltani}, {Smadja}, {Strafella}, {Venancio}, {Weiler}, {Amara},
  {Auphan}, {Auricchio}, {Balestra}, {Bender}, {Bonino}, {Branchini},
  {Brescia}, {Capobianco}, {Carbone}, {Carretero}, {Casas}, {Castellano},
  {Cavuoti}, {Cimatti}, {Cledassou}, {Congedo}, {Conselice}, {Conversi},
  {Copin}, {Corcione}, {Costille}, {Courbin}, {Da Silva}, {Degaudenzi},
  {Douspis}, {Dubath}, {Dupac}, {Dusini}, {Ealet}, {Farrens}, {Ferriol},
  {Fosalba}, {Frailis}, {Franceschi}, {Franzetti}, {Fumana}, {Garilli},
  {Gillard}, {Gillis}, {Giocoli}, {Grazian}, {Guzzo}, {Haugan}, {Hoekstra},
  {Holmes}, {Hornstrup}, {K{\"u}mmel}, {Kermiche}, {Kiessling}, {Kilbinger},
  {Kitching}, {Kohley}, {Kunz}, {Kurki-Suonio}, {Laureijs}, {Ligori}, {Lilje},
  {Lloro}, {Maciaszek}, {Maiorano}, {Mansutti}, {Marggraf}, {Markovic},
  {Marulli}, {Massey}, {Maurogordato}, {Mellier}, {Meneghetti}, {Merlin},
  {Meylan}, {Moresco}, {Moscardini}, {Munari}, {Nakajima}, {Nichol}, {Niemi},
  {Padilla}, {Pasian}, {Pedersen}, {Percival}, {Pettorino}, {Pires}, {Poncet},
  {Popa}, {Pozzetti}, {Prieto}, {Raison}, {Rhodes}, {Rix}, {Roncarelli},
  {Rossetti}, {Saglia}, {Sartoris}, {Scaramella}, {Schneider}, {Secroun},
  {Serrano}, {Sirignano}, {Sirri}, {Stanco}, {Tallada-Cresp{\'\i}}, {Taylor},
  {Teplitz}, {Tereno}, {Toledo-Moreo}, {Torradeflot}, {Trifoglio}, {Valentijn},
  {Valenziano}, {Wang}, {Weller}, {Zamorani}, {Zoubian}, {Andreon}, {Bardelli},
  {Boucaud}, {Camera}, {Farinelli}, {Graci{\'a}-Carpio}, {Maino}, {Medinaceli},
  {Mei}, {Morisset}, {Polenta}, {Renzi}, {Romelli}, {Tenti}, {Vassallo},
  {Zacchei}, {Zucca}, {Baccigalupi}, {Balaguera-Antol{\'\i}nez}, {Biviano},
  {Blanchard}, {Borgani}, {Bozzo}, {Burigana}, {Cabanac}, {Cappi}, {Carvalho},
  {Casas}, {Castignani}, {Colodro-Conde}, {Cooray}, {Courtois}, {Crocce},
  {Cuby}, {Davini}, {de la Torre}, {Di Ferdinando}, {Escartin}, {Farina},
  {Ferreira}, {Finelli}, {Fotopoulou}, {Galeotta}, {Garcia-Bellido},
  {Gaztanaga}, {George}, {Gozaliasl}, {Hook}, {Ili{\'c}}, {Kansal},
  {Kashlinsky}, {Keihanen}, {Kirkpatrick}, {Lindholm}, {Mainetti}, {Maoli},
  {Martinelli}, {Martinet}, \& {Maturi}}]{schirmer22}
{Euclid Collaboration}, {Schirmer}, M., {Jahnke}, K., {et~al.} 2022, 662, A92

\bibitem[{{Faherty} {et~al.}(2012){Faherty}, {Burgasser}, {Walter}, {Van der
  Bliek}, {Shara}, {Cruz}, {West}, {Vrba}, \& {Anglada-Escud{\'e}}}]{faherty12}
{Faherty}, J.~K., {Burgasser}, A.~J., {Walter}, F.~M., {et~al.} 2012, ApJ, 752,
  56

\bibitem[{{Faherty} {et~al.}(2016){Faherty}, {Riedel}, {Cruz}, {Gagne},
  {Filippazzo}, {Lambrides}, {Fica}, {Weinberger}, {Thorstensen}, {Tinney},
  {Baldassare}, {Lemonier}, \& {Rice}}]{faherty16a}
{Faherty}, J.~K., {Riedel}, A.~R., {Cruz}, K.~L., {et~al.} 2016, ApJS, 225, 10

\bibitem[{{Filippazzo} {et~al.}(2015){Filippazzo}, {Rice}, {Faherty}, {Cruz},
  {Van Gordon}, \& {Looper}}]{filippazzo15}
{Filippazzo}, J.~C., {Rice}, E.~L., {Faherty}, J., {et~al.} 2015, ApJ, 810, 158

\bibitem[{{Gagn{\'e}} {et~al.}(2015){Gagn{\'e}}, {Faherty}, {Cruz},
  {Lafreni{\'e}re}, {Doyon}, {Malo}, {Burgasser}, {Naud}, {Artigau},
  {Bouchard}, {Gizis}, \& {Albert}}]{gagne15c}
{Gagn{\'e}}, J., {Faherty}, J.~K., {Cruz}, K.~L., {et~al.} 2015, ApJS, 219, 33

\bibitem[{{Gagn{\'e}} {et~al.}(2020){Gagn{\'e}}, {Faherty}, \&
  {Popinchalk}}]{gagne20b}
{Gagn{\'e}}, J., {Faherty}, J.~K., \& {Popinchalk}, M. 2020, Research Notes of
  the American Astronomical Society, 4, 92

\bibitem[{{Gagn{\'e}} {et~al.}(2026{\natexlab{a}}){Gagn{\'e}}, {Faherty}, {Ruiz
  Diaz}, {Coulombe}, {Bickle}, {Schneider}, {Kirkpatrick}, {Kuchner},
  {Meisner}, {Caselden}, {Burgasser}, {Casewell}, {Honaker}, {Kiwy}, {Marocco},
  {Bardalez Gagliuffi}, {Stevnbak Andersen}, {Ruiz Arroyo}, {Baller},
  {Beaulieu}, {Bell}, {Bilsing}, {Bohling}, {Colin}, {Colombo}, {Deen},
  {Dereveanco}, {Dixon}, {Durantini Luca}, {Flores}, {Franck}, {Fulvi},
  {Gallmann}, {Gantier}, {Glebov}, {Gramaize}, {Hamlet}, {Hinckley},
  {Jablonski}, {Ja{\l}owiczor}, {Kabatnik}, {Kasprowitz}, {Ly}, {Martin},
  {Marzak}, {McColgan}, {McEwan}, {Michaels}, {Pendrill}, {Perlin}, {Pumphrey},
  {Rabe}, {Raway}, {Robledo}, {Roser}, {Roy}, {Sainio}, {Schindler}, {Schonau},
  {Sch{\"u}mann}, {Selg-Mann}, {Serio}, {Smith}, {Stenner}, {Tanner},
  {Th{\'e}venot}, {Thakur}, {Torres Guerrero}, {Ventura}, {Voloshin}, {Walla},
  {W{\c{e}}dracki}, {Weyandt}, {Wilhite}, \& {Zitouni}}]{gagne26b}
{Gagn{\'e}}, J., {Faherty}, J.~K., {Ruiz Diaz}, A., {et~al.}
  2026{\natexlab{a}}, arXiv e-prints, arXiv:2604.22012

\bibitem[{{Gagn{\'e}} {et~al.}(2014{\natexlab{a}}){Gagn{\'e}},
  {Lafreni{\`e}re}, {Doyon}, {Artigau}, {Malo}, {Robert}, \&
  {Nadeau}}]{gagne14d}
{Gagn{\'e}}, J., {Lafreni{\`e}re}, D., {Doyon}, R., {et~al.}
  2014{\natexlab{a}}, ApJL, 792, L17

\bibitem[{{Gagn{\'e}} {et~al.}(2014{\natexlab{b}}){Gagn{\'e}},
  {Lafreni{\`e}re}, {Doyon}, {Malo}, \& {Artigau}}]{gagne14a}
{Gagn{\'e}}, J., {Lafreni{\`e}re}, D., {Doyon}, R., {Malo}, L., \& {Artigau},
  {\'E}. 2014{\natexlab{b}}, ApJ, 783, 121

\bibitem[{{Gagn{\'e}} {et~al.}(2018{\natexlab{a}}){Gagn{\'e}}, {Mamajek},
  {Malo}, {Riedel}, {Rodriguez}, {Lafreni{\`e}re}, {Faherty}, {Roy-Loubier},
  {Pueyo}, {Robin}, \& {Doyon}}]{gagne18a}
{Gagn{\'e}}, J., {Mamajek}, E.~E., {Malo}, L., {et~al.} 2018{\natexlab{a}},
  ApJ, 856, 23

\bibitem[{{Gagn{\'e}} {et~al.}(2026{\natexlab{b}}){Gagn{\'e}}, {Moranta},
  {Faherty}, {Curtis}, {Bickle}, {Couture}, {Chiasson David}, {Christie},
  {Lambier}, {Leclerc}, {Poliquin}, {Belzile}, \& {Mamajek}}]{gagne26a}
{Gagn{\'e}}, J., {Moranta}, L., {Faherty}, J.~K., {et~al.} 2026{\natexlab{b}},
  287, 8

\bibitem[{{Gagn{\'e}} {et~al.}(2018{\natexlab{b}}){Gagn{\'e}}, {Roy-Loubier},
  {Faherty}, {Doyon}, \& {Malo}}]{gagne18b}
{Gagn{\'e}}, J., {Roy-Loubier}, O., {Faherty}, J.~K., {Doyon}, R., \& {Malo},
  L. 2018{\natexlab{b}}, ApJ, 860, 43

\bibitem[{{Gaia Collaboration} {et~al.}(2021{\natexlab{a}}){Gaia
  Collaboration}, {Brown}, {Vallenari}, {Prusti}, {de Bruijne}, {Babusiaux},
  {Biermann}, {Creevey}, {Evans}, {Eyer}, {Hutton}, {Jansen}, {Jordi},
  {Klioner}, {Lammers}, {Lindegren}, {Luri}, {Mignard}, {Panem}, {Pourbaix},
  {Randich}, {Sartoretti}, {Soubiran}, {Walton}, {Arenou}, {Bailer-Jones},
  {Bastian}, {Cropper}, {Drimmel}, {Katz}, {Lattanzi}, {van Leeuwen}, {Bakker},
  {Cacciari}, {Casta{\~n}eda}, {De Angeli}, {Ducourant}, {Fabricius},
  {Fouesneau}, {Fr{\'e}mat}, {Guerra}, {Guerrier}, {Guiraud}, {Jean-Antoine
  Piccolo}, {Masana}, {Messineo}, {Mowlavi}, {Nicolas}, {Nienartowicz},
  {Pailler}, {Panuzzo}, {Riclet}, {Roux}, {Seabroke}, {Sordo}, {Tanga},
  {Th{\'e}venin}, {Gracia-Abril}, {Portell}, {Teyssier}, {Altmann}, {Andrae},
  {Bellas-Velidis}, {Benson}, {Berthier}, {Blomme}, {Brugaletta}, {Burgess},
  {Busso}, {Carry}, {Cellino}, {Cheek}, {Clementini}, {Damerdji}, {Davidson},
  {Delchambre}, {Dell'Oro}, {Fern{\'a}ndez-Hern{\'a}ndez}, {Galluccio},
  {Garc{\'\i}a-Lario}, {Garcia-Reinaldos}, {Gonz{\'a}lez-N{\'u}{\~n}ez},
  {Gosset}, {Haigron}, {Halbwachs}, {Hambly}, {Harrison}, {Hatzidimitriou},
  {Heiter}, {Hern{\'a}ndez}, {Hestroffer}, {Hodgkin}, {Holl}, {Jan{\ss}en},
  {Jevardat de Fombelle}, {Jordan}, {Krone-Martins}, {Lanzafame},
  {L{\"o}ffler}, {Lorca}, {Manteiga}, {Marchal}, {Marrese}, {Moitinho}, {Mora},
  {Muinonen}, {Osborne}, {Pancino}, {Pauwels}, {Petit}, {Recio-Blanco},
  {Richards}, {Riello}, {Rimoldini}, {Robin}, {Roegiers}, {Rybizki}, {Sarro},
  {Siopis}, {Smith}, {Sozzetti}, {Ulla}, {Utrilla}, {van Leeuwen}, {van
  Reeven}, {Abbas}, {Abreu Aramburu}, {Accart}, {Aerts}, {Aguado}, {Ajaj},
  {Altavilla}, {{\'A}lvarez}, {{\'A}lvarez Cid-Fuentes}, {Alves}, {Anderson},
  {Anglada Varela}, {Antoja}, {Audard}, {Baines}, {Baker},
  {Balaguer-N{\'u}{\~n}ez}, {Balbinot}, {Balog}, {Barache}, {Barbato},
  {Barros}, {Barstow}, {Bartolom{\'e}}, {Bassilana}, {Bauchet},
  {Baudesson-Stella}, {Becciani}, {Bellazzini}, {Bernet}, {Bertone}, {Bianchi},
  {Blanco-Cuaresma}, {Boch}, {Bombrun}, {Bossini}, {Bouquillon}, {Bragaglia},
  {Bramante}, {Breedt}, {Bressan}, {Brouillet}, {Bucciarelli}, {Burlacu},
  {Busonero}, {Butkevich}, {Buzzi}, {Caffau}, {Cancelliere}, {C{\'a}novas},
  {Cantat-Gaudin}, {Carballo}, {Carlucci}, {Carnerero}, {Carrasco},
  {Casamiquela}, {Castellani}, {Castro-Ginard}, {Castro Sampol}, {Chaoul},
  {Charlot}, {Chemin}, {Chiavassa}, {Cioni}, {Comoretto}, {Cooper}, {Cornez},
  {Cowell}, {Crifo}, {Crosta}, {Crowley}, {Dafonte}, {Dapergolas}, {David}, \&
  {David}}]{Gaia_Brown21}
{Gaia Collaboration}, {Brown}, A.~G.~A., {Vallenari}, A., {et~al.}
  2021{\natexlab{a}}, 649, A1

\bibitem[{{Gaia Collaboration} {et~al.}(2021{\natexlab{b}}){Gaia
  Collaboration}, {Smart}, {Sarro}, {Rybizki}, {Reyl{\'e}}, {Robin}, {Hambly},
  {Abbas}, {Barstow}, {de Bruijne}, {Bucciarelli}, {Carrasco}, {Cooper},
  {Hodgkin}, {Masana}, {Michalik}, {Sahlmann}, {Sozzetti}, {Brown},
  {Vallenari}, {Prusti}, {Babusiaux}, {Biermann}, {Creevey}, {Evans}, {Eyer},
  {Hutton}, {Jansen}, {Jordi}, {Klioner}, {Lammers}, {Lindegren}, {Luri},
  {Mignard}, {Panem}, {Pourbaix}, {Randich}, {Sartoretti}, {Soubiran},
  {Walton}, {Arenou}, {Bailer-Jones}, {Bastian}, {Cropper}, {Drimmel}, {Katz},
  {Lattanzi}, {van Leeuwen}, {Bakker}, {Casta{\~n}eda}, {De Angeli},
  {Ducourant}, {Fabricius}, {Fouesneau}, {Fr{\'e}mat}, {Guerra}, {Guerrier},
  {Guiraud}, {Jean-Antoine Piccolo}, {Messineo}, {Mowlavi}, {Nicolas},
  {Nienartowicz}, {Pailler}, {Panuzzo}, {Riclet}, {Roux}, {Seabroke}, {Sordo},
  {Tanga}, {Th{\'e}venin}, {Gracia-Abril}, {Portell}, {Teyssier}, {Altmann},
  {Andrae}, {Bellas-Velidis}, {Benson}, {Berthier}, {Blomme}, {Brugaletta},
  {Burgess}, {Busso}, {Carry}, {Cellino}, {Cheek}, {Clementini}, {Damerdji},
  {Davidson}, {Delchambre}, {Dell'Oro}, {Fern{\'a}ndez-Hern{\'a}ndez},
  {Galluccio}, {Garc{\'\i}a-Lario}, {Garcia-Reinaldos},
  {Gonz{\'a}lez-N{\'u}{\~n}ez}, {Gosset}, {Haigron}, {Halbwachs}, {Harrison},
  {Hatzidimitriou}, {Heiter}, {Hern{\'a}ndez}, {Hestroffer}, {Holl},
  {Jan{\ss}en}, {Jevardat de Fombelle}, {Jordan}, {Krone-Martins}, {Lanzafame},
  {L{\"o}ffler}, {Lorca}, {Manteiga}, {Marchal}, {Marrese}, {Moitinho}, {Mora},
  {Muinonen}, {Osborne}, {Pancino}, {Pauwels}, {Recio-Blanco}, {Richards},
  {Riello}, {Rimoldini}, {Roegiers}, {Siopis}, {Smith}, {Ulla}, {Utrilla}, {van
  Leeuwen}, {van Reeven}, {Abreu Aramburu}, {Accart}, {Aerts}, {Aguado},
  {Ajaj}, {Altavilla}, {{\'A}lvarez}, {{\'A}lvarez Cid-Fuentes}, {Alves},
  {Anderson}, {Anglada Varela}, {Antoja}, {Audard}, {Baines}, {Baker},
  {Balaguer-N{\'u}{\~n}ez}, {Balbinot}, {Balog}, {Barache}, {Barbato},
  {Barros}, {Bartolom{\'e}}, {Bassilana}, {Bauchet}, {Baudesson-Stella},
  {Becciani}, {Bellazzini}, {Bernet}, {Bertone}, {Bianchi}, {Blanco-Cuaresma},
  {Boch}, {Bombrun}, {Bossini}, {Bouquillon}, {Bragaglia}, {Bramante},
  {Breedt}, {Bressan}, {Brouillet}, {Burlacu}, {Busonero}, {Butkevich},
  {Buzzi}, {Caffau}, {Cancelliere}, {C{\'a}novas}, {Cantat-Gaudin}, {Carballo},
  {Carlucci}, {Carnerero}, {Casamiquela}, {Castellani}, {Castro-Ginard},
  {Castro Sampol}, {Chaoul}, {Charlot}, {Chemin}, {Chiavassa}, {Cioni},
  {Comoretto}, {Cornez}, {Cowell}, {Crifo}, {Crosta}, {Crowley}, {Dafonte},
  {Dapergolas}, {David}, {David}, {de Laverny}, {De Luise}, {De March}, {De
  Ridder}, {de Souza}, {de Teodoro}, {de Torres}, {del Peloso}, {del Pozo},
  {Delgado}, {Delgado}, {Delisle}, {Di Matteo}, {Diakite}, {Diener},
  {Distefano}, {Dolding}, {Eappachen}, {Edvardsson}, {Enke}, {Esquej}, {Fabre},
  {Fabrizio}, {Faigler}, {Fedorets}, {Fernique}, {Fienga}, {Figueras},
  {Fouron}, {Fragkoudi}, {Fraile}, {Franke}, {Gai}, {Garabato},
  {Garcia-Gutierrez}, {Garc{\'\i}a-Torres}, {Garofalo}, {Gavras}, {Gerlach},
  {Geyer}, {Giacobbe}, {Gilmore}, {Girona}, {Giuffrida}, {Gomel}, {Gomez},
  {Gonzalez-Santamaria}, {Gonz{\'a}lez-Vidal}, {Granvik},
  {Guti{\'e}rrez-S{\'a}nchez}, {Guy}, {Hauser}, {Haywood}, {Helmi}, {Hidalgo},
  {Hilger}, {H{\l}adczuk}, {Hobbs}, {Holland}, {Huckle}, {Jasniewicz},
  {Jonker}, {Juaristi Campillo}, {Julbe}, {Karbevska}, {Kervella}, {Khanna},
  {Kochoska}, {Kontizas}, {Kordopatis}, {Korn}, {Kostrzewa-Rutkowska},
  {Kruszy{\'n}ska}, {Lambert}, {Lanza}, {Lasne}, {Le Campion}, {Le Fustec},
  {Lebreton}, {Lebzelter}, {Leccia}, {Leclerc}, {Lecoeur-Taibi}, {Liao},
  {Licata}, {Lindstr{\o}m}, {Lister}, {Livanou}, {Lobel}, {Madrero Pardo},
  {Managau}, {Mann}, {Marchant}, {Marconi}, {Marcos Santos}, {Marinoni},
  {Marocco}, {Marshall}, {Martin Polo}, {Mart{\'\i}n-Fleitas}, {Masip},
  {Massari}, {Mastrobuono-Battisti}, {Mazeh}, {McMillan}, {Messina}, {Millar},
  {Mints}, {Molina}, {Molinaro}, {Moln{\'a}r}, {Montegriffo}, {Mor},
  {Morbidelli}, {Morel}, {Morris}, {Mulone}, {Munoz}, {Muraveva}, {Murphy},
  {Musella}, {Noval}, {Ord{\'e}novic}, {Orr{\`u}}, {Osinde}, {Pagani},
  {Pagano}, {Palaversa}, {Palicio}, {Panahi}, {Pawlak}, {Pe{\~n}alosa
  Esteller}, {Penttil{\"a}}, {Piersimoni}, {Pineau}, {Plachy}, {Plum},
  {Poggio}, {Poretti}, {Poujoulet}, {Pr{\v{s}}a}, {Pulone}, {Racero},
  {Ragaini}, {Rainer}, {Raiteri}, {Rambaux}, {Ramos}, {Ramos-Lerate}, {Re
  Fiorentin}, {Regibo}, {Ripepi}, {Riva}, {Rixon}, {Robichon}, {Robin},
  {Roelens}, {Rohrbasser}, {Romero-G{\'o}mez}, {Rowell}, {Royer}, {Rybicki},
  {Sadowski}, {Sagrist{\`a} Sell{\'e}s}, {Salgado}, {Salguero}, {Samaras},
  {Sanchez Gimenez}, {Sanna}, {Santove{\~n}a}, {Sarasso}, {Schultheis},
  {Sciacca}, {Segol}, {Segovia}, {S{\'e}gransan}, {Semeux}, {Shahaf},
  {Siddiqui}, {Siebert}, {Siltala}, {Slezak}, {Solano}, {Solitro}, {Souami},
  {Souchay}, {Spagna}, {Spoto}, {Steele}, {Steidelm{\"u}ller}, {Stephenson},
  {S{\"u}veges}, {Szabados}, {Szegedi-Elek}, {Taris}, {Tauran}, {Taylor},
  {Teixeira}, {Thuillot}, {Tonello}, {Torra}, {Torra}, {Turon}, {Unger},
  {Vaillant}, {van Dillen}, {Vanel}, {Vecchiato}, {Viala}, {Vicente},
  {Voutsinas}, {Weiler}, {Wevers}, {Wyrzykowski}, {Yoldas}, {Yvard}, {Zhao},
  {Zorec}, {Zucker}, {Zurbach}, \& {Zwitter}}]{smart21a}
{Gaia Collaboration}, {Smart}, R.~L., {Sarro}, L.~M., {et~al.}
  2021{\natexlab{b}}, A\&A, 649, A6

\bibitem[{{Gauza} {et~al.}(2015){Gauza}, {B{\'e}jar}, {P{\'e}rez-Garrido},
  {Rosa Zapatero Osorio}, {Lodieu}, {Rebolo}, {Pall{\'e}}, \&
  {Nowak}}]{gauza15a}
{Gauza}, B., {B{\'e}jar}, V.~J.~S., {P{\'e}rez-Garrido}, A., {et~al.} 2015,
  ApJ, 804, 96

\bibitem[{{Gizis} {et~al.}(2015{\natexlab{a}}){Gizis}, {Allers}, {Liu},
  {Harris}, {Faherty}, {Burgasser}, \& {Kirkpatrick}}]{gizis15a}
{Gizis}, J.~E., {Allers}, K.~N., {Liu}, M.~C., {et~al.} 2015{\natexlab{a}},
  ApJ, 799, 203

\bibitem[{{Gizis} {et~al.}(2015{\natexlab{b}}){Gizis}, {Dettman}, {Burgasser},
  {Camnasio}, {Alam}, {Filippazzo}, {Cruz}, {Metchev}, {Berger}, \&
  {Williams}}]{gizis15b}
{Gizis}, J.~E., {Dettman}, K.~G., {Burgasser}, A.~J., {et~al.}
  2015{\natexlab{b}}, ApJ, 813, 104

\bibitem[{{Gizis} {et~al.}(2012){Gizis}, {Faherty}, {Liu}, {Castro}, {Shaw},
  {Vrba}, {Harris}, {Aller}, \& {Deacon}}]{gizis12}
{Gizis}, J.~E., {Faherty}, J.~K., {Liu}, M.~C., {et~al.} 2012, AJ, 144, 94

\bibitem[{{Gizis} {et~al.}(2000){Gizis}, {Monet}, {Reid}, {Kirkpatrick},
  {Liebert}, \& {Williams}}]{gizis00a}
{Gizis}, J.~E., {Monet}, D.~G., {Reid}, I.~N., {et~al.} 2000, AJ, 120, 1085

\bibitem[{{Golimowski} {et~al.}(2004){Golimowski}, {Leggett}, {Marley}, {Fan},
  {Geballe}, {Knapp}, {Vrba}, {Henden}, \& {11 authors}}]{golimowski04a}
{Golimowski}, D.~A., {Leggett}, S.~K., {Marley}, M.~S., {et~al.} 2004, AJ, 127,
  3516

\bibitem[{{Helling} {et~al.}(2008){Helling}, {Ackerman}, {Allard}, {Dehn},
  {Hauschildt}, {Homeier}, {Lodders}, {Marley}, {Rietmeijer}, {Tsuji}, \&
  {Woitke}}]{hellin08a}
{Helling}, C., {Ackerman}, A., {Allard}, F., {et~al.} 2008, MNRAS, 391, 1854

\bibitem[{{Hiranaka} {et~al.}(2016){Hiranaka}, {Cruz}, {Douglas}, {Marley}, \&
  {Baldassare}}]{hiranaka16}
{Hiranaka}, K., {Cruz}, K.~L., {Douglas}, S.~T., {Marley}, M.~S., \&
  {Baldassare}, V.~F. 2016, 830, 96

\bibitem[{{Jones} {et~al.}(2015){Jones}, {White}, {Boyajian}, {Schaefer},
  {Baines}, {Ireland}, {Patience}, {McAlister}, \& {Ten Brummelaar}}]{jones15}
{Jones}, J., {White}, R.~J., {Boyajian}, T.~S., {et~al.} 2015, in American
  Astronomical Society Meeting Abstracts, Vol. 225, American Astronomical
  Society Meeting Abstracts \#225, 112.03

\bibitem[{{Kirkpatrick}(2005)}]{kirkpatrick05}
{Kirkpatrick}, J.~D. 2005, ARA\&A, 43, 195

\bibitem[{{Kirkpatrick} {et~al.}(2006){Kirkpatrick}, {Barman}, {Burgasser},
  {McGovern}, {McLean}, {Tinney}, \& {Lowrance}}]{kirkpatrick06}
{Kirkpatrick}, J.~D., {Barman}, T.~S., {Burgasser}, A.~J., {et~al.} 2006, ApJ,
  639, 1120

\bibitem[{{Kirkpatrick} {et~al.}(2008){Kirkpatrick}, {Cruz}, {Barman},
  {Burgasser}, {Looper}, {Tinney}, {Gelino}, {Lowrance}, {Liebert},
  {Carpenter}, {Hillenbrand}, \& {Stauffer}}]{kirkpatrick08}
{Kirkpatrick}, J.~D., {Cruz}, K.~L., {Barman}, T.~S., {et~al.} 2008, ApJ, 689,
  1295

\bibitem[{{Kirkpatrick} {et~al.}(2011){Kirkpatrick}, {Cushing}, {Gelino},
  {Griffith}, {Skrutskie}, {Marsh}, {Wright}, {Mainzer}, {Eisenhardt},
  {McLean}, \& {30 co-authors}}]{kirkpatrick11}
{Kirkpatrick}, J.~D., {Cushing}, M.~C., {Gelino}, C.~R., {et~al.} 2011, ApJS,
  197, 19

\bibitem[{{Kirkpatrick} {et~al.}(2021){Kirkpatrick}, {Gelino}, {Faherty},
  {Meisner}, {Caselden}, {Schneider}, {Marocco}, {Cayago}, {Smart},
  {Eisenhardt}, {Kuchner}, {Wright}, {Cushing}, {Allers}, {Bardalez Gagliuffi},
  {Burgasser}, {Gagn{\'e}}, {Logsdon}, {Martin}, {Ingalls}, {Lowrance},
  {Abrahams}, {Aganze}, {Gerasimov}, {Gonzales}, {Hsu}, {Kamraj}, {Kiman},
  {Rees}, {Theissen}, {Ammar}, {Andersen}, {Beaulieu}, {Colin}, {Elachi},
  {Goodman}, {Gramaize}, {Hamlet}, {Hong}, {Jonkeren}, {Khalil}, {Martin},
  {Pendrill}, {Pumphrey}, {Rothermich}, {Sainio}, {Stenner}, {Tanner},
  {Th{\'e}venot}, {Voloshin}, {Walla}, {W{\k{e}}dracki}, \& {Backyard Worlds:
  Planet 9 Collaboration}}]{kirkpatrick21a}
{Kirkpatrick}, J.~D., {Gelino}, C.~R., {Faherty}, J.~K., {et~al.} 2021, ApJS,
  253, 7

\bibitem[{{Kirkpatrick} {et~al.}(2010){Kirkpatrick}, {Looper}, {Burgasser},
  {Schurr}, {Cutri}, {Cushing}, {Cruz}, {Sweet}, {Knapp}, {Barman},
  {Bochanski}, {Roellig}, {McLean}, {McGovern}, \& {Rice}}]{kirkpatrick10}
{Kirkpatrick}, J.~D., {Looper}, D.~L., {Burgasser}, A.~J., {et~al.} 2010, ApJS,
  190, 100

\bibitem[{{Kirkpatrick} {et~al.}(1999){Kirkpatrick}, {Reid}, {Liebert},
  {Cutri}, {Nelson}, {Beichman}, {Dahn}, {Monet}, {Gizis}, \&
  {Skrutskie}}]{kirkpatrick99}
{Kirkpatrick}, J.~D., {Reid}, I.~N., {Liebert}, J., {et~al.} 1999, ApJ, 519,
  802

\bibitem[{{Kirkpatrick} {et~al.}(2000){Kirkpatrick}, {Reid}, {Liebert},
  {Gizis}, {Burgasser}, {Monet}, {Dahn}, {Nelson}, \&
  {Williams}}]{kirkpatrick00}
{Kirkpatrick}, J.~D., {Reid}, I.~N., {Liebert}, J., {et~al.} 2000, AJ, 120, 447

\bibitem[{{Knapp} {et~al.}(2004){Knapp}, {Leggett}, {Fan}, {Marley}, {Geballe},
  {Golimowski}, {Finkbeiner}, {Gunn}, , \& {21 co-authors}}]{knapp04}
{Knapp}, G.~R., {Leggett}, S.~K., {Fan}, X., {et~al.} 2004, AJ, 127, 3553

\bibitem[{{Kounkel} \& {Covey}(2019)}]{kounkel19a}
{Kounkel}, M. \& {Covey}, K. 2019, 158, 122

\bibitem[{{Lagrange} {et~al.}(2009){Lagrange}, {Kasper}, {Boccaletti},
  {Chauvin}, {Gratadour}, {Fusco}, {Ehrenreich}, {Apai}, {Mouillet}, \&
  {Rouan}}]{lagrange09c}
{Lagrange}, A.~M., {Kasper}, M., {Boccaletti}, A., {et~al.} 2009, A\&A, 506,
  927

\bibitem[{{Lam} {et~al.}(2026){Lam}, {Vos}, {Su{\'a}rez}, {Hsu}, {Bickle},
  {Faherty}, {Gagn{\'e}}, {Bardalez Gagliuffi}, {Biller}, {Burningham}, {Cruz},
  {Morley}, {Luszcz-Cook}, {Lawsky}, {Phillips}, \& {Rothermich}}]{lam26a}
{Lam}, M.~B., {Vos}, J.~M., {Su{\'a}rez}, G., {et~al.} 2026, 709, A56

\bibitem[{{Laureijs} {et~al.}(2011){Laureijs}, {Amiaux}, {Arduini},
  {Augu{\`e}res}, {Brinchmann}, {Cole}, {Cropper}, {Dabin}, {Duvet}, {Ealet},
  {Garilli}, {Gondoin}, {Guzzo}, {Hoar}, {Hoekstra}, {Holmes}, {Kitching},
  {Maciaszek}, {Mellier}, {Pasian}, {Percival}, {Rhodes}, {Saavedra Criado},
  {Sauvage}, {Scaramella}, {Valenziano}, {Warren}, {Bender}, {Castander},
  {Cimatti}, {Le F{\`e}vre}, {Kurki-Suonio}, {Levi}, {Lilje}, {Meylan},
  {Nichol}, {Pedersen}, {Popa}, {Rebolo Lopez}, {Rix}, {Rottgering},
  {Zeilinger}, {Grupp}, {Hudelot}, {Massey}, {Meneghetti}, {Miller}, {Paltani},
  {Paulin-Henriksson}, {Pires}, {Saxton}, {Schrabback}, {Seidel}, {Walsh},
  {Aghanim}, {Amendola}, {Bartlett}, {Baccigalupi}, {Beaulieu}, {Benabed},
  {Cuby}, {Elbaz}, {Fosalba}, {Gavazzi}, {Helmi}, {Hook}, {Irwin}, {Kneib},
  {Kunz}, {Mannucci}, {Moscardini}, {Tao}, {Teyssier}, {Weller}, {Zamorani},
  {Zapatero Osorio}, {Boulade}, {Foumond}, {Di Giorgio}, {Guttridge}, {James},
  {Kemp}, {Martignac}, {Spencer}, {Walton}, {Bl{\"u}mchen}, {Bonoli},
  {Bortoletto}, {Cerna}, {Corcione}, {Fabron}, {Jahnke}, {Ligori}, {Madrid},
  {Martin}, {Morgante}, {Pamplona}, {Prieto}, {Riva}, {Toledo}, {Trifoglio},
  {Zerbi}, {Abdalla}, {Douspis}, {Grenet}, {Borgani}, {Bouwens}, {Courbin},
  {Delouis}, {Dubath}, {Fontana}, {Frailis}, {Grazian}, {Koppenh{\"o}fer},
  {Mansutti}, {Melchior}, {Mignoli}, {Mohr}, {Neissner}, {Noddle}, {Poncet},
  {Scodeggio}, {Serrano}, {Shane}, {Starck}, {Surace}, {Taylor},
  {Verdoes-Kleijn}, {Vuerli}, {Williams}, {Zacchei}, {Altieri}, {Escudero
  Sanz}, {Kohley}, {Oosterbroek}, {Astier}, {Bacon}, {Bardelli}, {Baugh},
  {Bellagamba}, {Benoist}, {Bianchi}, {Biviano}, {Branchini}, {Carbone},
  {Cardone}, {Clements}, {Colombi}, {Conselice}, {Cresci}, {Deacon}, {Dunlop},
  {Fedeli}, {Fontanot}, {Franzetti}, {Giocoli}, {Garcia-Bellido}, {Gow},
  {Heavens}, {Hewett}, {Heymans}, {Holland}, {Huang}, {Ilbert}, {Joachimi},
  {Jennins}, {Kerins}, {Kiessling}, {Kirk}, {Kotak}, {Krause}, {Lahav}, {van
  Leeuwen}, {Lesgourgues}, {Lombardi}, {Magliocchetti}, {Maguire}, {Majerotto},
  {Maoli}, {Marulli}, {Maurogordato}, {McCracken}, {McLure}, {Melchiorri},
  {Merson}, {Moresco}, {Nonino}, {Norberg}, {Peacock}, {Pello}, {Penny},
  {Pettorino}, {Di Porto}, {Pozzetti}, {Quercellini}, {Radovich}, {Rassat},
  {Roche}, {Ronayette}, {Rossetti}, {Sartoris}, {Schneider}, {Semboloni},
  {Serjeant}, {Simpson}, {Skordis}, {Smadja}, {Smartt}, {Spano}, {Spiro},
  {Sullivan}, {Tilquin}, {Trotta}, {Verde}, {Wang}, {Williger}, {Zhao},
  {Zoubian}, \& {Zucca}}]{laureijs11}
{Laureijs}, R., {Amiaux}, J., {Arduini}, S., {et~al.} 2011, arXiv e-prints,
  arXiv:1110.3193

\bibitem[{{Lawrence} {et~al.}(2007){Lawrence}, {Warren}, {Almaini}, {Edge},
  {Hambly}, \& {17 co-authors}}]{lawrence07}
{Lawrence}, A., {Warren}, S.~J., {Almaini}, O., {et~al.} 2007, MNRAS, 379, 1599

\bibitem[{{Liu} {et~al.}(2016){Liu}, {Dupuy}, \& {Allers}}]{liu16a}
{Liu}, M.~C., {Dupuy}, T.~J., \& {Allers}, K.~N. 2016, ApJ, 833, 96

\bibitem[{{Liu} {et~al.}(2013){Liu}, {Magnier}, {Deacon}, {Allers}, {Dupuy},
  {Kotson}, {Aller}, {Burgett}, {Chambers}, {Draper}, {Hodapp}, {Jedicke},
  {Kaiser}, {Kudritzki}, {Metcalfe}, {Morgan}, {Price}, {Tonry}, \&
  {Wainscoat}}]{liu13a}
{Liu}, M.~C., {Magnier}, E.~A., {Deacon}, N.~R., {et~al.} 2013, ApJL, 777, L20

\bibitem[{{Lodieu}(2013)}]{lodieu13c}
{Lodieu}, N. 2013, MNRAS, 431, 3222

\bibitem[{{Lodieu} {et~al.}(2013){Lodieu}, {Dobbie}, {Cross}, {Hambly}, {Read},
  {Blake}, \& {Floyd}}]{lodieu13d}
{Lodieu}, N., {Dobbie}, P.~D., {Cross}, N.~J.~G., {et~al.} 2013, MNRAS, 435,
  2474

\bibitem[{{Lodieu} {et~al.}(2011){Lodieu}, {Dobbie}, \& {Hambly}}]{lodieu11a}
{Lodieu}, N., {Dobbie}, P.~D., \& {Hambly}, N.~C. 2011, A\&A, 527, A24

\bibitem[{{Lodieu} {et~al.}(2021){Lodieu}, {Hambly}, \& {Cross}}]{lodieu21}
{Lodieu}, N., {Hambly}, N.~C., \& {Cross}, N.~J.~G. 2021, MNRAS, 503, 2265

\bibitem[{{Lodieu} {et~al.}(2008){Lodieu}, {Hambly}, {Jameson}, \&
  {Hodgkin}}]{lodieu08a}
{Lodieu}, N., {Hambly}, N.~C., {Jameson}, R.~F., \& {Hodgkin}, S.~T. 2008,
  MNRAS, 383, 1385

\bibitem[{{Lodieu} {et~al.}(2018){Lodieu}, {Zapatero Osorio}, {B{\'e}jar}, \&
  {Pe{\~n}a Ram{\'{\i}}rez}}]{lodieu18a}
{Lodieu}, N., {Zapatero Osorio}, M.~R., {B{\'e}jar}, V.~J.~S., \& {Pe{\~n}a
  Ram{\'{\i}}rez}, K. 2018, MNRAS, 473, 2020

\bibitem[{{Lodieu} {et~al.}(2009){Lodieu}, {Zapatero Osorio}, {Rebolo},
  {Mart{\'{\i}}n}, \& {Hambly}}]{lodieu09e}
{Lodieu}, N., {Zapatero Osorio}, M.~R., {Rebolo}, R., {Mart{\'{\i}}n}, E.~L.,
  \& {Hambly}, N.~C. 2009, A\&A, 505, 1115

\bibitem[{{Looper} {et~al.}(2008){Looper}, {Kirkpatrick}, {Cutri}, {Barman},
  {Burgasser}, {Cushing}, {Roellig}, {McGovern}, {McLean}, {Rice}, {Swift}, \&
  {Schurr}}]{looper08b}
{Looper}, D.~L., {Kirkpatrick}, J.~D., {Cutri}, R.~M., {et~al.} 2008, ApJ, 806, 1059

\bibitem[{{Lueber} {et~al.}(2024){Lueber}, {Heng}, {Bowler}, {Kitzmann}, {Vos},
  \& {Zhou}}]{lueber24}
{Lueber}, A., {Heng}, K., {Bowler}, B.~P., {et~al.} 2024, A\&A, 690, A357

\bibitem[{{Luhman}(2000)}]{luhman00a}
{Luhman}, K.~L. 2000, ApJ, 544, 1044

\bibitem[{{Luhman}(2004)}]{luhman04a}
{Luhman}, K.~L. 2004, ApJ, 617, 1216

\bibitem[{{Luhman}(2018)}]{luhman18b}
{Luhman}, K.~L. 2018, AJ, 156, 271

\bibitem[{{Luhman} {et~al.}(2003){Luhman}, {Brice{\~ n}o}, {Stauffer},
  {Hartmann}, {Barrado y Navascu{\' e}s}, \& {Caldwell}}]{luhman03a}
{Luhman}, K.~L., {Brice{\~ n}o}, C., {Stauffer}, J.~R., {et~al.} 2003, ApJ,
  590, 348

\bibitem[{{Luhman} {et~al.}(2018){Luhman}, {Herrmann}, {Mamajek}, {Esplin}, \&
  {Pecaut}}]{luhman18a}
{Luhman}, K.~L., {Herrmann}, K.~A., {Mamajek}, E.~E., {Esplin}, T.~L., \&
  {Pecaut}, M.~J. 2018, AJ, 156, 76

\bibitem[{{Luhman} {et~al.}(2017){Luhman}, {Mamajek}, {Shukla}, \&
  {Loutrel}}]{luhman17a}
{Luhman}, K.~L., {Mamajek}, E.~E., {Shukla}, S.~J., \& {Loutrel}, N.~P. 2017,
  AJ, 153, 46

\bibitem[{{Luhman} {et~al.}(2005){Luhman}, {Stauffer}, \&
  {Mamajek}}]{luhman05b}
{Luhman}, K.~L., {Stauffer}, J.~R., \& {Mamajek}, E.~E. 2005, 628, L69

\bibitem[{{Mace} {et~al.}(2013){Mace}, {Kirkpatrick}, {Cushing}, {Gelino},
  {Griffith}, {Skrutskie}, {Marsh}, {Wright}, {Eisenhardt}, {McLean},
  {Thompson}, {Mix}, {Bailey}, {Beichman}, {Bloom}, {Burgasser}, {Fortney},
  {Hinz}, {Knox}, {Lowrance}, {Marley}, {Morley}, {Rodigas}, {Saumon},
  {Sheppard}, \& {Stock}}]{mace13}
{Mace}, G.~N., {Kirkpatrick}, J.~D., {Cushing}, M.~C., {et~al.} 2013, ApJS,
  205, 6

\bibitem[{{Macintosh} {et~al.}(2015){Macintosh}, {Graham}, {Barman}, {De Rosa},
  {Konopacky}, {Marley}, {Marois}, {Nielsen}, {Pueyo}, {Rajan}, {Rameau},
  {Saumon}, {Wang}, {Patience}, {Ammons}, {Arriaga}, {Artigau}, {Beckwith},
  {Brewster}, {Bruzzone}, {Bulger}, {Burningham}, {Burrows}, {Chen}, {Chiang},
  {Chilcote}, {Dawson}, {Dong}, {Doyon}, {Draper}, {Duch{\^e}ne}, {Esposito},
  {Fabrycky}, {Fitzgerald}, {Follette}, {Fortney}, {Gerard}, {Goodsell},
  {Greenbaum}, {Hibon}, {Hinkley}, {Cotten}, {Hung}, {Ingraham},
  {Johnson-Groh}, {Kalas}, {Lafreniere}, {Larkin}, {Lee}, {Line}, {Long},
  {Maire}, {Marchis}, {Matthews}, {Max}, {Metchev}, {Millar-Blanchaer},
  {Mittal}, {Morley}, {Morzinski}, {Murray-Clay}, {Oppenheimer}, {Palmer},
  {Patel}, {Perrin}, {Poyneer}, {Rafikov}, {Rantakyr{\"o}}, {Rice}, {Rojo},
  {Rudy}, {Ruffio}, {Ruiz}, {Sadakuni}, {Saddlemyer}, {Salama}, {Savransky},
  {Schneider}, {Sivaramakrishnan}, {Song}, {Soummer}, {Thomas}, {Vasisht},
  {Wallace}, {Ward-Duong}, {Wiktorowicz}, {Wolff}, \&
  {Zuckerman}}]{macintosh15}
{Macintosh}, B., {Graham}, J.~R., {Barman}, T., {et~al.} 2015, Science, 350, 64

\bibitem[{{Madhusudhan} {et~al.}(2011){Madhusudhan}, {Burrows}, \&
  {Currie}}]{madhusudhan11a}
{Madhusudhan}, N., {Burrows}, A., \& {Currie}, T. 2011, 737, 34

\bibitem[{{Malo} {et~al.}(2013){Malo}, {Doyon}, {Lafreni{\`e}re}, {Artigau},
  {Gagn{\'e}}, {Baron}, \& {Riedel}}]{malo13}
{Malo}, L., {Doyon}, R., {Lafreni{\`e}re}, D., {et~al.} 2013, ApJ, 762, 88

\bibitem[{{Mamajek} \& {Bell}(2014)}]{mamajek14}
{Mamajek}, E.~E. \& {Bell}, C. P.~M. 2014, 445, 2169

\bibitem[{{Manjavacas} {et~al.}(2014){Manjavacas}, {Bonnefoy}, {Schlieder},
  {Allard}, {Rojo}, {Goldman}, {Chauvin}, {Homeier}, {Lodieu}, \&
  {Henning}}]{manjavacas14}
{Manjavacas}, E., {Bonnefoy}, M., {Schlieder}, J.~E., {et~al.} 2014, A\&A, 564,
  A55

\bibitem[{{Manjavacas} {et~al.}(2020){Manjavacas}, {Lodieu}, {B{\'e}jar},
  {Zapatero-Osorio}, {Boudreault}, \& {Bonnefoy}}]{manjavacas20a}
{Manjavacas}, E., {Lodieu}, N., {B{\'e}jar}, V.~J.~S., {et~al.} 2020, MNRAS,
  491, 5925

\bibitem[{{Marley} {et~al.}(2010){Marley}, {Saumon}, \& {Goldblatt}}]{marley10}
{Marley}, M.~S., {Saumon}, D., \& {Goldblatt}, C. 2010, ApJL, 723, L117

\bibitem[{{Marley} {et~al.}(2002){Marley}, {Seager}, {Saumon}, {Lodders},
  {Ackerman}, {Freedman}, \& {Fan}}]{marley02}
{Marley}, M.~S., {Seager}, S., {Saumon}, D., {et~al.} 2002, ApJ, 568, 335

\bibitem[{{Marocco} {et~al.}(2014){Marocco}, {Day-Jones}, {Lucas}, {Jones},
  {Smart}, {Zhang}, {Gomes}, {Burningham}, {Pinfield}, {Raddi}, \&
  {Smith}}]{marocco14}
{Marocco}, F., {Day-Jones}, A.~C., {Lucas}, P.~W., {et~al.} 2014, MNRAS, 439,
  372

\bibitem[{{Marocco} {et~al.}(2021){Marocco}, {Eisenhardt}, {Fowler},
  {Kirkpatrick}, {Meisner}, {Schlafly}, {Stanford}, {Garcia}, {Caselden},
  {Cushing}, {Cutri}, {Faherty}, {Gelino}, {Gonzalez}, {Jarrett}, {Koontz},
  {Mainzer}, {Marchese}, {Mobasher}, {Schlegel}, {Stern}, {Teplitz}, \&
  {Wright}}]{marocco21}
{Marocco}, F., {Eisenhardt}, P. R.~M., {Fowler}, J.~W., {et~al.} 2021, \apjs,
  253, 8

\bibitem[{{Marocco} {et~al.}(2015){Marocco}, {Jones}, {Day-Jones}, {Pinfield},
  {Lucas}, {Burningham}, {Zhang}, {Smart}, {Gomes}, \& {Smith}}]{marocco15}
{Marocco}, F., {Jones}, H.~R.~A., {Day-Jones}, A.~C., {et~al.} 2015, \mnras,
  449, 3651

\bibitem[{{Marois} {et~al.}(2008){Marois}, {Macintosh}, {Barman}, {Zuckerman},
  {Song}, {Patience}, {Lafreni{\`e}re}, \& {Doyon}}]{marois08}
{Marois}, C., {Macintosh}, B., {Barman}, T., {et~al.} 2008, 322, 1348

\bibitem[{{Marois} {et~al.}(2010){Marois}, {Zuckerman}, {Konopacky},
  {Macintosh}, \& {Barman}}]{marois10}
{Marois}, C., {Zuckerman}, B., {Konopacky}, Q.~M., {Macintosh}, B., \&
  {Barman}, T. 2010, 468, 1080

\bibitem[{{Mart\'{\i}n} {et~al.}(1999){Mart\'{\i}n}, {Delfosse}, {Basri},
  {Goldman}, {Forveille}, \& {Zapatero Osorio}}]{martin99a}
{Mart\'{\i}n}, E.~L., {Delfosse}, X., {Basri}, G., {et~al.} 1999, AJ, 118, 2466

\bibitem[{{Mart{\'{\i}}n} {et~al.}(2001{\natexlab{a}}){Mart{\'{\i}}n},
  {Dougados}, {Magnier}, {M{\' e}nard}, {Magazz{\` u}}, {Cuillandre}, \&
  {Delfosse}}]{martin01b}
{Mart{\'{\i}}n}, E.~L., {Dougados}, C., {Magnier}, E., {et~al.}
  2001{\natexlab{a}}, ApJL, 561, L195

\bibitem[{{Mart{\'{\i}}n} \& {Zapatero Osorio}(2003)}]{martin03b}
{Mart{\'{\i}}n}, E.~L. \& {Zapatero Osorio}, M.~R. 2003, ApJL, 593, L113

\bibitem[{{Mart{\'{\i}}n} {et~al.}(2001{\natexlab{b}}){Mart{\'{\i}}n},
  {Zapatero Osorio}, {Barrado y Navascu{\' e}s}, {B{\' e}jar}, \&
  {Rebolo}}]{martin01a}
{Mart{\'{\i}}n}, E.~L., {Zapatero Osorio}, M.~R., {Barrado y Navascu{\' e}s},
  D., {B{\' e}jar}, V.~J.~S., \& {Rebolo}, R. 2001{\natexlab{b}}, ApJL, 558,
  L117

\bibitem[{{Mellier}(2016)}]{mellier16a}
{Mellier}, Y. 2016, in COSPAR Meeting, Vol.~41, 41st COSPAR Scientific
  Assembly, abstracts from the meeting that was to be held 30 July - 7 August
  at the Istanbul Congress Center (ICC), Turkey, but was cancelled. See
  http://cospar2016.tubitak.gov.tr/en/, Abstract H0.2-1-16.

\bibitem[{{Metchev} \& {Hillenbrand}(2006)}]{metchev06}
{Metchev}, S.~A. \& {Hillenbrand}, L.~A. 2006, ApJ, 651, 1166

\bibitem[{{Miles} {et~al.}(2023){Miles}, {Biller}, {Patapis}, {Worthen},
  {Rickman}, {Hoch}, {Skemer}, {Perrin}, {Whiteford}, {Chen}, {Sargent},
  {Mukherjee}, {Morley}, {Moran}, {Bonnefoy}, {Petrus}, {Carter}, {Choquet},
  {Hinkley}, {Ward-Duong}, {Leisenring}, {Millar-Blanchaer}, {Pueyo}, {Ray},
  {Sallum}, {Stapelfeldt}, {Stone}, {Wang}, {Absil}, {Balmer}, {Boccaletti},
  {Bonavita}, {Booth}, {Bowler}, {Chauvin}, {Christiaens}, {Currie},
  {Danielski}, {Fortney}, {Girard}, {Grady}, {Greenbaum}, {Henning}, {Hines},
  {Janson}, {Kalas}, {Kammerer}, {Kennedy}, {Kenworthy}, {Kervella}, {Lagage},
  {Lew}, {Liu}, {Macintosh}, {Marino}, {Marley}, {Marois}, {Matthews},
  {Matthews}, {Mawet}, {McElwain}, {Metchev}, {Meyer}, {Molliere}, {Pantin},
  {Quirrenbach}, {Rebollido}, {Ren}, {Schneider}, {Vasist}, {Wyatt}, {Zhou},
  {Briesemeister}, {Bryan}, {Calissendorff}, {Cantalloube}, {Cugno}, {De
  Furio}, {Dupuy}, {Factor}, {Faherty}, {Fitzgerald}, {Franson}, {Gonzales},
  {Hood}, {Howe}, {Kraus}, {Kuzuhara}, {Lagrange}, {Lawson}, {Lazzoni}, {Liu},
  {Llop-Sayson}, {Lloyd}, {Martinez}, {Mazoyer}, {Quanz}, {Redai}, {Samland},
  {Schlieder}, {Tamura}, {Tan}, {Uyama}, {Vigan}, {Vos}, {Wagner}, {Wolff},
  {Ygouf}, {Zhang}, {Zhang}, \& {Zhang}}]{miles23a}
{Miles}, B.~E., {Biller}, B.~A., {Patapis}, P., {et~al.} 2023, ApJL, 946, L6

\bibitem[{{Miret-Roig} {et~al.}(2024){Miret-Roig}, {Alves}, {Barrado},
  {Burkert}, {Ratzenb{\"o}ck}, \& {Konietzka}}]{miret_roig24}
{Miret-Roig}, N., {Alves}, J., {Barrado}, D., {et~al.} 2024, Nature Astronomy,
  8, 216

\bibitem[{{Miret-Roig} {et~al.}(2022){Miret-Roig}, {Galli}, {Olivares}, {Bouy},
  {Alves}, \& {Barrado}}]{miret_roig22}
{Miret-Roig}, N., {Galli}, P.~A.~B., {Olivares}, J., {et~al.} 2022, 667, A163

\bibitem[{{Moranta} {et~al.}(2022){Moranta}, {Gagn{\'e}}, {Couture}, \&
  {Faherty}}]{moranta22}
{Moranta}, L., {Gagn{\'e}}, J., {Couture}, D., \& {Faherty}, J.~K. 2022, 939,
  94

\bibitem[{{Morley} {et~al.}(2012){Morley}, {Fortney}, {Marley}, {Visscher},
  {Saumon}, \& {Leggett}}]{morley12}
{Morley}, C.~V., {Fortney}, J.~J., {Marley}, M.~S., {et~al.} 2012, ApJ, 756,
  172

\bibitem[{{Pe{\~n}a Ram{\'{\i}}rez} {et~al.}(2016){Pe{\~n}a Ram{\'{\i}}rez},
  {B{\'e}jar}, \& {Zapatero Osorio}}]{penya16a}
{Pe{\~n}a Ram{\'{\i}}rez}, K., {B{\'e}jar}, V.~J.~S., \& {Zapatero Osorio},
  M.~R. 2016, A\&A, 586, A157

\bibitem[{{Pe{\~n}a Ram{\'{\i}}rez} {et~al.}(2012){Pe{\~n}a Ram{\'{\i}}rez},
  {B{\'e}jar}, {Zapatero Osorio}, {Petr-Gotzens}, \&
  {Mart{\'{\i}}n}}]{penya12a}
{Pe{\~n}a Ram{\'{\i}}rez}, K., {B{\'e}jar}, V.~J.~S., {Zapatero Osorio}, M.~R.,
  {Petr-Gotzens}, M.~G., \& {Mart{\'{\i}}n}, E.~L. 2012, ApJ, 754, 30

\bibitem[{{Petrus} {et~al.}(2024){Petrus}, {Whiteford}, {Patapis}, {Biller},
  {Skemer}, {Hinkley}, {Su{\'a}rez}, {Palma-Bifani}, {Morley}, {Tremblin},
  {Charnay}, {Vos}, {Wang}, {Stone}, {Bonnefoy}, {Chauvin}, {Miles}, {Carter},
  {Lueber}, {Helling}, {Sutlieff}, {Janson}, {Gonzales}, {Hoch}, {Absil},
  {Balmer}, {Boccaletti}, {Bonavita}, {Booth}, {Bowler}, {Briesemeister},
  {Bryan}, {Calissendorff}, {Cantalloube}, {Chen}, {Choquet}, {Christiaens},
  {Cugno}, {Currie}, {Danielski}, {De Furio}, {Dupuy}, {Factor}, {Faherty},
  {Fitzgerald}, {Fortney}, {Franson}, {Girard}, {Grady}, {Henning}, {Hines},
  {Hood}, {Howe}, {Kalas}, {Kammerer}, {Kennedy}, {Kenworthy}, {Kervella},
  {Kim}, {Kitzmann}, {Kraus}, {Kuzuhara}, {Lagage}, {Lagrange}, {Lawson},
  {Lazzoni}, {Leisenring}, {Lew}, {Liu}, {Liu}, {Llop-Sayson}, {Lloyd},
  {Macintosh}, {M{\^a}lin}, {Manjavacas}, {Marino}, {Marley}, {Marois},
  {Martinez}, {Matthews}, {Matthews}, {Mawet}, {Mazoyer}, {McElwain},
  {Metchev}, {Meyer}, {Millar-Blanchaer}, {Molli{\`e}re}, {Moran}, {Mukherjee},
  {Pantin}, {Perrin}, {Pueyo}, {Quanz}, {Quirrenbach}, {Ray}, {Rebollido},
  {Adams Redai}, {Ren}, {Rickman}, {Sallum}, {Samland}, {Sargent}, {Schlieder},
  {Stapelfeldt}, {Tamura}, {Tan}, {Theissen}, {Uyama}, {Vasist}, {Vigan},
  {Wagner}, {Ward-Duong}, {Wolff}, {Worthen}, {Wyatt}, {Ygouf}, {Zurlo},
  {Zhang}, {Zhang}, {Zhang}, \& {Zhou}}]{petrus24}
{Petrus}, S., {Whiteford}, N., {Patapis}, P., {et~al.} 2024, ApJL, 966, L11

\bibitem[{{Phillips} {et~al.}(2024){Phillips}, {Faherty}, {Burningham}, {Vos},
  {Gonzales}, {Griffith}, {Alejandro Merchan}, {Calamari}, {Visscher},
  {Morley}, {Whiteford}, {Gaarn}, {Ilyin}, {Strassmeier}, \&
  {Wang}}]{phillips24}
{Phillips}, C.~L., {Faherty}, J.~K., {Burningham}, B., {et~al.} 2024, 972, 172

\bibitem[{{Reyl{\'e}} {et~al.}(2021){Reyl{\'e}}, {Jardine}, {Fouqu{\'e}},
  {Caballero}, {Smart}, \& {Sozzetti}}]{reyle21}
{Reyl{\'e}}, C., {Jardine}, K., {Fouqu{\'e}}, P., {et~al.} 2021, A\&A, 650,
  A201

\bibitem[{{Riedel} {et~al.}(2017){Riedel}, {Blunt}, {Lambrides}, {Rice},
  {Cruz}, \& {Faherty}}]{riedel17}
{Riedel}, A.~R., {Blunt}, S.~C., {Lambrides}, E.~L., {et~al.} 2017, AJ, 153, 95

\bibitem[{{Rodriguez} {et~al.}(2013){Rodriguez}, {Zuckerman}, {Kastner},
  {Bessell}, {Faherty}, \& {Murphy}}]{rodriguez13}
{Rodriguez}, D.~R., {Zuckerman}, B., {Kastner}, J.~H., {et~al.} 2013, 774, 101

\bibitem[{{Saumon} \& {Marley}(2008)}]{saumon08}
{Saumon}, D. \& {Marley}, M.~S. 2008, ApJ, 689, 1327

\bibitem[{{Schmidt} {et~al.}(2010){Schmidt}, {West}, {Burgasser}, {Bochanski},
  \& {Hawley}}]{schmidt10a}
{Schmidt}, S.~J., {West}, A.~A., {Burgasser}, A.~J., {Bochanski}, J.~J., \&
  {Hawley}, S.~L. 2010, AJ, 139, 1045

\bibitem[{{Schneider} {et~al.}(2014){Schneider}, {Cushing}, {Kirkpatrick},
  {Mace}, {Gelino}, {Faherty}, {Fajardo-Acosta}, \& {Sheppard}}]{schneider14a}
{Schneider}, A.~C., {Cushing}, M.~C., {Kirkpatrick}, J.~D., {et~al.} 2014, 147,
  34

\bibitem[{{Schneider} {et~al.}(2026){Schneider}, {Kuchner}, {Faherty},
  {Meisner}, {Kirkpatrick}, {Burgasser}, {Bardalez Gagliuffi}, {Bickle},
  {Caselden}, {Casewell}, \& et~al.}]{schneider26}
{Schneider}, A.~C., {Kuchner}, M.~J., {Faherty}, J.~K., {et~al.} 2026, 171, 341

\bibitem[{{Schneider} {et~al.}(2017){Schneider}, {Windsor}, {Cushing},
  {Kirkpatrick}, \& {Shkolnik}}]{schneider17a}
{Schneider}, A.~C., {Windsor}, J., {Cushing}, M.~C., {Kirkpatrick}, J.~D., \&
  {Shkolnik}, E.~L. 2017, AJ, 153, 196

\bibitem[{{Schneider} {et~al.}(2016){Schneider}, {Windsor}, {Cushing},
  {Kirkpatrick}, \& {Wright}}]{schneider16b}
{Schneider}, A.~C., {Windsor}, J., {Cushing}, M.~C., {Kirkpatrick}, J.~D., \&
  {Wright}, E.~L. 2016, ApJL, 822, L1

\bibitem[{{Scholz} {et~al.}(2020){Scholz}, {Pangsy}, \& {Cioni}}]{scholz20a}
{Scholz}, R.-D., {Pangsy}, L., \& {Cioni}, M.-R.~L. 2020, Research Notes of the
  American Astronomical Society, 4, 20

\bibitem[{{Skrutskie} {et~al.}(2006){Skrutskie}, {Cutri}, {Stiening},
  {Weinberg}, {Schneider}, {Carpenter}, \& {25 co-authors}}]{skrutskie06}
{Skrutskie}, M.~F., {Cutri}, R.~M., {Stiening}, R., {et~al.} 2006, AJ, 131,
  1163

\bibitem[{{Slesnick} {et~al.}(2006){Slesnick}, {Carpenter}, {Hillenbrand}, \&
  {Mamajek}}]{slesnick06b}
{Slesnick}, C.~L., {Carpenter}, J.~M., {Hillenbrand}, L.~A., \& {Mamajek},
  E.~E. 2006, AJ, 132, 2665

\bibitem[{{Solano} {et~al.}(2021){Solano}, {G{\'a}lvez-Ortiz}, {Mart{\'\i}n},
  {G{\'o}mez Mu{\~n}oz}, {Rodrigo}, {Burgasser}, {Lodieu}, {B{\'e}jar},
  {Hu{\'e}lamo}, {Morales-Calder{\'o}n}, \& {Bouy}}]{solano21}
{Solano}, E., {G{\'a}lvez-Ortiz}, M.~C., {Mart{\'\i}n}, E.~L., {et~al.} 2021,
  MNRAS, 501, 281

\bibitem[{{Stauffer} {et~al.}(1998){Stauffer}, {Schultz}, \&
  {Kirkpatrick}}]{stauffer98}
{Stauffer}, J.~R., {Schultz}, G., \& {Kirkpatrick}, J.~D. 1998, ApJL, 499, 199

\bibitem[{{Stephens} {et~al.}(2009){Stephens}, {Leggett}, {Cushing}, {Marley},
  {Saumon}, {Geballe}, {Golimowski}, {Fan}, \& {Noll}}]{stephens09}
{Stephens}, D.~C., {Leggett}, S.~K., {Cushing}, M.~C., {et~al.} 2009, ApJ, 702,
  154

\bibitem[{{Su{\'a}rez} \& {Metchev}(2022)}]{suarez22}
{Su{\'a}rez}, G. \& {Metchev}, S. 2022, \mnras, 513, 5701

\bibitem[{{Todorov} {et~al.}(2010){Todorov}, {Luhman}, \&
  {McLeod}}]{todorov10a}
{Todorov}, K., {Luhman}, K.~L., \& {McLeod}, K.~K. 2010, ApJL, 714, L84

\bibitem[{{Torres} {et~al.}(2006){Torres}, {Quast}, {da Silva}, {de La Reza},
  {Melo}, \& {Sterzik}}]{torres06}
{Torres}, C.~A.~O., {Quast}, G.~R., {da Silva}, L., {et~al.} 2006, A\&A, 460,
  695

\bibitem[{{Torres} {et~al.}(2008){Torres}, {Quast}, {Melo}, \&
  {Sterzik}}]{torres08}
{Torres}, C.~A.~O., {Quast}, G.~R., {Melo}, C.~H.~F., \& {Sterzik}, M.~F. 2008,
  {Young Nearby Loose Associations}, ed. B.~{Reipurth}, 757

\bibitem[{{Tremblin} {et~al.}(2015){Tremblin}, {Amundsen}, {Mourier},
  {Baraffe}, {Chabrier}, {Drummond}, {Homeier}, \& {Venot}}]{tremblin15}
{Tremblin}, P., {Amundsen}, D.~S., {Mourier}, P., {et~al.} 2015, ApJL, 804, L17

\bibitem[{{Tsuji}(2002)}]{tsuji02}
{Tsuji}, T. 2002, ApJ, 575, 264

\bibitem[{{Tsuji}(2005)}]{tsuji05}
{Tsuji}, T. 2005, ApJ, 621, 1033

\bibitem[{{Tsuji} {et~al.}(2004){Tsuji}, {Nakajima}, \& {Yanagisawa}}]{tsuji04}
{Tsuji}, T., {Nakajima}, T., \& {Yanagisawa}, K. 2004, ApJ, 607, 511

\bibitem[{{Vrba} {et~al.}(2004){Vrba}, {Henden}, {Luginbuhl}, {Guetter},
  {Munn}, {Canzian}, {Burgasser}, {Kirkpatrick}, {Fan}, {Geballe},
  {Golimowski}, {Knapp}, {Leggett}, {Schneider}, \& {Brinkmann}}]{vrba04}
{Vrba}, F.~J., {Henden}, A.~A., {Luginbuhl}, C.~B., {et~al.} 2004, AJ, 127,
  2948

\bibitem[{{{\v{Z}}erjal} {et~al.}(2026){{\v{Z}}erjal}, {Dominguez-Tagle},
  {Vitas}, {Sedighi}, {Mart{\'\i}n}, {Zapatero Osorio}, {Olivares}, {Mu{\~n}oz
  Torres}, {Tsilia}, {Zhang}, {Barrado}, {B{\'e}jar}, {Bouy}, {Burgasser},
  {Cruz}, {Lodieu}, {Mas Buitrago}, {Phan-Bao}, {Solano}, {Tata}, {Goldman},
  {Mohandasan}, {Reyl{\'e}}, {Smart}, {Aghanim}, {Altieri}, {Amara}, {Andreon},
  {Auricchio}, {Baccigalupi}, {Baldi}, {Balestra}, {Bardelli}, {Battaglia},
  {Biviano}, {Bonchi}, {Branchini}, {Brescia}, {Brinchmann}, {Camera},
  {Ca{\~n}as-Herrera}, {Capobianco}, {Carbone}, {Carretero}, {Casas},
  {Castellano}, {Castignani}, {Cavuoti}, {Chambers}, {Cimatti},
  {Colodro-Conde}, {Congedo}, {Conselice}, {Conversi}, {Copin}, {Courbin},
  {Courtois}, {Cropper}, {Cuby}, {Da Silva}, {Degaudenzi}, {De Lucia},
  {Dolding}, {Dole}, {Douspis}, {Dubath}, {Dupac}, {Dusini}, {Escoffier},
  {Farina}, {Faustini}, {Ferriol}, {Fotopoulou}, {Frailis}, {Franceschi},
  {Galeotta}, {George}, {Gillis}, {Giocoli}, {G{\'o}mez-Alvarez},
  {Gracia-Carpio}, {Granett}, {Grazian}, {Grupp}, {Haugan}, {Hoar}, {Holmes},
  {Hormuth}, {Hornstrup}, {Jahnke}, {Jhabvala}, {Keih{\"a}nen}, {Kermiche},
  {Kiessling}, {Kubik}, {Kuijken}, {K{\"u}mmel}, {Kunz}, {Kurki-Suonio}, {Le
  Boulc'h}, {Le Brun}, {Ligori}, {Lilje}, {Lindholm}, {Lloro}, {Mainetti},
  {Maino}, {Maiorano}, {Mansutti}, {Marggraf}, {Martinelli}, {Martinet},
  {Marulli}, {Massey}, {Medinaceli}, {Mei}, {Mellier}, {Meneghetti}, {Merlin},
  {Meylan}, {Mora}, {Moresco}, {Moscardini}, {Nakajima}, {Neissner}, {Niemi},
  {Padilla}, {Paltani}, {Pasian}, {Pedersen}, {Percival}, {Pettorino}, {Pires},
  {Polenta}, {Poncet}, {Popa}, {Pozzetti}, {Raison}, {Rebolo}, {Renzi},
  {Rhodes}, {Riccio}, {Romelli}, {Roncarelli}, {Saglia}, {Sakr}, {Sapone},
  {Sartoris}, {Schewtschenko}, {Schirmer}, {Schneider}, {Secroun}, {Seidel},
  {Seiffert}, {Serrano}, {Simon}, {Sirignano}, {Sirri}, {Stanco},
  {Steinwagner}, {Tallada-Cresp{\'\i}}, {Taylor}, {Tereno}, {Toft},
  {Toledo-Moreo}, {Torradeflot}, {Tsyganov}, {Tutusaus}, {Valenziano},
  {Valiviita}, {Vassallo}, {Verdoes Kleijn}, {Veropalumbo}, {Wang}, {Weller},
  {Zacchei}, {Zamorani}, {Zerbi}, {Zucca}, {Mart{\'\i}n-Fleitas}, \&
  {Scottez}}]{zerjal26}
{{\v{Z}}erjal}, M., {Dominguez-Tagle}, C., {Vitas}, N., {et~al.} 2026, 711, A37

\bibitem[{{White} \& {Basri}(2003)}]{white03}
{White}, R.~J. \& {Basri}, G. 2003, ApJ, 582, 1109

\bibitem[{{Wright} {et~al.}(2010){Wright}, {Eisenhardt}, {Mainzer}, {Ressler},
  {Cutri}, {Jarrett}, {Kirkpatrick}, \& {31 co-authors}}]{wright10}
{Wright}, E.~L., {Eisenhardt}, P.~R.~M., {Mainzer}, A.~K., {et~al.} 2010, AJ,
  140, 1868

\bibitem[{{York} {et~al.}(2000){York}, {Adelman}, {Anderson}, {Anderson},
  {Annis}, {Bahcall}, {Bakken}, {Barkhouser}, {Bastian}, {Berman}, \& {134
  coauthors}}]{york00}
{York}, D.~G., {Adelman}, J., {Anderson}, J.~E., {et~al.} 2000, AJ, 120, 1579

\bibitem[{{Zapatero Osorio} {et~al.}(2000){Zapatero Osorio}, {B{\' e}jar},
  {Mart{\'{\i}}n}, {Rebolo}, {Barrado y Navascu{\' e}s}, {Bailer-Jones}, \&
  {Mundt}}]{zapatero00}
{Zapatero Osorio}, M.~R., {B{\' e}jar}, V.~J.~S., {Mart{\'{\i}}n}, E.~L.,
  {et~al.} 2000, Science, 290, 103

\bibitem[{{Zapatero Osorio} {et~al.}(1999){Zapatero Osorio}, {B{\' e}jar},
  {Rebolo}, {Mart{\'{\i}}n}, \& {Basri}}]{zapatero99a}
{Zapatero Osorio}, M.~R., {B{\' e}jar}, V.~J.~S., {Rebolo}, R.,
  {Mart{\'{\i}}n}, E.~L., \& {Basri}, G. 1999, ApJL, 524, L115

\bibitem[{{Zapatero Osorio} {et~al.}(2014{\natexlab{a}}){Zapatero Osorio},
  {B{\'e}jar}, {Miles-P{\'a}ez}, {Pe{\~n}a Ram{\'{\i}}rez}, {Rebolo}, \&
  {Pall{\'e}}}]{zapatero14a}
{Zapatero Osorio}, M.~R., {B{\'e}jar}, V.~J.~S., {Miles-P{\'a}ez}, P.~A.,
  {et~al.} 2014{\natexlab{a}}, A\&A, 568, A6

\bibitem[{{Zapatero Osorio} {et~al.}(2017){Zapatero Osorio}, {B{\'e}jar}, \&
  {Pe{\~n}a Ram{\'{\i}}rez}}]{zapatero17}
{Zapatero Osorio}, M.~R., {B{\'e}jar}, V.~J.~S., \& {Pe{\~n}a Ram{\'{\i}}rez},
  K. 2017, ApJ, 842, 65

\bibitem[{{Zapatero Osorio} {et~al.}(2014{\natexlab{b}}){Zapatero Osorio},
  {G{\'a}lvez Ortiz}, {Bihain}, {Bailer-Jones}, {Rebolo}, {Henning},
  {Boudreault}, {B{\'e}jar}, {Goldman}, {Mundt}, \& {Caballero}}]{zapatero14b}
{Zapatero Osorio}, M.~R., {G{\'a}lvez Ortiz}, M.~C., {Bihain}, G., {et~al.}
  2014{\natexlab{b}}, A\&A, 568, A77

\bibitem[{{Zhang} {et~al.}(2024){Zhang}, {Lodieu}, \& {Mart{\'\i}n}}]{zhang24a}
{Zhang}, J.~Y., {Lodieu}, N., \& {Mart{\'\i}n}, E.~L. 2024, A\&A, 686, A171

\bibitem[{{Zhang} {et~al.}(2022){Zhang}, {Liu}, {Morley}, {Magnier}, {Tucker},
  {Vanderbosch}, {Do}, \& {Shappee}}]{zhang22a}
{Zhang}, Z., {Liu}, M.~C., {Morley}, C.~V., {et~al.} 2022, 935, 15

\bibitem[{{Zuckerman} \& {Song}(2004)}]{zuckerman04}
{Zuckerman}, B. \& {Song}, I. 2004, ARA\&A, 42, 685

\bibitem[{{Zuckerman} {et~al.}(2001{\natexlab{a}}){Zuckerman}, {Song},
  {Bessell}, \& {Webb}}]{zuckerman01b}
{Zuckerman}, B., {Song}, I., {Bessell}, M.~S., \& {Webb}, R.~A.
  2001{\natexlab{a}}, 562, L87

\bibitem[{{Zuckerman} {et~al.}(2001{\natexlab{b}}){Zuckerman}, {Song}, \&
  {Webb}}]{zuckerman01a}
{Zuckerman}, B., {Song}, I., \& {Webb}, R.~A. 2001{\natexlab{b}}, 559, 388

\end{thebibliography}

%
%%%%%%%%%%%%%%%%%%%%%%%%%%%%%%%%%%%%%%%%%%%%
%%%%% Appendix %%%%%
%%%%%%%%%%%%%%%%%%%%%%%%%%%%%%%%%%%%%%%%%%%%
%
\newpage
\begin{appendix} %First appendix
%
%
%%%%%%%%%%%%%%%%%%%%%%%%%%%%%%%%%%%%%%%%%%%%
%%%%% Section: list URLDs %%%%%
%%%%%%%%%%%%%%%%%%%%%%%%%%%%%%%%%%%%%%%%%%%%
%
\section{List of URLD candidates in {\it{Euclid}} DR1}
\label{tab_redLdwarfs:Euclid_cand_redL}
%

%
%%%%%%%%%%%%%%%%%%%%%%%%%%%%%%%%%%%%%%%%%%%%
%%%%% Table: 45 red L dwarfs %%%%%
%%%%%%%%%%%%%%%%%%%%%%%%%%%%%%%%%%%%%%%%%%%%
%
\begin{table*}[!h]
\centering
\caption{Photometry of 45 URLD with {\it{Euclid}} red grism spectra with their tentative spectral types.}
\begin{tabular}{@{\hspace{0mm}}l@{\hspace{2mm}}c@{\hspace{2mm}}c@{\hspace{2mm}}c@{\hspace{2mm}}c@{\hspace{2mm}}c@{\hspace{2mm}}c@{\hspace{2mm}}c@{\hspace{0mm}}}
\hline
\hline
ID & ra & dec & $I_E$ & $Y_E$ & $J_E$ & $H_E$ & SpT \\
\hline
$-$45941666611808960 & 00:18:22.60 & $-$61:10:51.2 & 24.684 & 21.471 & 20.640 & 19.963 & L6 slightly red \\
1357448441671416566 & 09:02:58.76 & $+$67:08:29.9 & 21.815 & 18.397 & 17.570 & 16.845 & L8 young \\
1376728593687371903 & 09:10:41.48 & $+$68:44:13.8 & 24.844 & 21.336 & 20.601 & 19.948 & L8 normal \\
$-$145501753568921593 & 00:58:12.04 & $-$56:53:31.7 & 21.805 & 18.517 & 17.881 & 17.203 & L8 red \\
1452300871691204635 & 09:40:55.22 & $+$69:07:13.6 & 24.642 & 21.520 & 20.795 & 20.043 & L9 slightly red \\
1495492300718620996 & 09:58:11.81 & $+$71:51:43.5 & 23.047 & 19.647 & 19.000 & 18.283 & L9 red \\
1534759777701437823 & 10:13:54.23 & $+$70:08:37.6 & 24.839 & 21.509 & 20.772 & 20.045 & L8 normal \\
1573556887707328508 & 10:29:25.36 & $+$70:43:58.2 & 25.080 & 21.764 & 20.934 & 20.347 & L2 normal \\
1881542060667865073 & 12:32:37.00 & $+$66:47:11.4 & 22.960 & 19.693 & 18.950 & 18.397 & L6 normal \\
2019439721665852300 & 13:27:46.55 & $+$66:35:06.8 & 25.219 & 21.907 & 21.202 & 20.507 & L8 normal/slightly red \\
2116140242727283477 & 14:06:27.36 & $+$72:43:42.0 & 24.664 & 21.305 & 20.593 & 20.033 & L7 normal \\
2575039597586286417 & 17:10:00.95 & $+$58:37:43.1 & 25.077 & 21.676 & 20.880 & 20.225 & L7 normal \\
$-$265131383570597754 & 01:46:03.15 & $-$57:03:35.1 & 24.369 & 20.940 & 20.201 & 19.575 & L7 slightly red \\
2647470739617127439 & 17:38:59.29 & $+$61:42:45.8 & 23.371 & 19.817 & 18.995 & 18.109 & L7 red \\
2775981445631192462 & 18:30:23.55 & $+$63:07:09.2 & 24.481 & 21.057 & 20.332 & 19.782 & L7 normal \\
$-$287391936539877439 & 01:54:57.40 & $-$53:59:15.8 & 23.604 & 20.318 & 19.619 & 18.944 & L8 slightly red \\
2808558722681698256 & 18:43:25.41 & $+$68:10:11.4 & 23.628 & 20.442 & 19.754 & 19.176 & L3 slightly red \\
$-$319061547567939513 & 02:07:37.48 & $-$56:47:38.2 & 22.745 & 19.420 & 18.681 & 18.094 & L3 slightly red \\
$-$348561891518913923 & 02:19:25.45 & $-$51:53:29.0 & 23.850 & 20.508 & 19.656 & 18.828 & L7 red, young \\
$-$372851399530883137 & 02:29:08.43 & $-$53:05:17.9 & 25.022 & 21.587 & 20.726 & 19.870 & L7 red \\
$-$418558596493423172 & 02:47:25.40 & $-$49:20:32.3 & 23.706 & 20.188 & 19.319 & 18.423 & L7 red \\
$-$487845417438547042 & 03:15:08.26 & $-$43:51:16.9 & 25.458 & 22.060 & 21.287 & 20.681 & L6 normal \\
$-$497285970343578614 & 03:18:54.85 & $-$34:21:28.3 & 20.705 & 17.357 & 16.704 & 16.092 & L7 slightly red \\
$-$550087388455256481 & 03:40:02.09 & $-$45:31:32.3 & 23.712 & 20.480 & 19.837 & 19.172 & L9 young \\
$-$583627745460707400 & 03:53:27.06 & $-$46:04:14.7 & 23.781 & 20.563 & 19.801 & 19.132 & L7 red, young \\
$-$603532600301276456 & 04:01:24.78 & $-$30:07:39.5 & 24.397 & 21.054 & 20.098 & 19.188 & L7 red \\
$-$606978226381899004 & 04:02:47.48 & $-$38:11:23.6 & 24.748 & 21.414 & 20.470 & 19.591 & L9 red, young \\
$-$621465574267028128 & 04:08:35.17 & $-$26:42:10.1 & 23.508 & 20.181 & 19.421 & 18.861 & L6 doubtful \\
$-$654189936225947427 & 04:21:40.56 & $-$22:35:41.1 & 23.939 & 20.681 & 19.961 & 19.426 & L8 red \\
$-$663300524440992964 & 04:25:19.21 & $-$44:05:57.5 & 25.723 & 22.470 & 21.608 & 20.744 & L9 normal \\
$-$673433146567177770 & 04:29:22.39 & $-$56:43:04.0 & 24.546 & 21.153 & 20.367 & 19.673 & L7 normal \\
$-$693289112551618214 & 04:37:18.94 & $-$55:09:42.6 & 22.169 & 18.807 & 18.033 & 17.382 & L7 slightly red \\
$-$696342844478904353 & 04:38:32.23 & $-$47:53:25.6 & 22.827 & 19.560 & 18.895 & 18.246 & L7 normal \\
$-$710855464391437625 & 04:44:20.51 & $-$39:08:37.5 & 23.822 & 20.495 & 19.667 & 18.983 & L6 red \\
$-$717096746540523855 & 04:46:50.32 & $-$54:03:08.6 & 24.122 & 20.806 & 20.048 & 19.442 & L2 slightly red \\
$-$747520247288936772 & 04:59:00.49 & $-$28:53:37.2 & 22.941 & 19.687 & 18.902 & 18.061 & L7 red \\
$-$782067599436445676 & 05:12:49.62 & $-$43:38:40.4 & 22.749 & 19.415 & 18.720 & 18.150 & L9 slightly red \\
$-$786968280389151351 & 05:14:47.24 & $-$38:54:54.5 & 23.166 & 19.746 & 18.975 & 18.245 & L9 red young \\
$-$808155239328489359 & 05:23:15.73 & $-$32:50:56.2 & 23.501 & 20.077 & 19.293 & 18.722 & L6 normal \\
$-$842932247408362401 & 05:37:10.38 & $-$40:50:10.5 & 22.470 & 19.246 & 18.547 & 17.987 & L6 slightly red \\
$-$880739624365670640 & 05:52:17.75 & $-$36:34:01.4 & 23.344 & 19.983 & 19.320 & 18.690 & L8 blue \\
$-$882638652410764584 & 05:53:03.33 & $-$41:04:35.3 & 24.397 & 21.180 & 20.231 & 19.356 & L7 slightly red \\
$-$902636607384468600 & 06:01:03.28 & $-$38:26:48.7 & 24.122 & 20.740 & 20.104 & 19.489 & L9 blue \\
$-$3580308253582308249 & 23:52:07.40 & $-$58:13:51.0 & 24.729 & 21.546 & 20.790 & 20.184 & L9 normal? \\
\hline
\end{tabular}
\label{tab_redLdwarfs:Euclid_45_redL}
\end{table*}

%
%%%%%%%%%%%%%%%%%%%%%%%%%%%%%%%%%%%%%%%%%%%
%%%%% Figure: Euclid images %%%%%
%%%%%%%%%%%%%%%%%%%%%%%%%%%%%%%%%%%%%%%%%%%
%
% Plot: Euclid cutout images
% Plot made with programme program_Euclid_cutouts_with_input_list.py
%
\begin{figure*}[!h]
   \centering
   \includegraphics[width=\linewidth, angle=0]{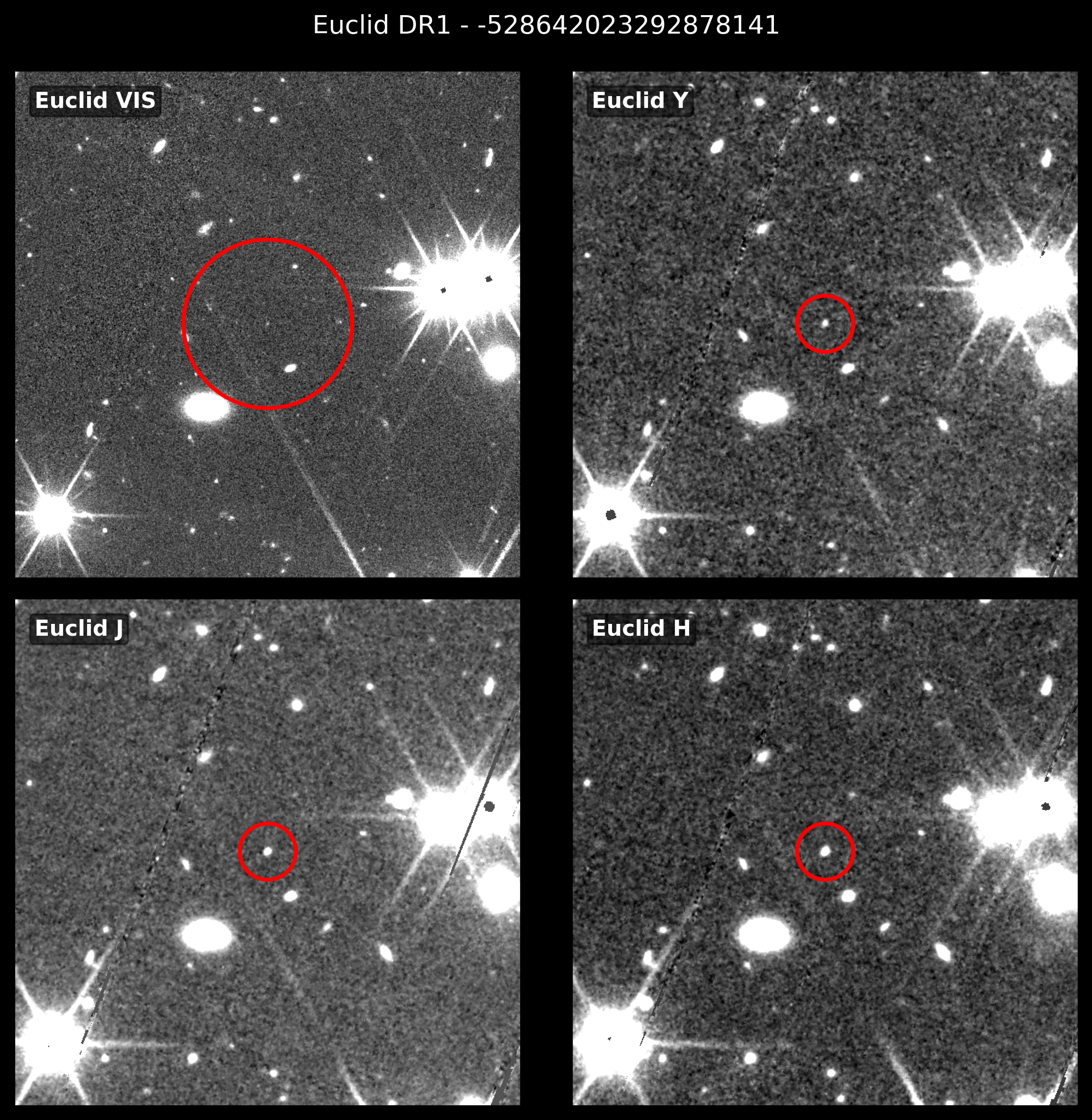}
   \caption{{\it{Euclid}} images in the four filters ($I_E$, $Y_E$, $J_E$, and $H_E$ of one URLD candidate. Image size is 2$\times$2 arcmin across.
   We show one example in this appendix; the other will be made available online.
   }
   \label{fig_redLdwarfs:Euclid_images}
\end{figure*}
%

%
%%%%%%%%%%%%%%%%%%%%%%%%%%%%%%%%%%%%%%%%%%%
%%%%% Figure: Euclid spectra %%%%%
%%%%%%%%%%%%%%%%%%%%%%%%%%%%%%%%%%%%%%%%%%%
%
% Plot: Euclid spectra provided by Carlos
%
\begin{figure*}[!h]
   \centering
   \includegraphics[width=0.44\linewidth, angle=0]{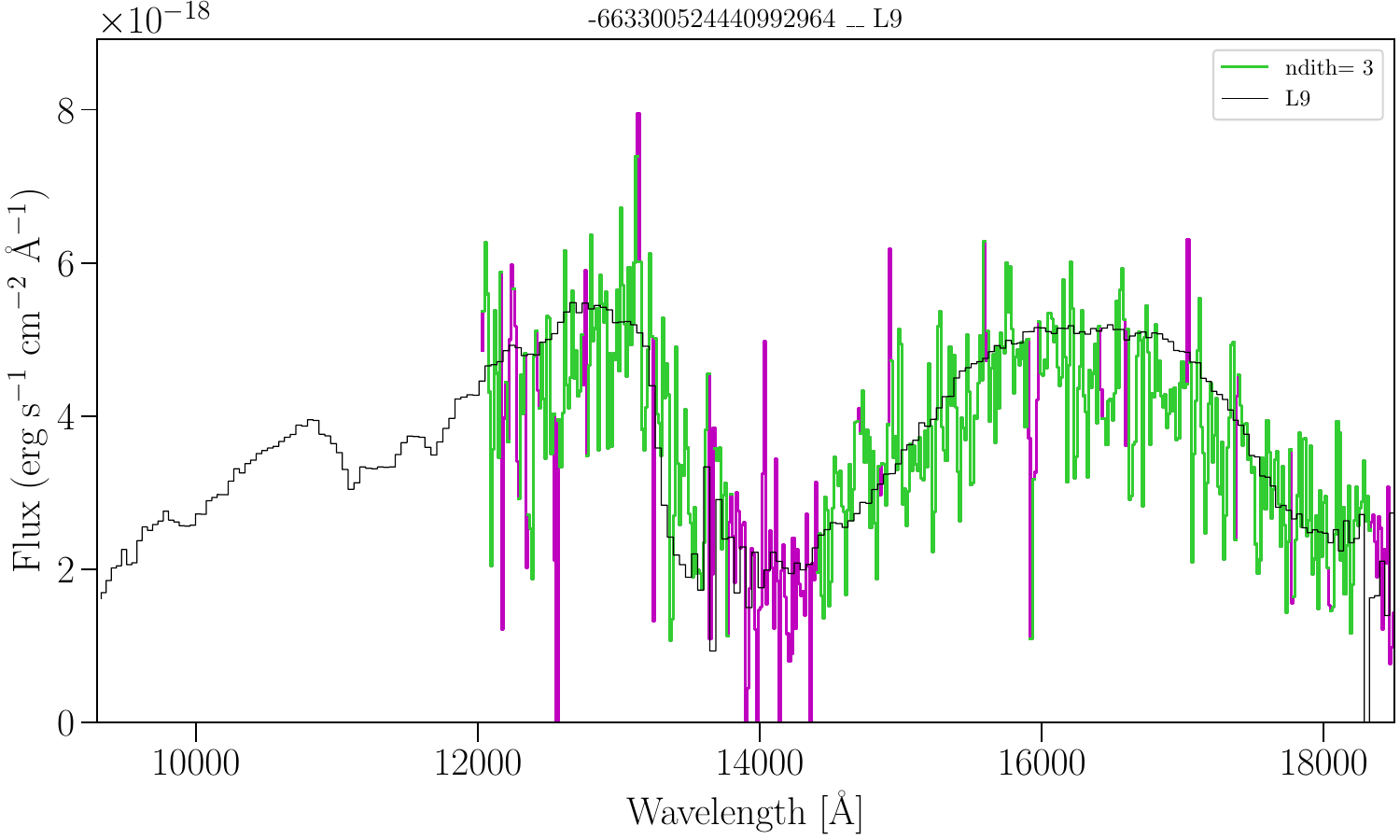}
   \includegraphics[width=0.44\linewidth, angle=0]{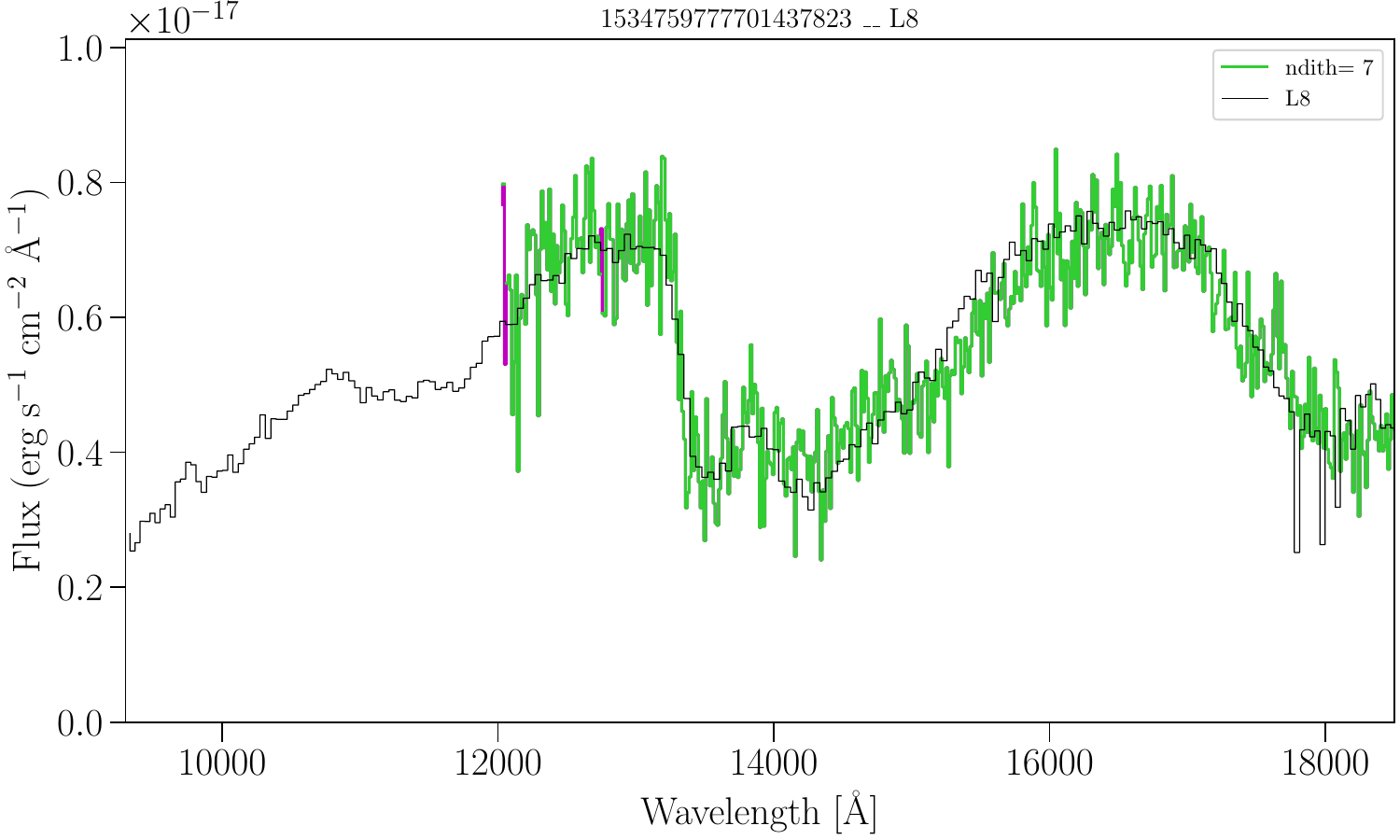}
   \includegraphics[width=0.44\linewidth, angle=0]{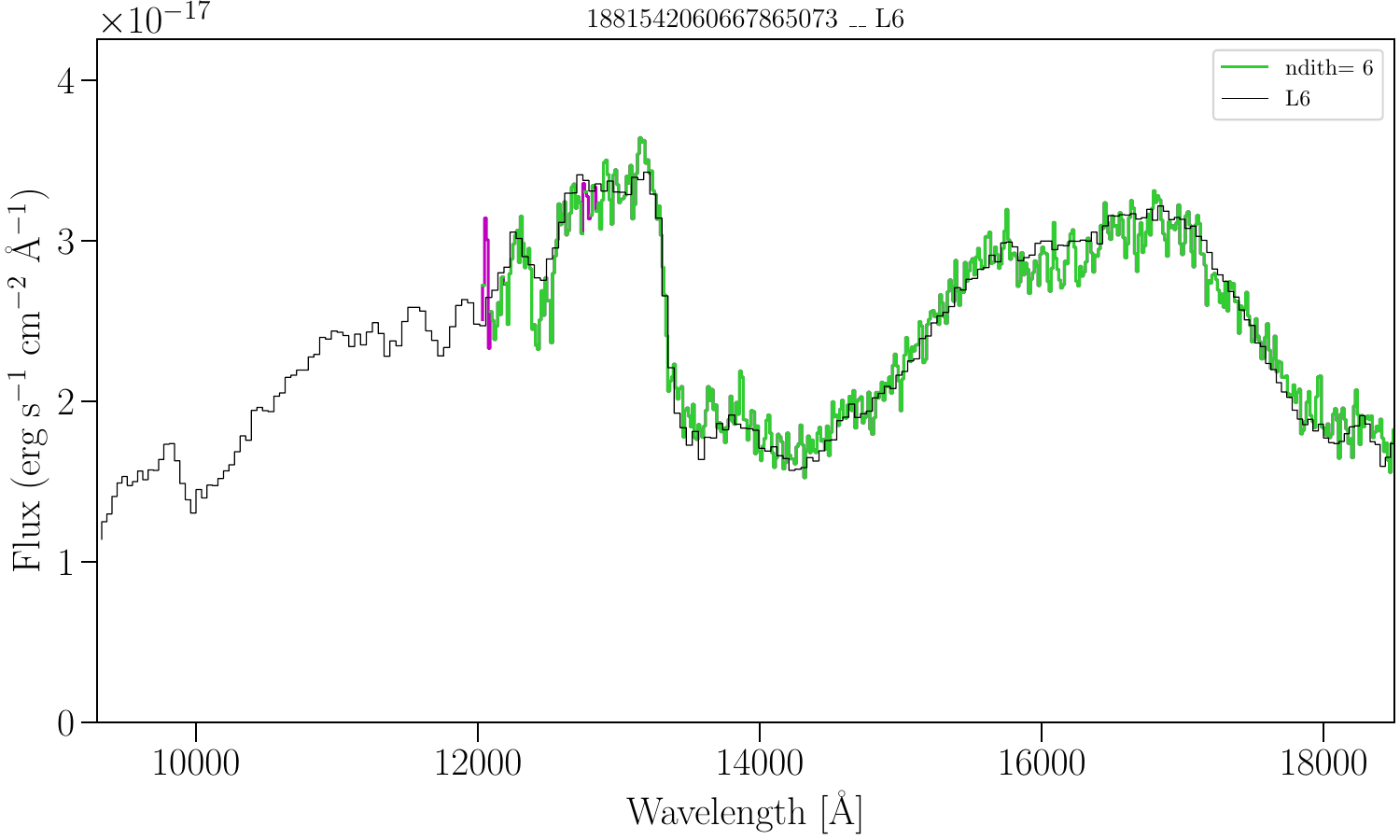}
   \includegraphics[width=0.44\linewidth, angle=0]{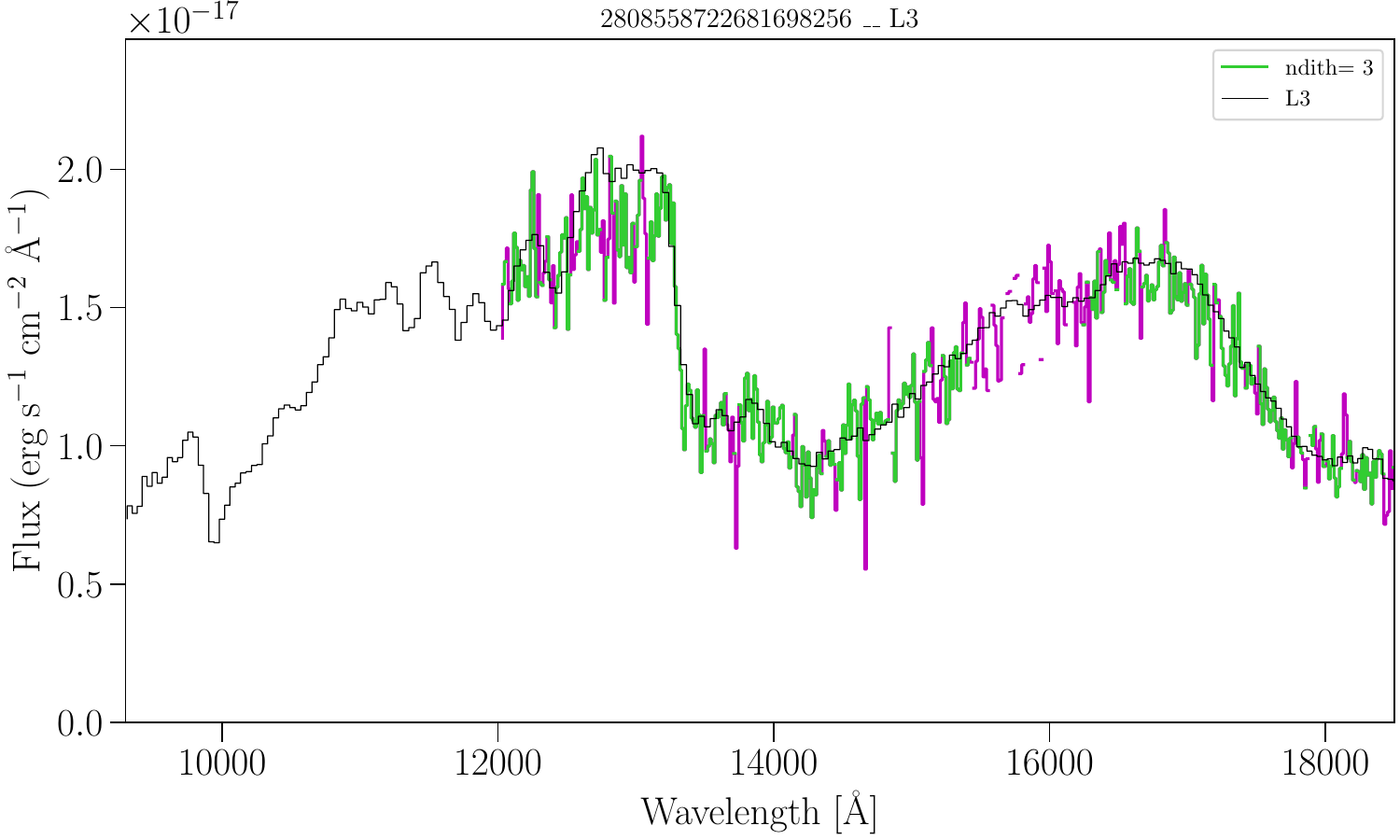}
   \caption{{\it{Euclid}} red grism spectra for URLD candidates. The green lines are good quality regions ({\tt{NDITH}}\,$>$\,2) of the spectra while the purple one are not ({\tt{QUALITY}}\,$<$\,0.7). Overplotted are old field L dwarf template with the spectral type shown at the top right. 
   We show a few examples in this figure; the other will be made available online.
   }
   \label{fig_redLdwarfs:Euclid_spectra}
\end{figure*}
%

%
%%%%%%%%%%%%%%%%%%%%%%%%%%%%%%%%%%%%%%%%%%%%
%%%%% Table: candidates in WIDE %%%%%
%%%%%%%%%%%%%%%%%%%%%%%%%%%%%%%%%%%%%%%%%%%%
%
\begin{table*}[!h]
\centering
\caption{{\it{Euclid}} photometry of URLD candidates in the {\it{Euclid}} DR1 wide survey (showing the first two and last two entries of the full sample).}
%\begin{tabular}{lcccccc}
\begin{tabular}{@{\hspace{0mm}}l@{\hspace{2mm}}c@{\hspace{2mm}}c@{\hspace{2mm}}c@{\hspace{2mm}}c@{\hspace{2mm}}c@{\hspace{2mm}}c@{\hspace{0mm}}}
\hline
\hline
ID & ra &  dec & $I_E$ & $Y_E$ & $J_E$ & $H_E$ \\
\hline
$-$13739033596159249  & 00:05:29.7 & $-$59:36:57 & 27.544 & 24.133 & 23.510 & 22.785 \\
$-$37665307604336489  & 00:15:03.9 & $-$60:26:01 & 25.843 & 22.405 & 21.717 & 21.119 \\
---                 &    ---     &    ---      &  ---   &  ---   &  ---   &  ---   \\
$-$299455197550436043 & 01:59:46.9 & $-$55:02:37 & 27.224 & 24.170 & 23.332 & 22.801 \\
$-$317163585497306650 & 02:06:51.9 & $-$49:43:50 & 27.387 & 24.045 & 23.378 & 22.822 \\
\hline
\end{tabular}
\label{tab_redLdwarfs:Euclid_cand_redL_WIDE}
\end{table*}

%
%%%%%%%%%%%%%%%%%%%%%%%%%%%%%%%%%%%%%%%%%%%%
%%%%% Table: candidates in DEEP %%%%%
%%%%%%%%%%%%%%%%%%%%%%%%%%%%%%%%%%%%%%%%%%%%
%
\begin{table*}[!h]
\centering
\caption{{\it{Euclid}} photometry of URLD candidates in the {\it{Euclid}} DR1 deep survey (showing the first two and last two entries of the full sample).}
%\begin{tabular}{lcccccc}
\begin{tabular}{@{\hspace{0mm}}l@{\hspace{2mm}}c@{\hspace{2mm}}c@{\hspace{2mm}}c@{\hspace{2mm}}c@{\hspace{2mm}}c@{\hspace{2mm}}c@{\hspace{0mm}}}
\hline
\hline
ID & ra &  dec & $I_E$ & $Y_E$ & $J_E$ & $H_E$ \\
\hline
$-$528642023292878141  & 03:31:27.41 & $-$29:17:16.1 & 27.174 & 24.044 & 23.294 & 22.617 \\
$-$531611583284931105  & 03:32:38.68 & $-$28:29:35.2 & 25.694 & 22.443 & 21.761 & 21.188 \\
---                  &    ---      &    ---        &  ---   &  ---   &  ---   &  ---   \\
2757331876665593862  & 18:22:55.97 & $+$66:33:33.8 & 28.194 & 24.940 & 24.209 & 23.702 \\
2757447401655108047  & 18:22:58.74 & $+$65:30:38.9 & 28.229 & 24.975 & 24.306 & 23.592 \\
\hline
\end{tabular}
\label{tab_redLdwarfs:Euclid_cand_redL_DEEP}
\end{table*}
%

%
%%%%%%%%%%%%%%%%%%%%%%%%%%%%%%%%%%%%%%%%%%%%
%%%%% Section: SPHEREx spectra %%%%%
%%%%%%%%%%%%%%%%%%%%%%%%%%%%%%%%%%%%%%%%%%%%
%
\section{SPHEREx spectra of URLDs}
\label{redLdwarfs:append_SPHEREx}
%

%
%%%%%%%%%%%%%%%%%%%%%%%%%%%%%%%%%%%%%%%%%%%%%
%%%%% Figure: SPHEREx spec KNOWN URLDs %%%%%
%%%%%%%%%%%%%%%%%%%%%%%%%%%%%%%%%%%%%%%%%%%%%
%
\begin{figure*}[!h]
   \centering
   \includegraphics[width=0.43\linewidth, angle=0]{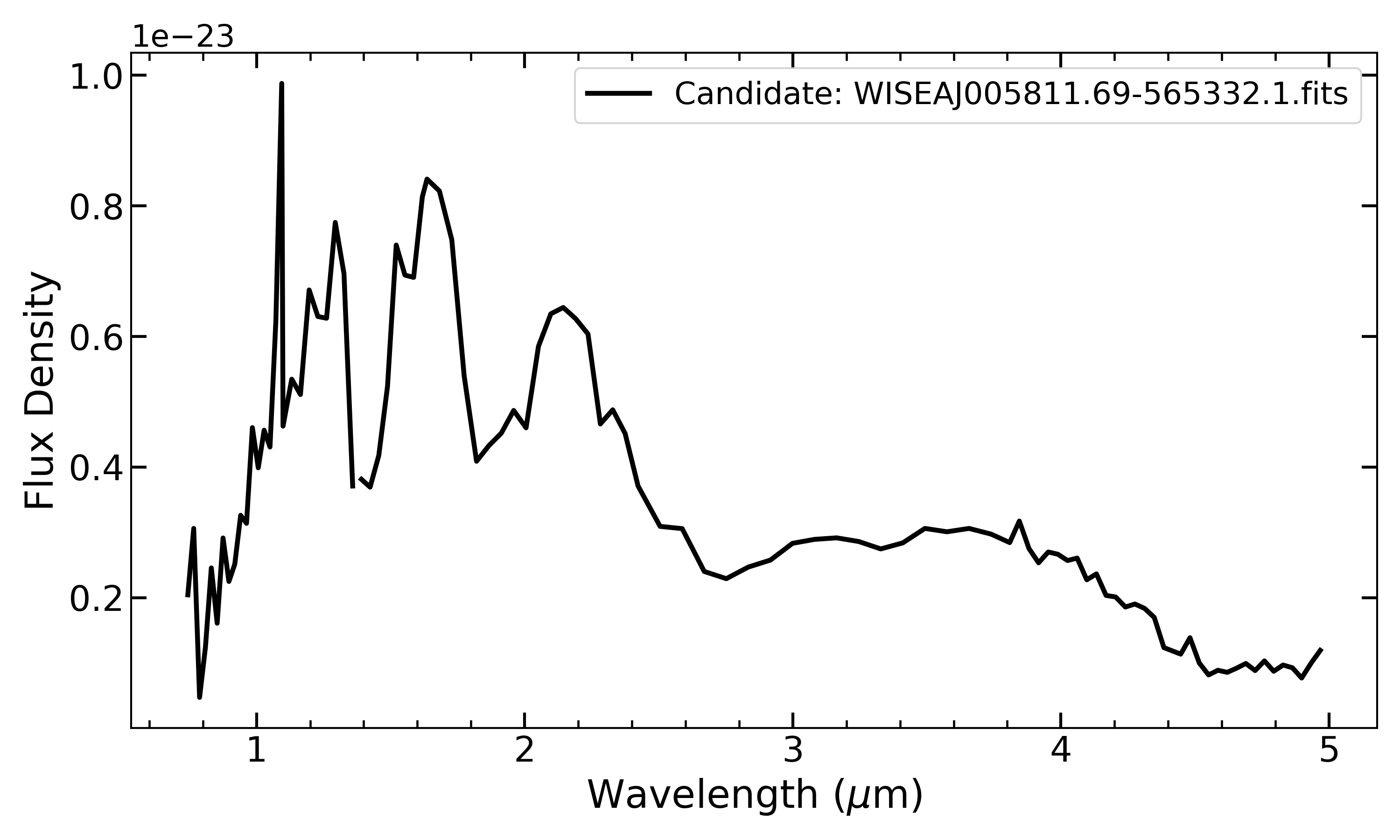}
   \includegraphics[width=0.43\linewidth, angle=0]{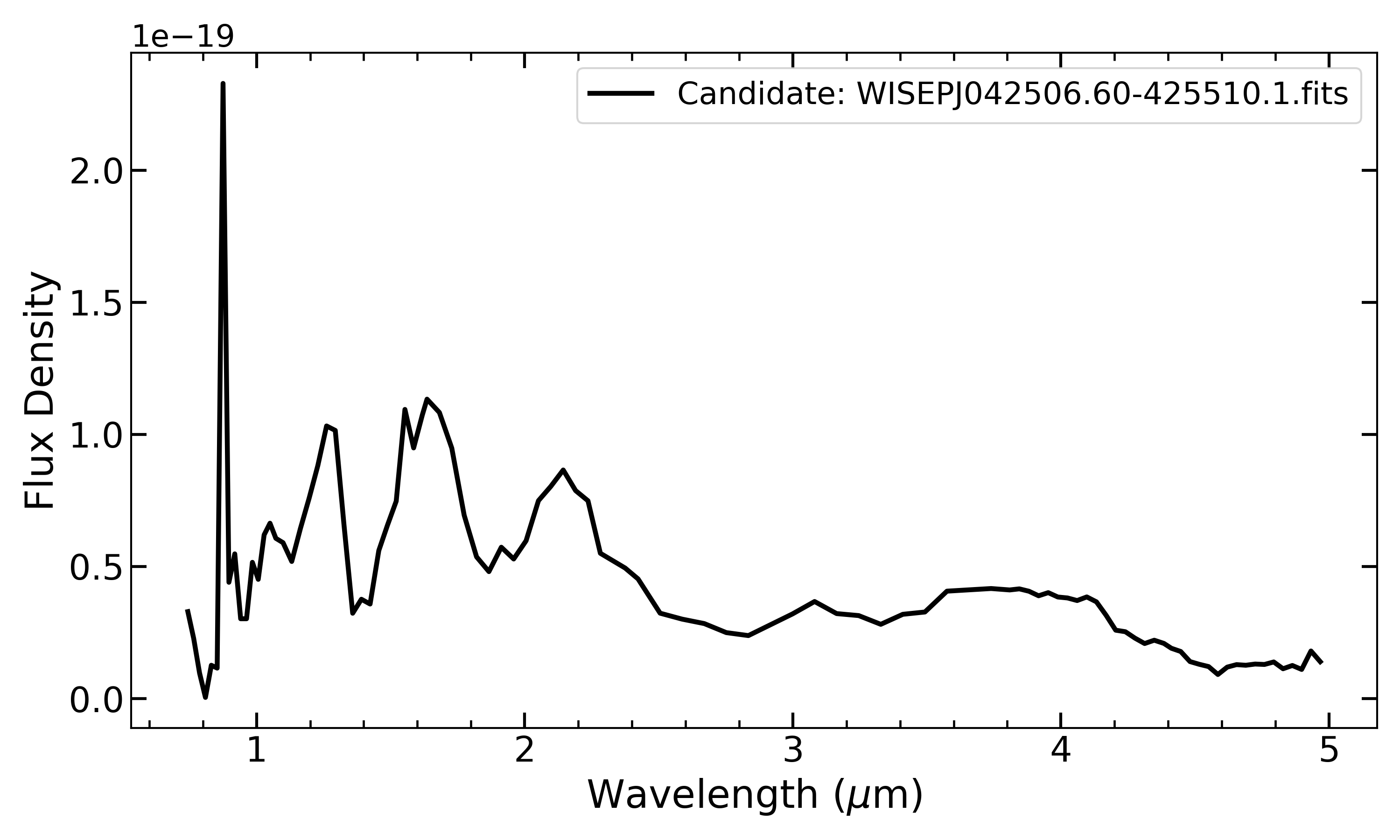}
   \includegraphics[width=0.43\linewidth, angle=0]{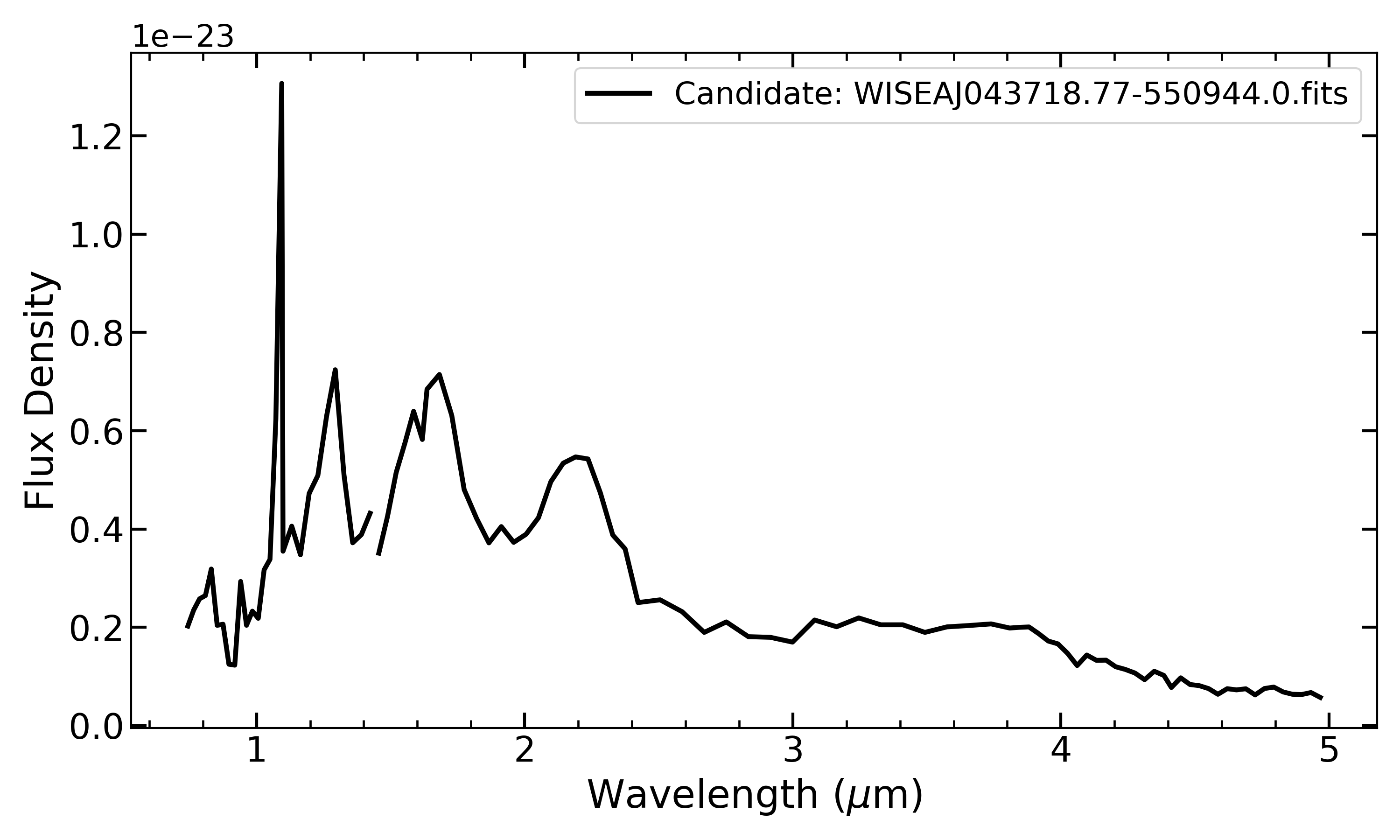}
   \includegraphics[width=0.43\linewidth, angle=0]{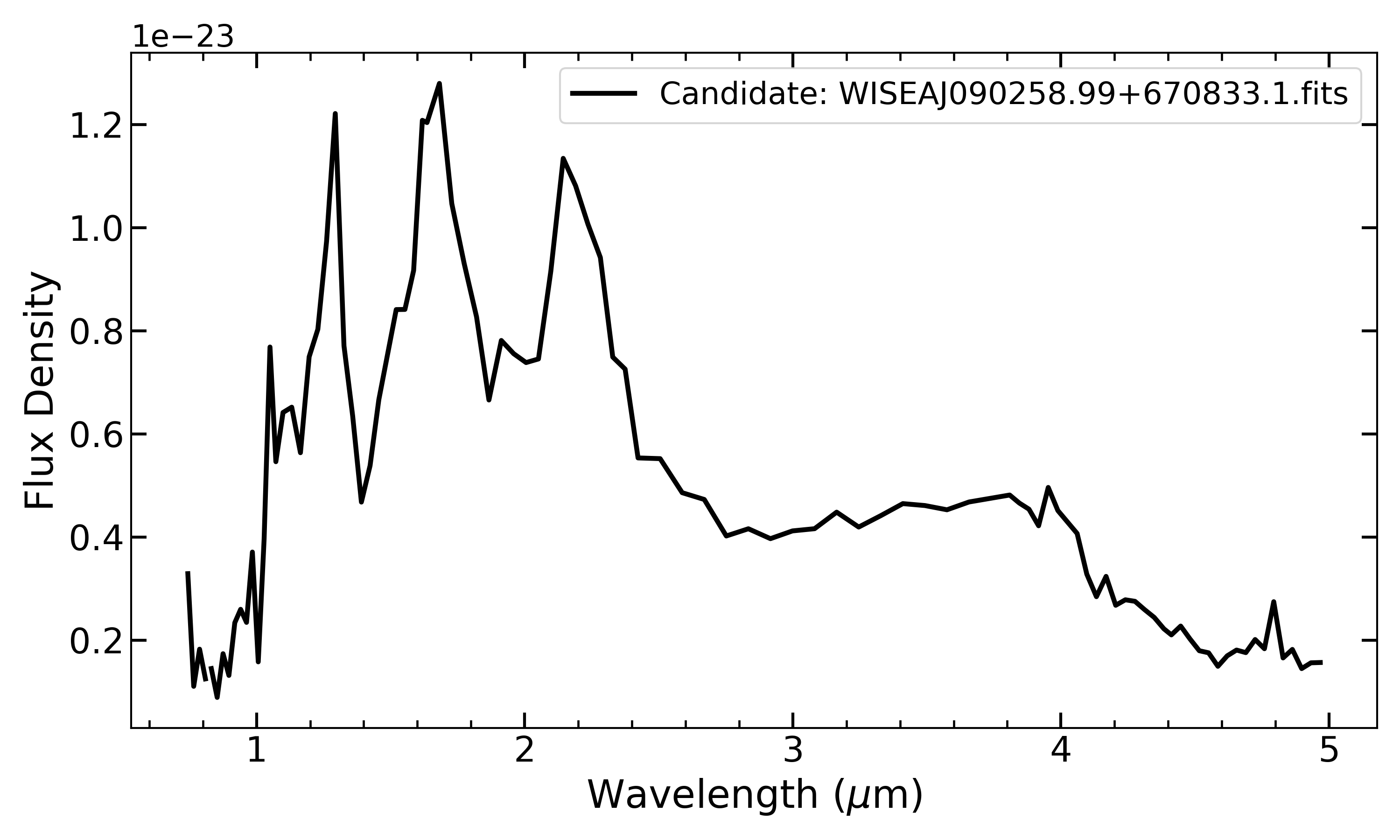}
   \includegraphics[width=0.43\linewidth, angle=0]{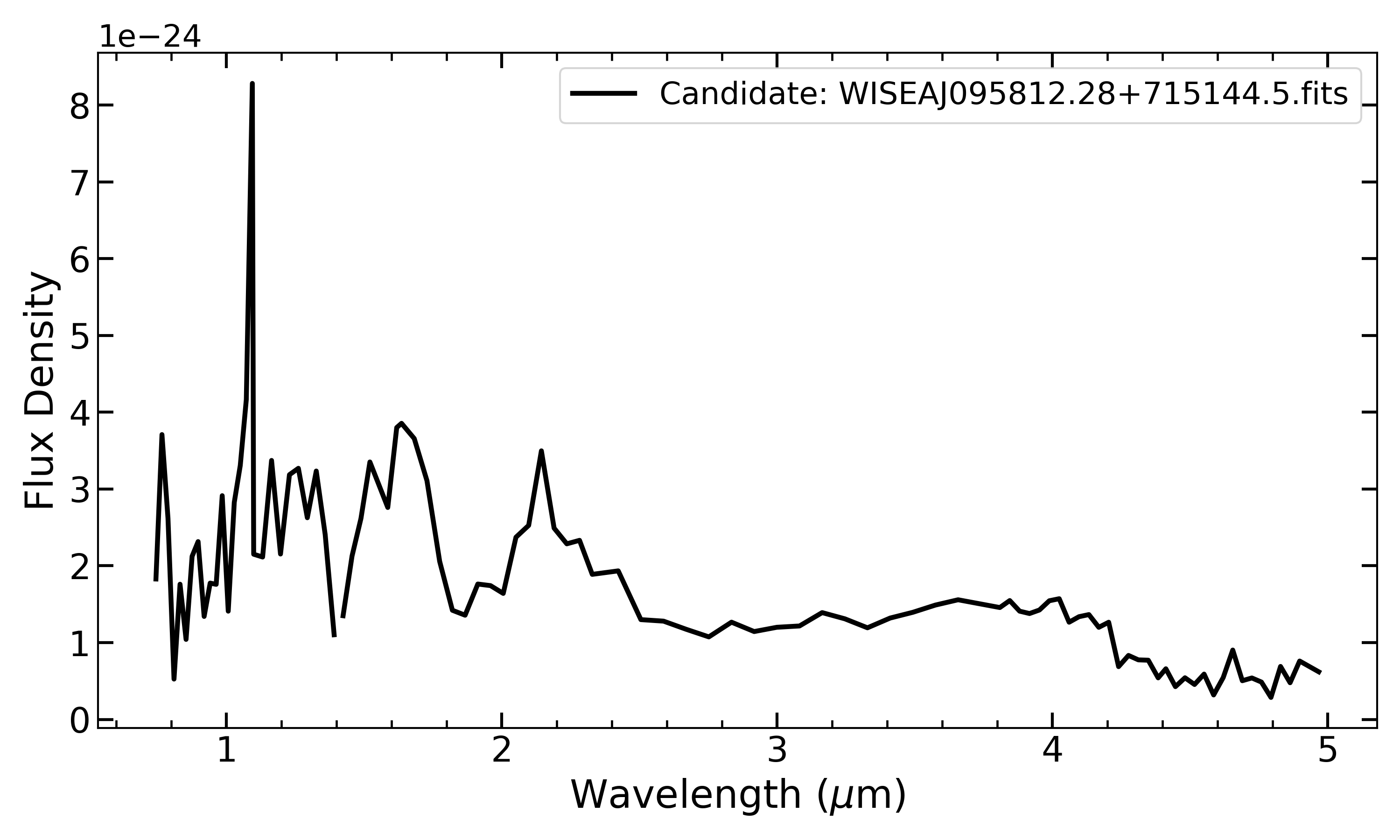}
   \includegraphics[width=0.43\linewidth, angle=0]{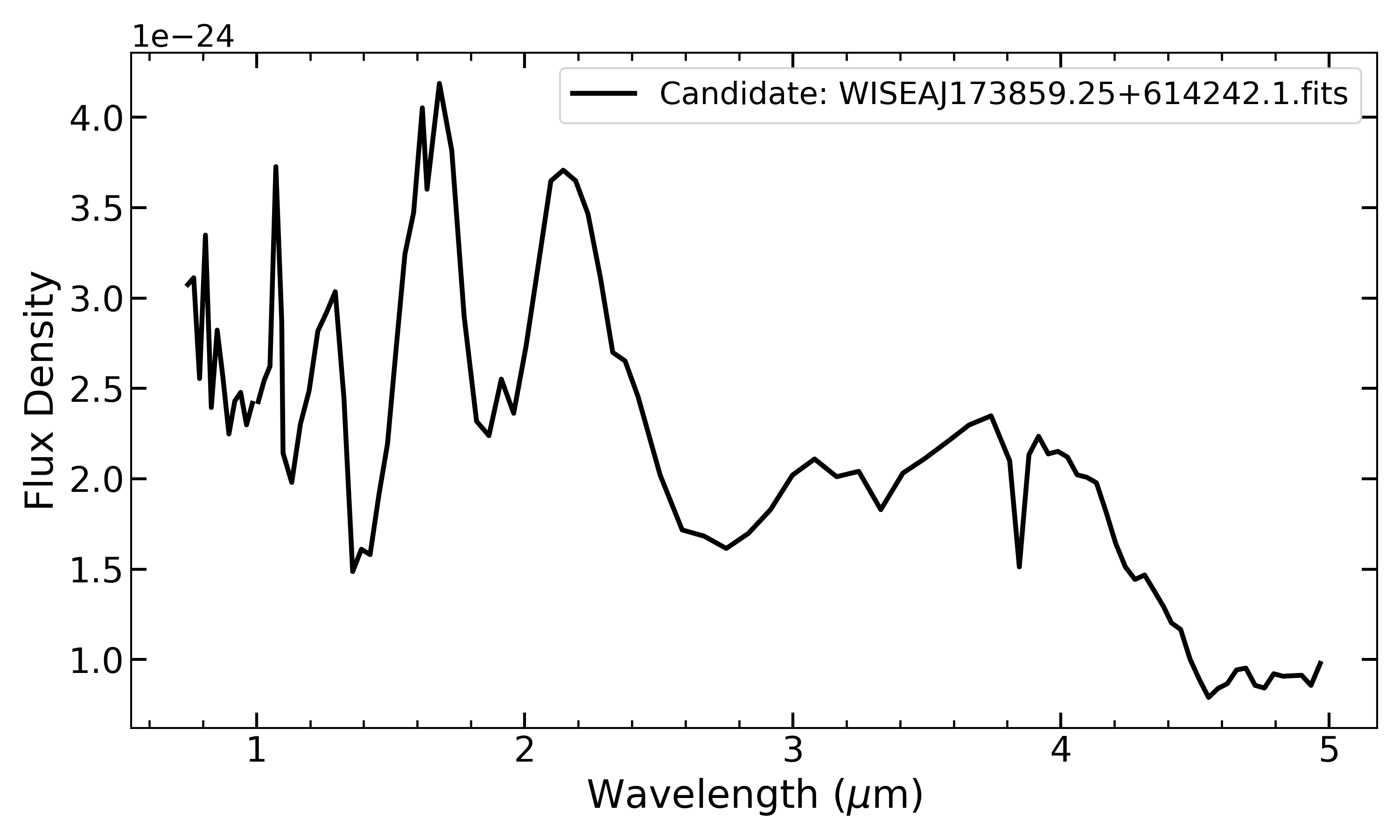}
   \caption{SPHEREx spectra of known URLD candidates recovered in {\it{Euclid}} DR1.   }
   \label{fig_redLdwarfs:SPHEREX_spectra_known}
\end{figure*}
%

%
%%%%%%%%%%%%%%%%%%%%%%%%%%%%%%%%%%%%%%%%%%%%%
%%%%% Figure: SPHEREx spec NEW URLDs %%%%%
%%%%%%%%%%%%%%%%%%%%%%%%%%%%%%%%%%%%%%%%%%%%%
%
\begin{figure*}[!h]
   \centering
   \includegraphics[width=0.43\linewidth, angle=0]{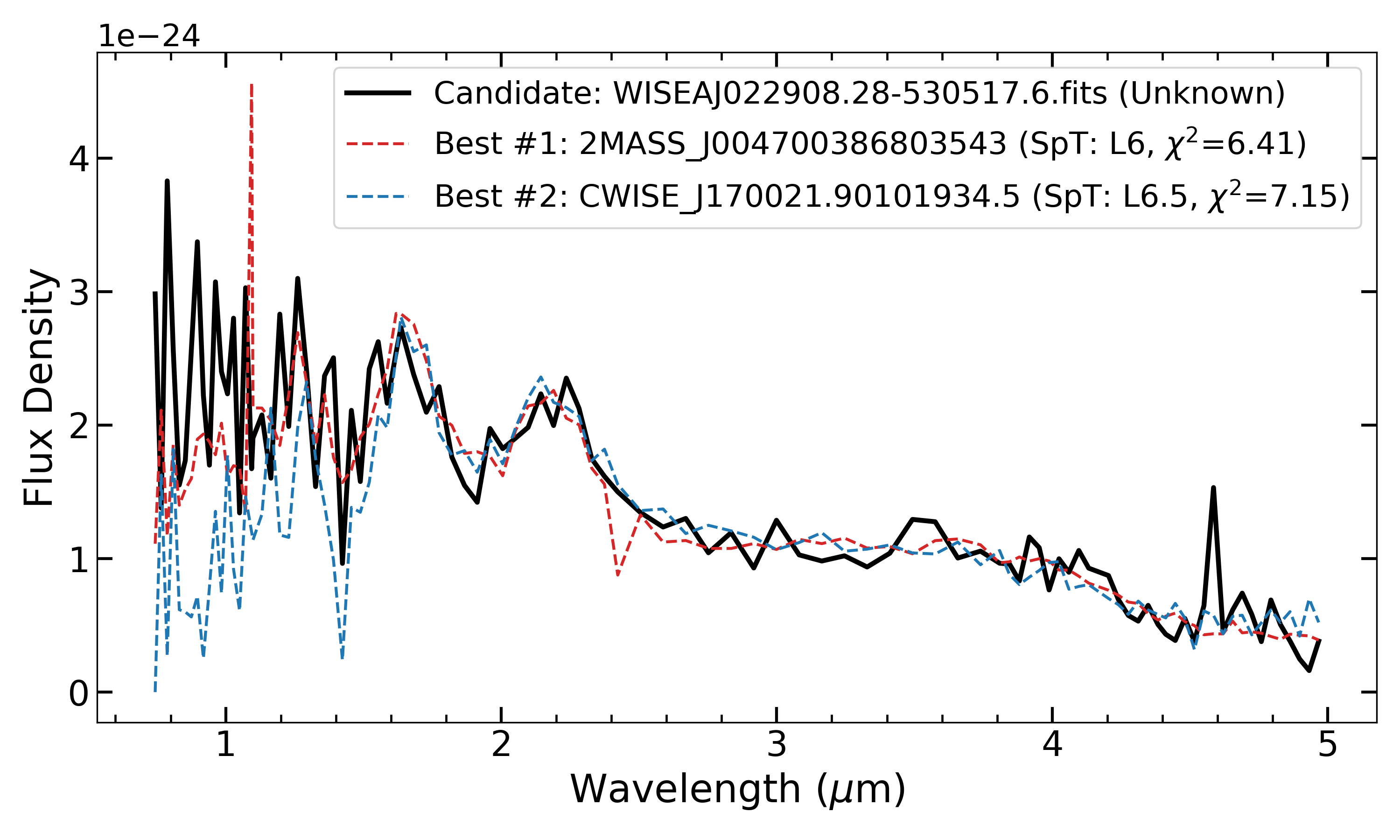}
   \includegraphics[width=0.43\linewidth, angle=0]{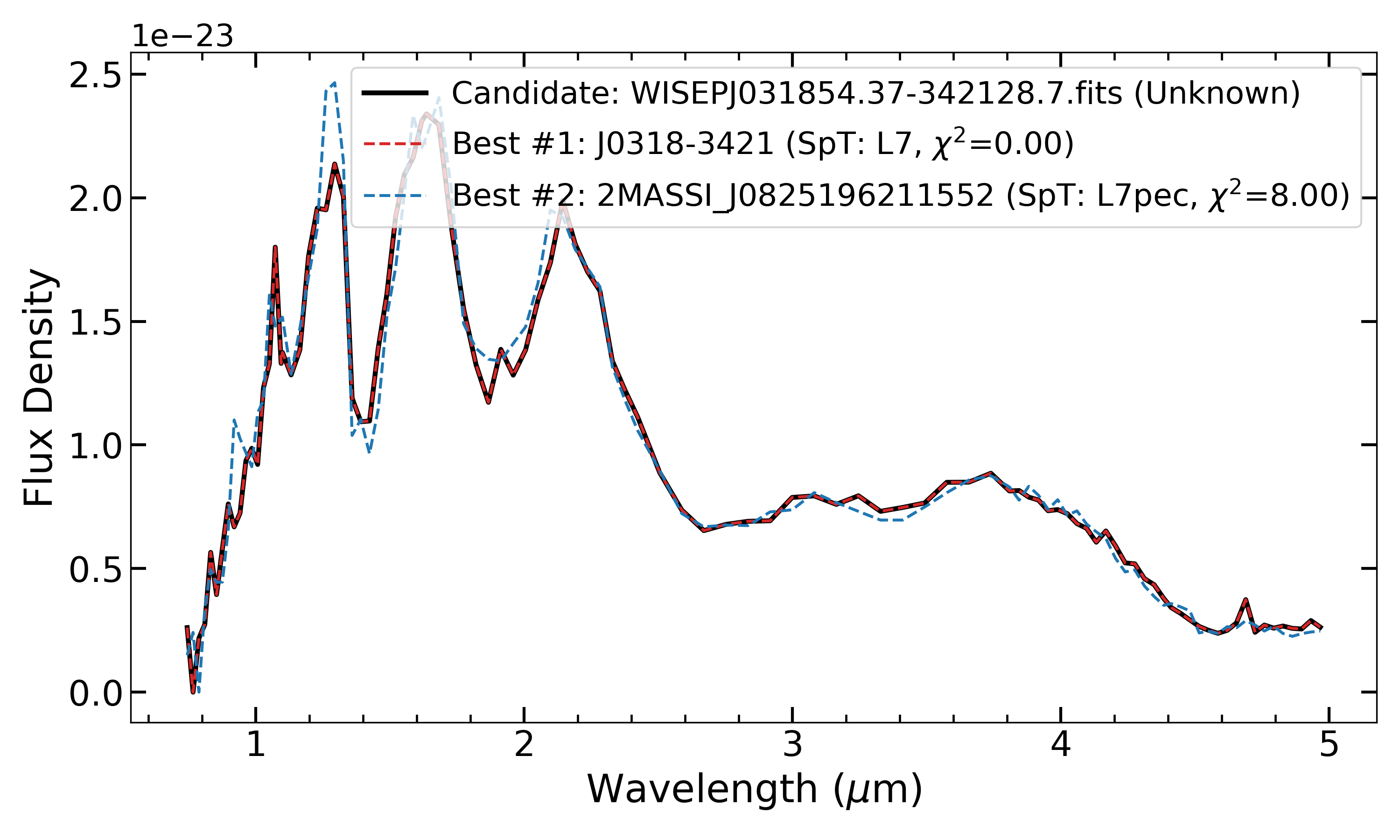}
   \includegraphics[width=0.43\linewidth, angle=0]{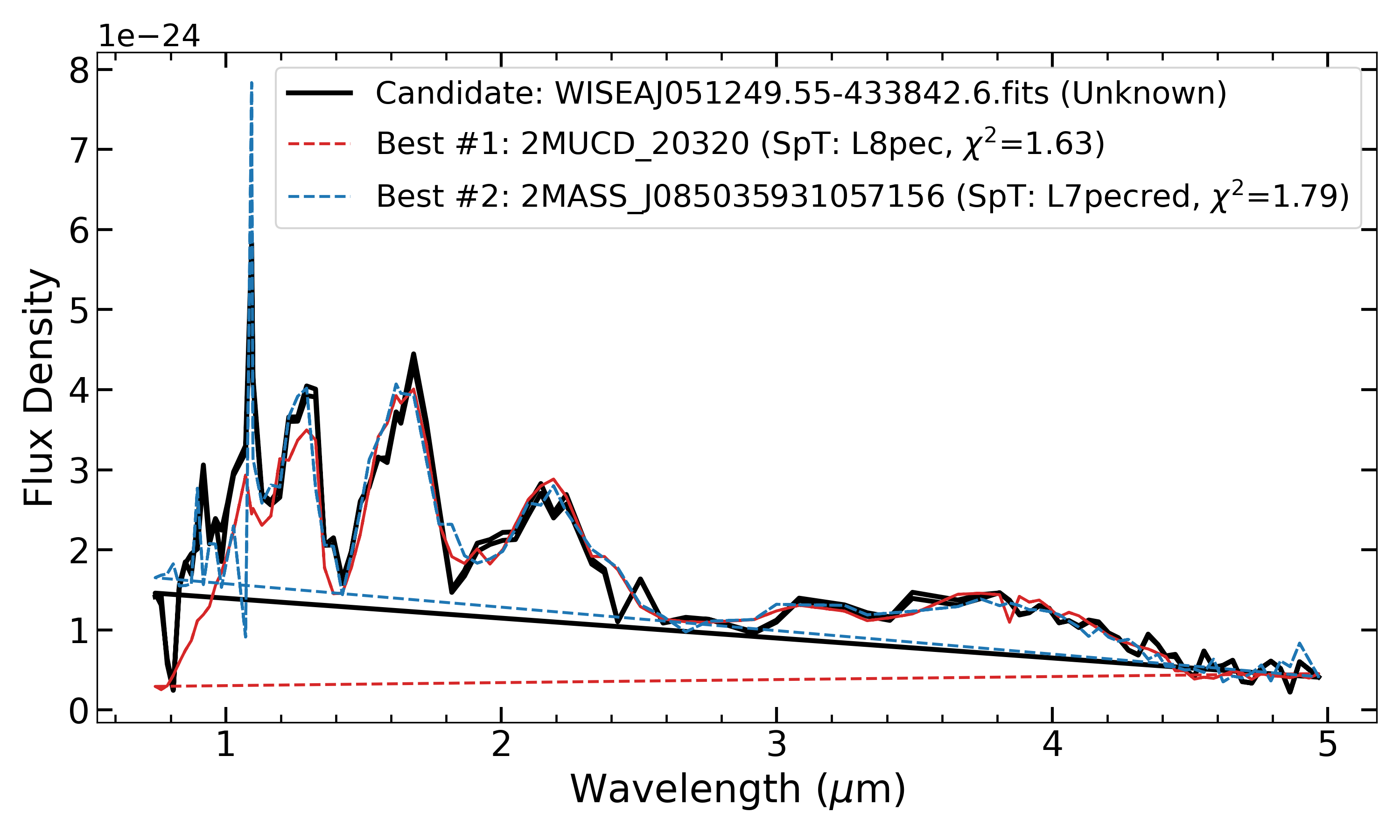}
   \includegraphics[width=0.43\linewidth, angle=0]{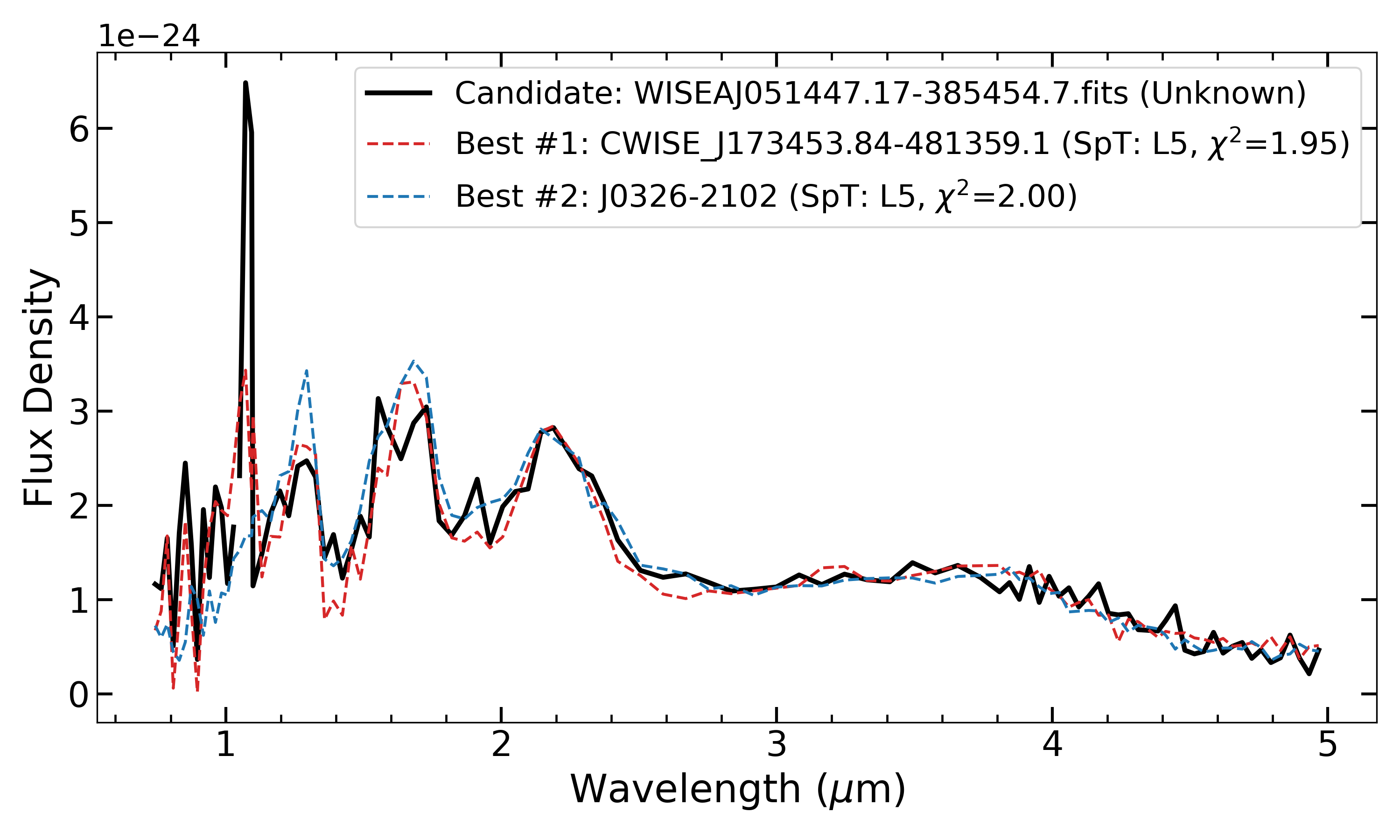}
   \includegraphics[width=0.43\linewidth, angle=0]{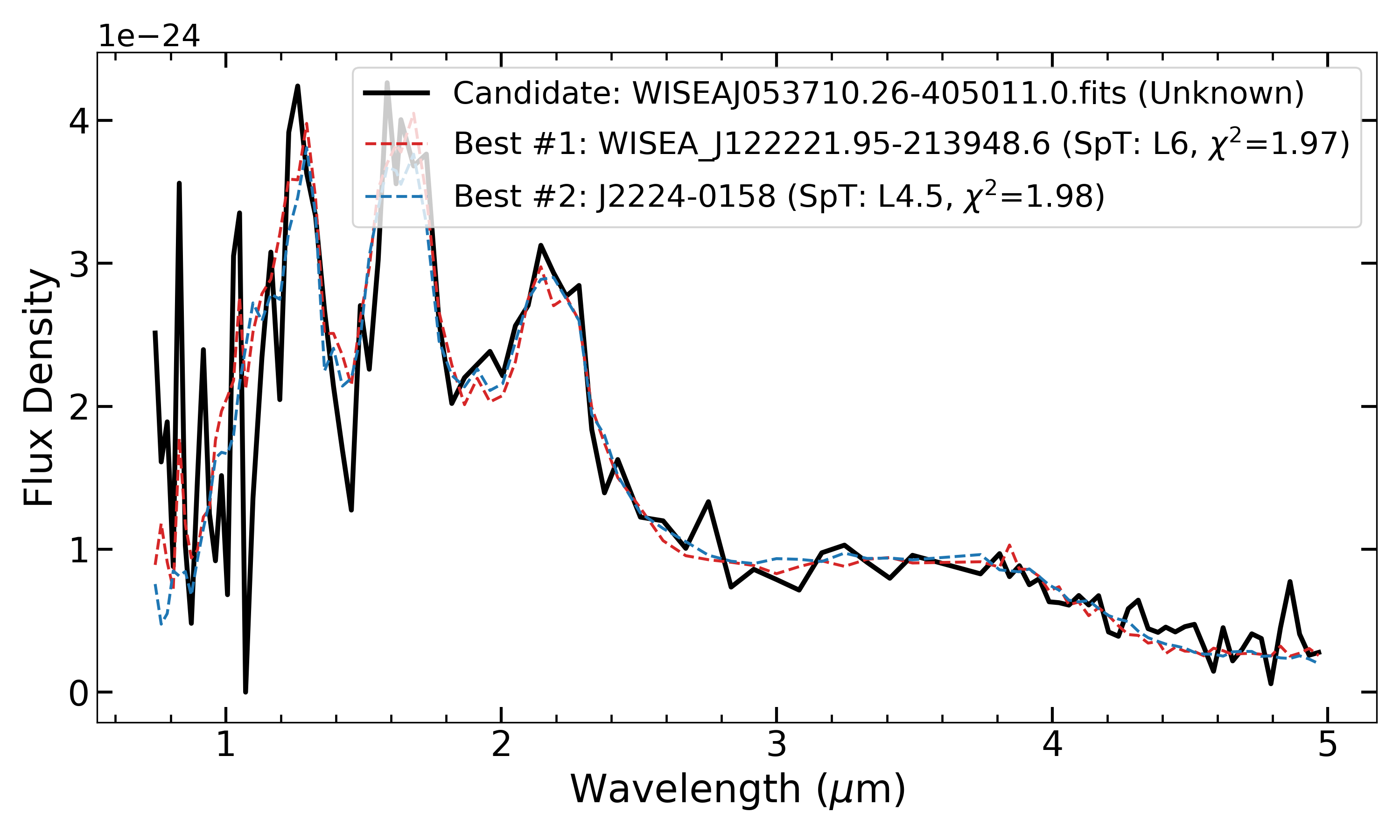}
   \caption{SPHEREx spectra of new URLD candidates identified in {\it{Euclid}} DR1. Best two fits are indicated and come from the sample of known ultracool dwarfs with SPHEREx data \citep{gagne26b}
   }
   \label{fig_redLdwarfs:SPHEREX_spectra_new}
\end{figure*}

\end{appendix}

%%%% End of aa.dem
\end{document}